\documentclass[a4paper,11pt]{article}

\usepackage{jheppub} 
\usepackage[T1]{fontenc}
\usepackage{comment}
\usepackage{pdflscape}
\usepackage{array}
\usepackage{pgfplots}
\usepackage{float}
\usepackage{tabularray}
\usepackage{booktabs}
\usepackage{graphicx}
\pgfplotsset{width=10cm,compat=1.9}
\usepgfplotslibrary{fillbetween}

\newcommand{\be}{\begin{equation}}
\newcommand{\ee}{\end{equation}}
\newcommand{\bea}{\begin{eqnarray}}
\newcommand{\eea}{\end{eqnarray}}
\numberwithin{equation}{section}

\definecolor{darkgreen}{rgb}{0, 0.6, 0}

\newcommand{\dd}{{\text{d}}}
\newcommand{\M}{M_{\rm P}}
\newcommand{\eff}{\text{eff}}
\newcommand{\kin}{\text{kin}}
\newcommand{\fp}{\text{FP}}
\newcommand{\eq}{\text{eq}}
\newcommand{\DE}{\text{DE}}
\newcommand{\lcdm}{$\Lambda$CDM~}
\newcommand{\mc}{\mathcal}

\title{Multifield dark energy: Interplay between curved field space and curved spacetime}

\author[a]{Diego Gallego,}
\author[b]{J. Bayron Orjuela-Quintana}

\affiliation[a]{Grupo de F\'isica de Altas Energ\'ias\\
	Escuela de F\'isica, Universidad Pedag\'ogica y Tecnol\'ogica de Colombia UPTC,\\ Avenida Central de Norte, Tunja, Colombia.}
\affiliation[b]{Departamento de F\'isica, Universidad del Valle, Ciudad Universitaria Mel\'endez, 760032, Cali, Colombia}

\emailAdd{diego.gallego@uptc.edu.co}
\emailAdd{john.orjuela@correounivalle.edu.co}

\abstract{Exponential quintessence models motivated by string compactifications naturally involve both a dilatonic scalar and its axionic partner evolving on a curved field space, while spatial curvature enlarges the cosmological phase space and may affect late-time dynamics. We perform a systematic analysis of the minimal two-field exponential system in a curved FLRW background including radiation and matter components, combining a complete dynamical systems classification with analytical approximations and numerical integration. In the scalar-dominated limit, nongeodesic trajectories can sustain accelerated expansion on steep potentials, and in curved universes a scaling–curvature fixed point can in principle soften the requirements for acceleration. However, we show that these mechanisms arise in distinct invariant manifolds and cannot be simultaneously realized in the presence of a background fluid: no nongeodesic scaling fixed point exists within an open region of parameter space. As a consequence, in the observationally viable thawing regime the axion does not track the background fluid and spatial curvature becomes dynamically subdominant, leading to an effectively single-field evolution. We further identify a degeneracy between curvature effects and scalar-field dynamics that limits their independent impact on late-time cosmology. Confronting the model with current cosmological background data (Planck 2018 distance priors, Pantheon+, BAO, and cosmic chronometers), we obtain an upper bound $\lambda \lesssim 0.75$ (95\% C.L.) on the potential slope. Our results demonstrate that even in the minimal multifield setup with spatial curvature, sustained late-time acceleration requires a sufficiently flat potential, so that the tension between cosmic acceleration and quantum gravity expectations persists within this framework.}

\begin{document} 

\maketitle

\flushbottom

\section{Introduction}
\label{Sec: Intro}

The origin of the present cosmic acceleration~\cite{SupernovaSearchTeam:1998fmf, SupernovaCosmologyProject:1998vns,WMAP:2012nax, Komatsu:2014ioa, Planck:2018vyg} remains one of the central open problems in cosmology. While the $\Lambda$-Cold Dark Matter (\lcdm) model provides an excellent phenomenological description of current observations, its theoretical interpretation in terms of a small cosmological constant~\cite{Peebles:2002gy, Padmanabhan:2002ji} raises well-known conceptual challenges~\cite{Martin:2012bt}. These difficulties, together with persistent observational tensions~\cite{Riess:2016jrr, Riess:2020fzl, Verde:2019ivm, Schoneberg:2021qvd, Perivolaropoulos:2021jda} and the prospect of forthcoming high-precision surveys (e.g., Refs.~\cite{DESI:2016fyo, Amendola:2016saw, Gebhardt:2021vfo}), continue to motivate the exploration of dynamical alternatives to vacuum energy~\cite{DESI:2025fii, DESI:2025wyn}.

Among such alternatives, models based on dynamical scalar fields~\cite{Wetterich:1987fm, Peebles:1987ek, Wetterich:1994bg}, collectively known as \textit{quintessence}~\cite{Caldwell:1997ii, Zlatev:1998tr} (see Refs.~\cite{Copeland:2006wr, Bamba:2012cp} for reviews), provide a well-motivated framework. Within this class, \textit{exponential quintessence}, characterized by a potential of the scalar field $\phi$ given by $V(\phi)\sim e^{-\lambda \phi}$ with $\lambda$ a constant, constitutes a particularly simple and well-studied setup, where an accelerated expansion epoch can be realized for sufficiently small slopes, i.e., $\lambda<\sqrt{2}$~\cite{Ratra:1987rm, Wetterich:1994bg, Copeland:1997et}. \footnote{See also, e.g., Ref.~\cite{Elizalde:2004mq} for alternative realizations involving exponential potentials in scalar-field dark energy models.} From a string-theoretic perspective, exponential potentials arise naturally from dimensional reduction and moduli dynamics. In these, however, the scalar modulus is generically accompanied by its axionic partner, implying that single-field quintessence should be regarded as a truncation of a multifield system evolving on a curved field space. Accordingly, multifield dark energy models have been extensively investigated (see, e.g., Refs.~\cite{Kaloper:2005aj, Kaloper:2008qs, Cicoli:2012tz, Blaback:2013fca, kamionkowskiDarkEnergyString2014b, Christodoulidis:2019jsx, Alexander:2019rsc, McDonough:2022pku, Bernardo:2022ztc, Revello:2023hro}), and are increasingly viewed as a natural framework for constructing viable late-time cosmologies.

The cosmological implications of exponential models augmented by axionic fields have been explored in a variety of contexts, including early studies~\cite{Sonner:2006yn, Catena:2007jf} and more recent analyses~\cite{Brown:2017osf, vandeBruck:2019vzd, Cicoli:2020cfj, Cicoli:2020noz}, as well as generalizations involving additional vector fields~\cite{Gallego:2024gay}. A distinctive feature of multifield scenarios is the possibility of sustaining accelerated expansion along nongeodesic trajectories\footnote{
Throughout this work, the terms geodesic and nongeodesic refer to whether the field space trajectory exhibits a nonvanishing turning rate. They should not be interpreted as geodesic motion in the strict affine sense, since the scalar potential and Hubble friction generally
induce acceleration along the trajectory in field space.} in curved field space~\cite{Achucarro:2018vey, Akrami:2020zfz}, where turning effects can effectively modify the background dynamics even for relatively steep potentials. These properties make multifield exponential models a particularly interesting arena for exploring the robustness of late-time acceleration beyond the single-field approximation.

A complementary ingredient that can significantly affect late-time dynamics is the geometry of the Universe itself. While spatial flatness is a cornerstone of the standard \lcdm model, spatial curvature remains observationally constrained rather than theoretically fixed. Current observations provide stringent bounds within the minimal curved $\Lambda\text{CDM}$ framework, with the combination of \textit{Planck} CMB data with Baryon Acoustic Oscillations (BAO) yielding a consensus constraint of $\Omega_k = 0.0007 \pm 0.0019$ for the curvature density parameter~\cite{Planck:2018vyg}. This flatness is robustly confirmed by recent DESI results ($\Omega_k = 0.0024 \pm 0.0016$)~\cite{DESI:2024mwx, DESI:2025zgx} and independent galaxy power spectrum analyses~\cite{Vagnozzi:2020rcz}, approaching the ``curvature floor'' of cosmic variance~\cite{Takada:2015mma, Leonard:2016evk}, with complementary late-time probes such as cosmic chronometers and model-independent reconstructions of the expansion history yielding consistent, albeit less precise, determinations of $\Omega_k$~\cite{Vagnozzi:2020dfn, Dhawan:2021mel, Jiang:2024xnu}.\footnote{A nuanced picture remains, however. \textit{Planck} spectra alone exhibit a persistent preference for a closed universe ($\Omega_k \approx -0.04$)~\cite{DiValentino:2019qzk, Handley:2019tkm}, a tension that is alleviated by CMB lensing~\cite{SPT:2019fqo} but highlights the dataset dependence of curvature constraints. Furthermore, independent measurements from ACT and SPT, while close to flatness, exhibit slight central-value deviations~\cite{AtacamaCosmologyTelescope:2025nti, SPT-3G:2025bzu}.}

However, this stringent picture changes once the minimal framework is extended to include dynamical dark energy. The well-known correlation between $w(z)$ and $\Omega_k$ leads to enlarged uncertainties in both sectors~\cite{Dossett:2012kd,Gonzalez:2021ojp,Yang:2022kho, deCruzPerez:2024shj, DESI:2025fii}. In particular, studies highlighted that even small curvatures, $|\Omega_k| \sim 0.01$, can induce dramatic variations (of order 50\%-100\%) in the reconstructed Equation of State (EoS) parameter $w(z)$ at redshifts $z \gtrsim 0.9$, revealing a potentially misleading degeneracy~\cite{Clarkson:2007bc}. This issue persists in full parameter inference: Markov Chain Monte Carlo (MCMC) analyses demonstrate that without restrictive assumptions on late-time dynamics, reliable constraints on curvature degrade significantly, typically relaxing to $|\Omega_k| \lesssim 0.03$~\cite{Gong:2007wx, Virey:2008nu}.\footnote{A closely related degeneracy between spatial curvature and neutrino masses has been discussed in the literature (see e.g.~\cite{Chen:2016eyp} and also \cite{Vagnozzi:2017ovm}).} This motivates the persistent inclusion of curvature in any detailed study of dynamical dark energy.

The interplay between spatial curvature and exponential quintessence was already investigated in early dynamical analyses~\cite{Halliwell:1986ja, vandenHoogen:1999qq} and revisited in modern form in Ref.~\cite{Gosenca:2015qha}. More recently, studies have explored whether curvature alone can enlarge the viable parameter space of single-field exponential models~\cite{Andriot:2023wvg, Andriot:2024jsh}.\footnote{A further motivation for including curvature in string-inspired scenarios stems from early predictions that string theory favours a negatively curved, open universe via Coleman–De Luccia tunneling. However, this conclusion was shown to be nonrobust, as it is highly sensitive to landscape properties \cite{Buniy:2006ed}. Moreover, Lorentzian treatments indicate that closed universes with positive curvature are equally probable \cite{Cespedes:2020xpn}. Explicit models further demonstrate that semiclassical tunneling can generate inflating regions with residual positive curvature \cite{Horn:2017kmv}, while future measurements of $|\Omega_k| \gtrsim 10^{-4}$ may constrain specific eternal inflation scenarios without falsifying the landscape paradigm \cite{Guth:2012ww, Kleban:2012ph}.} Although analytical considerations suggest a mild extension of the allowed $\lambda$ range, statistical analyses typically recover small values for both $\lambda$ and $\Omega_k$~\cite{Bhattacharya:2024hep, Alestas:2024gxe}, even when dilatonic couplings are included~\cite{Patil:2024mno}. A common feature of these reassessments, however, is that they restrict the dark energy sector to a single scalar field, neglecting the axionic partner expected in UV-complete constructions.

This question becomes especially sharp in the context of ultraviolet completion arguments, commonly formulated through the \textit{swampland conjectures}~\cite{Vafa:2005ui, Ooguri:2006in, Ooguri:2016pdq, Brennan:2017rbf, Palti:2019pca, VanRiet:2023pnx}. These conjectures disfavour parametrically flat scalar potentials and therefore challenge the standard slow-roll realization of quintessence. In this perspective, steep exponential potentials arise naturally, raising the question of whether the combined effects of multifield dynamics and spatial curvature can sustain a viable accelerated phase without resorting to unnaturally small slopes.

In this work, we perform a comprehensive analysis of the minimal fully fledged setup: a two-field exponential system comprising a scalar modulus and its axionic partner evolving on a curved field space, embedded in a curved, homogeneous and isotropic spacetime with radiation and matter components. Our goal is to investigate, within a unified dynamical and statistical framework, the extent to which nongeodesic motion and spatial curvature can modify the late-time cosmological evolution and potentially enlarge the viable parameter space of exponential quintessence.

We determine the global phase-space structure of the system, identify its invariant manifolds and fixed points, and clarify how curvature-induced scaling solutions and nongeodesic trajectories interplay in realistic cosmological histories. In particular, we show that although these mechanisms arise separately in distinct dynamical regimes, their simultaneous realization in the presence of a background fluid is highly constrained and occurs only on the bifurcation surface of the system. As a consequence, multifield effects remain subdominant during the present era, and the phenomenologically viable region of parameter space is more restricted than naïvely expected.

We finally confront the model with current cosmological data, mapping the degeneracies between curvature, the dark energy EoS, and field space dynamics, and assessing the implications for steep exponential potentials motivated by ultraviolet considerations.

The remainder of this paper is organized as follows. In Sec.~\ref{Sec: Setup}, we introduce the cosmological dynamics for the two-field system on top of a background universe filled with a generic barotropic fluid. Then, in Sec.~\ref{Sec: Single Barotropic}, we define the dimensionless variables necessary for the autonomous system analysis, and presents the complete classification of the fixed points and their stability properties. Section~\ref{Sec: Curved Multifield} focuses on the novel features introduced by the nonvanishing spatial curvature and the role of the axion field in the cosmological dynamics. In Sec.~\ref{Sec: Constraints}, we perform a numerical scan of the parameter space to verify the analytical findings and confront the model with basic observational viability criteria. Finally, Sec.~\ref{Sec: Conclusions} summarizes our findings and discusses their implications for Swampland phenomenology.

\section{General setup: Two scalars and background fluids}	
\label{Sec: Setup}

We consider a general, homogeneous, and isotropic Friedman-Lema\^itre-Robertson-Walker (FLRW) spacetime, which in spherical coordinates $\{t, r, \theta, \varphi\}$ reads:
\begin{equation}
	\dd s^2 = - \dd t^2 + a^2(t) \left(\frac{\dd r^2}{1-k r^2} + r^2 \dd \Omega^2 \right),
\end{equation}
where $a(t)$ is the cosmic time-dependent scale factor, $\dd \Omega^2$ is the differential solid angle on the unit 2-sphere, and $k=0, \pm 1$ encodes the spatial curvature corresponding to flat ($k = 0$), closed ($+$), and open ($-$) geometries. 	

A minimal field content motivated by supersymmetric compactifications consists of a scalar modulus $\phi_1$ and its axionic partner $\phi_2$, evolving on a curved field space and interacting through a potential depending solely on $\phi_1$. The corresponding Lagrangian reads
\begin{equation} 
\label{eq:lagrangianscalar alone}
	{\cal L}= -\frac12 \partial_\mu \phi_1 \partial^\mu \phi_1 - \frac12 f^2(\phi_1) \partial_\mu \phi_2 \partial^\mu \phi_2 - V(\phi_1)\,,
\end{equation}
which captures the essential geometric structure of a modulus–axion system while retaining the minimal set of interactions required to study the interplay between field space curvature and cosmological dynamics.

The corresponding equations of motion for homogeneous fields in
the FLRW background are
\begin{equation}
\label{Eq: phi eom}
		\ddot{\phi}_1 + 3 H \dot{\phi}_1 + V_{\phi_1} - f f_{\phi_1} \dot{\phi}_2^2 = 0\, \qquad 
        \ddot{\phi}_2 + 3 H \dot{\phi} _2 + 2 \frac{f_{\phi_1}}{f} \dot{\phi}_1 \dot{\phi}_2 = 0\,,
\end{equation}
where the overdot denotes time derivative, $H \equiv \dot{a}/a$ is the Hubble rate function, and the subscript $\phi_1$ denotes derivatives with respect to this field. Spatial curvature enters the dynamics only indirectly through the Hubble rate in the gravitational sector, and does not modify the field space geometry nor the intrinsic scalar equations of motion. More precisely, curvature appears explicitly in Einstein equations, also known as, Friedman equations,
\begin{equation}
\label{eq: Friedman FLRW}
	3 \M^2 H^2 = \rho_\phi + \rho_\alpha - 3\M^2 \frac{k}{a(t)^2}\,,  \qquad 
    -2 \M^2 \dot H = \rho_\phi + p_\phi + \rho_\alpha + p_\alpha - 2\M^2 \frac{k}{a(t)^2}\,, 
\end{equation}
where $\M$ is the reduced Planck mass. The density and pressure of the scalar sector are:
\begin{equation}
	\rho_\phi = \frac{1}{2} \dot{\phi}_1^2 + \frac{1}{2} f^2(\phi_1) \dot{\phi}_2^2 + V\,, \qquad 
    p_\phi = \frac{1}{2} \dot{\phi}_1^2 + \frac{1}{2} f^2(\phi_1) \dot{\phi}_2^2 - V\,,
\end{equation}
defining the field EoS parameter as $ w_\phi \equiv p_\phi/\rho_\phi$. Additionally, we consider the presence of a general barotropic fluid, with density $\rho_\alpha$ and pressure $p_\alpha \equiv (\alpha-1)\rho_\alpha$, obeying the continuity equation,
\begin{equation} 
\label{eq:genbarocont}
	\dot \rho_\alpha + 3\alpha H \rho_\alpha = 0\,.
\end{equation}
The constant $\alpha$ is related to the EoS parameter of the fluid by $\alpha = 1 + w_\alpha$. 

It is convenient to treat the curvature term as an effective fluid component with density $\rho_k \equiv -3\M^2 k/a(t)^2$ and pressure $p_k \equiv \M^2 k/a(t)^2$. This corresponds to an EoS $w_k = -1/3$ or $\alpha_k = 2/3$. For typical radiation and matter fluids we have $w_r = 1/3$ and $w_m = 0$, which lead to $\alpha_r = 4/3$ and $\alpha_m = 1$, respectively.

Friedman equations can be conveniently expressed in terms of the dimensionless density parameters $\Omega_i \equiv \rho_i / (3\M^2 H^2)$ as
\begin{equation}
\label{eq: Friedman constraints}
	1 = \Omega_\phi + \sum_{i} \Omega_i\,, 
\end{equation}
with the index $i$ running over radiation, matter, and curvature, such that the Friedman constraint explicitly includes all nonscalar contributions to the energy budget. The dynamics of the expansion are characterized by the deceleration parameter $q \equiv -a\ddot{a}/\dot{a}^2$, which relates to the abundances via
\begin{equation}
    1 + q = 3 \, \Omega _{\phi,\text{kin}} + \frac{3}{2} \sum_{i} \alpha_i \, \Omega_i, \quad \Omega_{\phi, \text{kin}}\equiv\frac{\dot\phi_1^2+f(\phi_1)^2\dot\phi_2^2}{6H^2 \M^2}\,.
\end{equation} 
Neglecting radiation, the condition for accelerated expansion ($q < 0$) becomes
\begin{equation}
\label{Eq: Acc constraint I}
	6 \Omega_{\phi,\kin} + 3\Omega_m < 2(1 - \Omega_k)\,.
\end{equation}
This relation illustrates how spatial curvature can effectively shift the balance between kinetic and potential contributions required for acceleration. In particular, a closed universe ($\Omega_k < 0$) enlarges the allowed kinetic fraction compatible with $q<0$, suggesting that curvature may, at the level of background kinematics, partially relax the conditions required for acceleration. Conversely, in an open universe ($\Omega_k > 0$), the kinetic energy must be further suppressed to achieve acceleration.
	
In the analysis that follows, we focus on the specific class of models defined by exponential potentials and couplings,
\begin{equation}
\label{Eq: Exponential Params}
    V(\phi_1) = V_0 e^{-\lambda \phi_1/\M}\,, \qquad f(\phi_1) = f_0 e^{-\nu \phi_1/\M}\,,
\end{equation}
where $\lambda$ and $\nu$ are dimensionless constants characterizing the slopes of the functions. Such exponential forms naturally emerge in dimensional reduction scenarios, where canonically normalized moduli fields controlling geometric or string-theoretic couplings acquire runaway potentials, and their axionic partners inherit field-dependent kinetic terms dictated by the underlying Kähler structure (see, e.g.,~\cite{Gallego:2024gay}). The following then can be understood as an extension and further detailed study to the work done in Ref.~\cite{Cicoli:2020cfj}.

\section{Dynamical Systems Analysis}
\label{Sec: Single Barotropic}

Before specializing to particular cosmological backgrounds, it is useful to analyze the two-field model in the presence of a generic barotropic component. This allows us to characterize the fixed-point structure and stability properties of the multifield system in a unified manner. The results can then be specialized to particular backgrounds, including matter, radiation, and spatial curvature.

\subsection{Fixed points and their stability}

In order to analyze the asymptotic behavior of the system, we follow the standard procedure of Ref.~\cite{Copeland:1997et} and introduce the following dimensionless variables:
\begin{equation} 
\label{eq:dynamicalvariables}
    x_1 \equiv \frac{\dot \phi_1}{\sqrt{6} H \M}\,, \quad 
    x_2 \equiv \frac{f(\phi_1)\dot \phi_2}{\sqrt{6} H \M}\,, \quad
	y \equiv \frac{\sqrt{V}}{\sqrt{3}H \M}\,, \quad 
    \Omega_\alpha \equiv \frac{\rho_\alpha}{3 \M^2 H^2}\,,
\end{equation}
related to the scalar fields kinetic energy, the potential energy, and a single general barotropic fluid, respectively. In terms of these variables, Eq.~\eqref{eq: Friedman constraints} becomes a constraint from which we can reduce the dimensionality of the system by one degree of freedom. We choose $\Omega_\alpha$ as the dependent variable,
\begin{equation}
    \Omega_\alpha = 1 - x_1^2 - x_2^2 - y^2\,,
\end{equation}
where $\Omega_\alpha=0$ forms an invariant manifold. For matter and radiation we must further impose the physical constraint $0 \le \Omega_\alpha$ such that the physical phase space corresponds to the unit 3-ball $x_1^2+x_2^2+y^2\le1$. In terms of these variables, the deceleration parameter and the effective EoS read
\begin{equation}
    q = \frac{1}{2} \left[3 \alpha - 2 + 3 (2 - \alpha) \left( x_1^2 + x_2^2 \right) - 3 \alpha y^2 \right]
\end{equation}
and
\begin{equation}
    w_\eff \equiv \frac{1}{3} (2q - 1) = \alpha - 1 + (2 - \alpha) \left( x_1^2 + x_2^2 \right) - \alpha y^2\,.
\end{equation}
Consequently, the condition for accelerated expansion ($q<0$) becomes the constraint derived in Eq.~\eqref{Eq: Acc constraint I}, which in these variables takes the form,
\begin{equation} 
\label{eq:acc-constraint axio-dil baro}
    3\alpha y^2 > 3\alpha-2 + 3 (2 - \alpha) \left(x_1^2 + x_2^2 \right)\,.
\end{equation}

The cosmological equations of motion can be recast as an autonomous dynamical system,
\begin{align}
	x_1' &= (q+1) x_1 + \sqrt{\frac{3}{2}} \left( \lambda y^2 -2 \nu x_2^2 - \sqrt{6} x_1 \right)\,, \label{Eq: x1 Eq} \\
	x_2' &= x_2 \left(q - 2 + \sqrt{6} \nu x_1 \right)\,, \\
	y'   &= y \left( q + 1 - \frac{\sqrt{6}}{2} \lambda x_1 \right)\,, \label{Eq: y Eq}
\end{align}
where the prime denotes the derivative with respect to the number of $e$-folds, $\dd N \equiv H \dd t$. This system is invariant under the transformation,
\begin{equation}
\label{eq:parametricsymmetries}
    (x_1,\lambda,\nu) \to -(x_1,\lambda,\nu)\,, \quad x_2 \to - x_2\,, \quad y \to-y\,,
\end{equation}
thus we can restrict the study to positive $\lambda$, $\nu$, $x_2$ and $y$. The dynamical study of this system was briefly done by Cicoli {\it et al.} in Ref.~\cite{Cicoli:2020cfj}. 

The fixed points of the autonomous system correspond to the locations, in the phase space $\{x_1, x_2, y\}$, where $x_1' = x_2' = y' = 0$. The stability properties at the linear level follow from the eigenvalues of the Jacobian matrix evaluated at each fixed point: negative (positive) real parts correspond to attractors (repellers), while mixed signs signal saddle behavior. Complex eigenvalues indicate spiral trajectories. Null real part eigenvalues require beyond the linear level analysis.
\begin{center}
\begin{table}[t]
\begin{centering}
\begin{tblr}{ c | c | c | c | c | c | c | c }
    \hline[1.2pt]
    \hline
	FP & $x_1$ & $x_2$ & $y$ & $\Omega_\alpha$ & $w_\eff$ & Existence & Acceleration \\
    \hline
    \hline[1.2pt]
	$\fp^{-}_{\kin}$ & $-1$ & 0 & 0 & 0 & 1 & $\forall \lambda, \nu, \alpha$ & No \\
    \hline
	$\fp_\kin^+$ & 1 & 0 & 0 & 0 & 1 & $\forall \lambda, \nu, \alpha$ & No \\
    \hline
	$\fp_{{\cal G}}$ & $\frac{\lambda}{\sqrt{6}}$ & 0 & $\sqrt{1-\frac{\lambda^2}{6}}$ & 0 & $\frac{1}{3} \left(\lambda^2 - 3\right)$ & $\lambda < \sqrt{6}$ & $\lambda < \sqrt{2}$ \\
    \hline
	$\fp_{{\cal NG}}$ & $\frac{\sqrt{6}}{\lambda + 2\nu}$ & $\frac{\sqrt{\lambda^2 + 2\lambda \nu - 6}}{\lambda + 2\nu}$ & $\frac{1}{\sqrt{\frac{\lambda}{2\nu} + 1}}$ & 0 & $\frac{\lambda - 2\nu}{\lambda + 2\nu}$ & $\frac{6 - \lambda^2}{2\lambda} \leq \nu$ & For $\lambda < \nu$ \\
    \hline
	$\fp_\alpha$ & 0 & 0 & 0 & 1 & $\alpha -1$ & $\forall \lambda, \nu, \alpha$ & $\alpha < 2/3$ \\
    \hline
	$\fp_\mc{S}$ & $\frac{\sqrt{\frac{3}{2}} \alpha}{\lambda}$ & 0 & $\frac{\sqrt{\frac{3}{2}} \sqrt{(2 - \alpha)\alpha}}{\lambda}$ & $1-\frac{3 \alpha}{\lambda^2}$ & $\alpha -1$ & $\alpha < 2$ & $\alpha < 2/3$ \\
    \hline[1.2pt]
\end{tblr}
\caption{Fixed points for the system of two scalar fields with exponential scalar potential and kinetic mixing, characterized by parameters $\lambda$ and $\nu$, and a barotropic fluid with EoS parameter $w_\alpha=\alpha-1$. The last column summarises the restriction in Eq.~\eqref{eq:acc-constraint axio-dil baro}.}
\label{tab:FPtwofieldsandbarotropic}
\end{centering}
\end{table}
\end{center}


The obtained fixed points are listed in
Table~\ref{tab:FPtwofieldsandbarotropic}. Four of them involve only the
scalar fields: the \textit{kination} points $\fp_\kin^\pm$, the
\textit{geodesic} point $\fp_\mc{G}$, and the \textit{nongeodesic}
point $\fp_\mc{NG}$. The other two fixed points have contribution from
the barotropic fluid: the $\alpha$-\textit{dominated} point $\fp_\alpha$
and the \textit{scaling} point $\fp_\mc{S}$.

Interestingly, the dynamical interplay between the scalar sector and the
background fluid is mediated primarily through the $x_1$ direction,
since the existence of a genuine scaling solution requires $x_2=0$.
The $x_2$ direction, instead, enters nontrivially only via the
nongeodesic solution $\fp_{\cal NG}$. The latter is characterized by a
nonvanishing transverse component of the field space acceleration,
which sustains a turning trajectory despite the presence of a single
exponential potential. By contrast, the geodesic solution
$\fp_{\cal G}$ evolves without transverse acceleration and therefore
without turning in field space.

\begin{center}
\begin{table}[t]
\begin{centering}
\begin{tblr}{ c | c | p{3.5cm} }
\hline[1.2pt]
\hline
 FP & Eigenvalues $(x_1,x_2,y)$ directions & \centering Nature \\
\hline    
\hline[1.2pt]
	$\fp_\kin^-$ & $\left\{3 (2-\alpha),\, -\sqrt{6} \nu, \, \frac{1}{2} \left(6+\sqrt{6} \lambda\right)\right\}$ & Saddle \\
	\hline
	$\fp_\kin^+$ & $\left\{3 (2-\alpha),\, \sqrt{6} \nu,\, \frac{1}{2} \left(6-\sqrt{6} \lambda\right)\right\}$ & Unstable. $y$ is stable for $\lambda > \sqrt{6}$ \\
    \hline
	$\fp_{\cal G}$ & $\left\{\lambda^2 - 3 \alpha,\, \frac{1}{2} \left(\lambda^2 + 2\lambda \nu - 6\right),\, -3 + \frac{\lambda^2}{2} \right\}$ & Stable for $\lambda \leq \sqrt{3\alpha}$ and $\nu < \frac{6 - \lambda^2}{2\lambda}$ \\
    \hline
	$\fp_{\cal NG}$  & Analytically cumbersome expressions & Stable for $\lambda < \frac{2\alpha \nu}{2 - \alpha}$ \\
    \hline
	$\fp_\alpha$ & $\left\{-\frac{3}{2} (2 - \alpha),\, -\frac{3}{2} (2 - \alpha),\, \frac{3\alpha}{2}\right\}$ & Saddle \\
    \hline
	$\fp_\mc{S}$ & $\left\{-\frac{3}{4} \left(2-\alpha + {\cal A} \right),\, \alpha \left(\frac{3\nu}{\lambda} + \frac{3}{2}\right) - 3,\, -\frac{3}{4} \left(2 - \alpha - {\cal A}\right) \right\}$ & Stable for $\lambda > \sqrt{3\alpha}$ and $\lambda > \frac{2\alpha \nu}{2 - \alpha}$ \\
\hline[1.2pt]
\end{tblr}
\caption{Linear stability analysis based on the eigenvalues of the Jacobian evaluated along the $(x_1,x_2,y)$ directions. Here ${\cal A}=\sqrt{(2-\alpha)(2-9 \alpha + 24 \alpha^2/\lambda ^2)}$, and the classification assumes $\alpha<2$. For the fixed point $\fp_{\cal NG}$, the explicit eigenvalue expressions are analytically cumbersome; nevertheless, its stability region can be determined numerically. We identify its basin of attraction through a parameter-space scan, following the methodology of~\cite{Garcia-Serna:2023xfw,Gallego:2024gay}, and verify that it coincides with the complement of the attraction regions associated with the other stable fixed points.}
\label{tab:stabilitymultifieldandbaro}
\end{centering}
\end{table}
\end{center}
The results of the linear analysis are summarised in Table~\ref{tab:stabilitymultifieldandbaro}: three of the fixed points, $\fp_{\cal G}$, $\fp_{\cal NG}$ and $\fp_\mc{S}$, are potential attractors. These stability properties can be qualitatively understood from the dilution rates $\rho \propto a^{-3(1+w_\eff)}$: kination ($a^{-6}$) in general dilutes fastest, while the relative hierarchy between the barotropic ($a^{-3\alpha}$), geodesic ($a^{-\lambda^2}$), and nongeodesic ($a^{-6\lambda/(\lambda+2\nu)}$) solutions is controlled by the parameter relations shown in Table~\ref{tab:stabilitymultifieldandbaro}.
   
Therefore, in general terms, we expect a cosmological evolution starting from a kination phase,\footnote{More precisely, a kination source phase requires the scalar potential contribution to increase backwards in time slower than the kinetic counterpart. This explains a possible attractor direction revealed for $\fp_\kin^+$, as shown in Ref.~\cite{Andriot:2024sif}.} followed by different scenarios depending on the chosen parameters: 
\begin{itemize}
    \item For $\lambda^2 < 3\alpha$ and small $\nu$, we expect the system to evolve towards $\fp_{\cal G}$ after passing by a barotropic fluid domination and a possible scaling.
    \item For $\nu \gg \lambda$, the system evolves towards $\fp_{\cal NG}$ approaching before the barotropic domination.
    \item For sufficiently large $\lambda$ (and moderate $\nu$), both $\fp_{\cal G}$ and $\fp_{\cal NG}$ dilute faster than the scaling solution and the system evolves towards this situation.
    \item The smaller $\alpha$ is the slower $\rho_\alpha$ dilutes and the larger the attractor region for $\fp_\mc{S}$ is.
\end{itemize}    

\begin{figure}[t!]
\centering
{\includegraphics[width=0.32\textwidth]{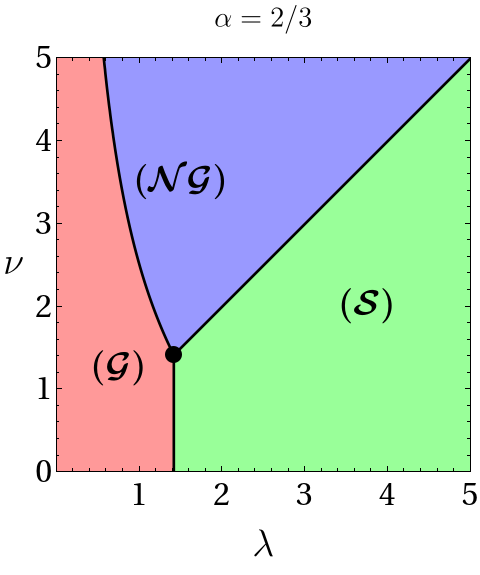} \hfill
\includegraphics[width=0.32\textwidth]{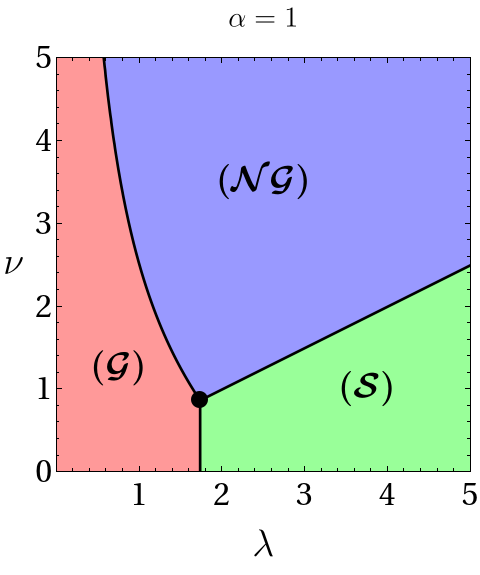} \hfill
\includegraphics[width=0.32\textwidth]{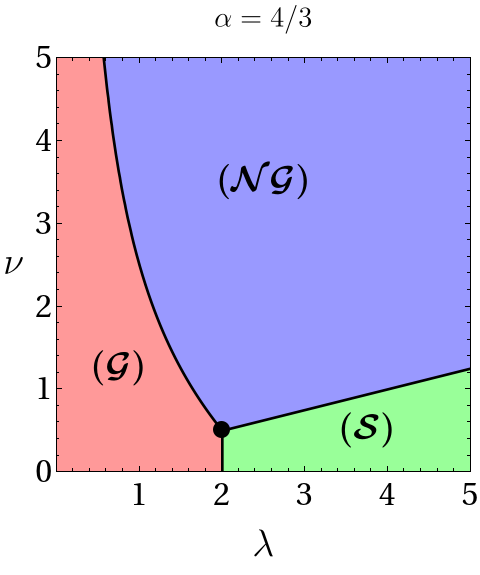}}
\caption{Attractor regions for the geodesic (${\cal G}$), nongeodesic (${\cal NG}$) and scaling (${\cal S}$) fixed points for the common realizations of the barotropic fluid: curvature (left), matter (middle), and radiation (right), respectively. These regions meet at the triple point.}
\label{fig:phasespacemultiandbaro}
\end{figure}

These situations are graphically summarised in the parameter phase spaces shown in Fig.~\ref{fig:phasespacemultiandbaro}, where the three possible attraction regions are separated by the bifurcation curves, \footnote{As an example, the $(\mc{G}-\mc{NG})$ boundary is obtained by setting to zero the eigenvalue associated with perturbations along the $x_2$ direction of the geodesic fixed point $\fp_\mc{NG}$, or equivalently by saturating the second stability condition of $\fp_\mc{NG}$ in Table~\ref{tab:stabilitymultifieldandbaro}, namely \(\lambda^2+2\lambda\nu-6=0\).}
\begin{equation}
    \nu = \frac{6 - \lambda^2}{2\lambda} ~~(\mc{G}-\mc{NG})\,, \quad \nu = \lambda \frac{2-\alpha}{2\alpha} ~~(\mc{NG}-\mc{S})\,, \quad \lambda = \sqrt{3\alpha} ~~(\mc{S}-\mc{G})\,.
\end{equation}

\begin{figure}[t]
\centering 
\rule{\textwidth}{0.5pt} \\ 
\textbf{Curvature fluid: $\alpha = 2/3$} \\ 
{\includegraphics[width=0.31\textwidth]{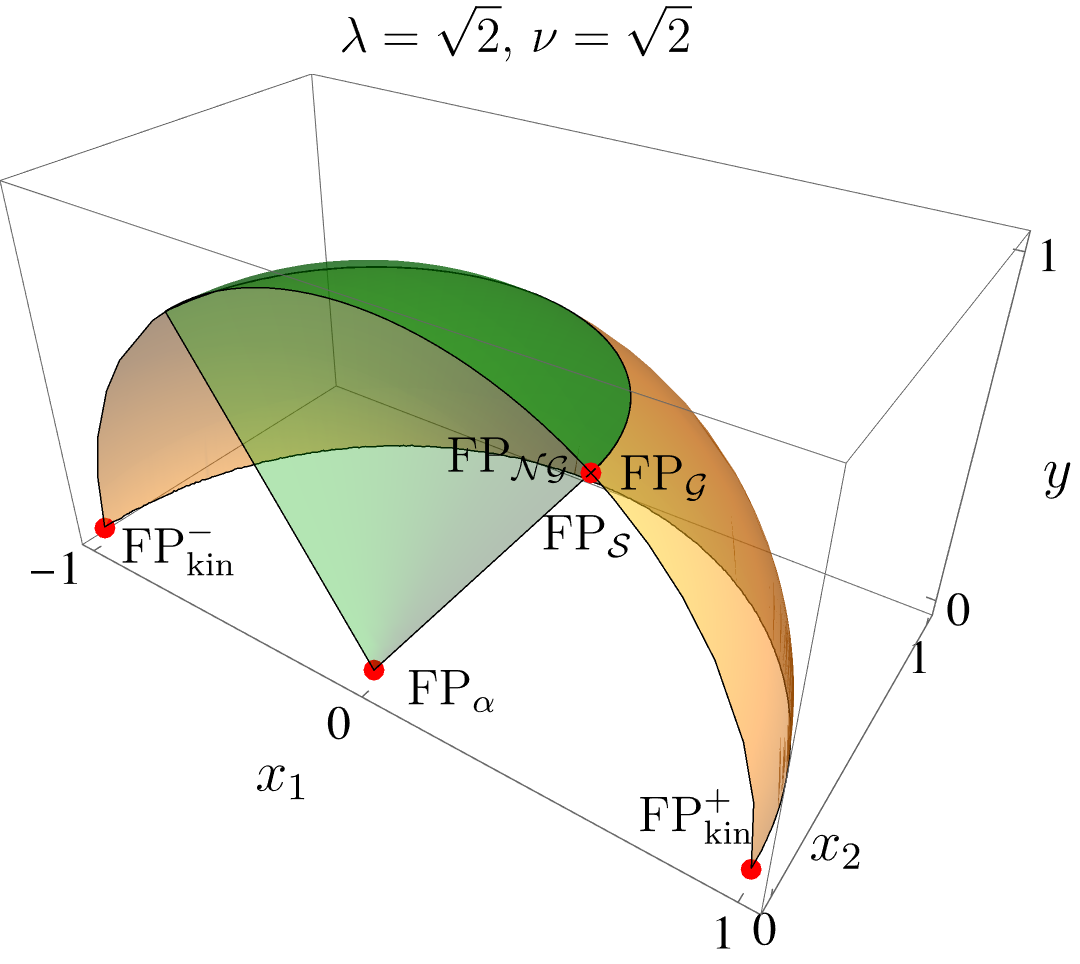} \hfill 
\includegraphics[width=0.31\textwidth]{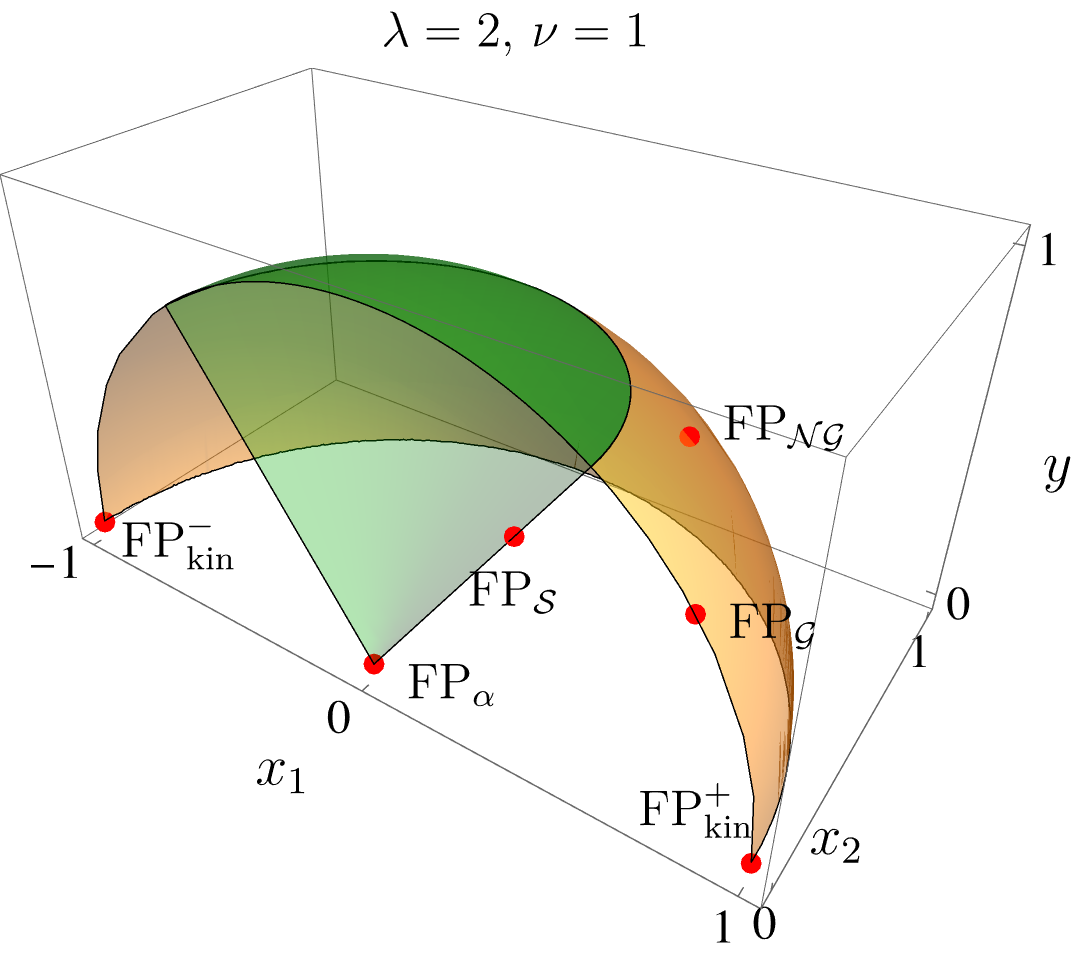} \hfill
\includegraphics[width=0.31\textwidth]{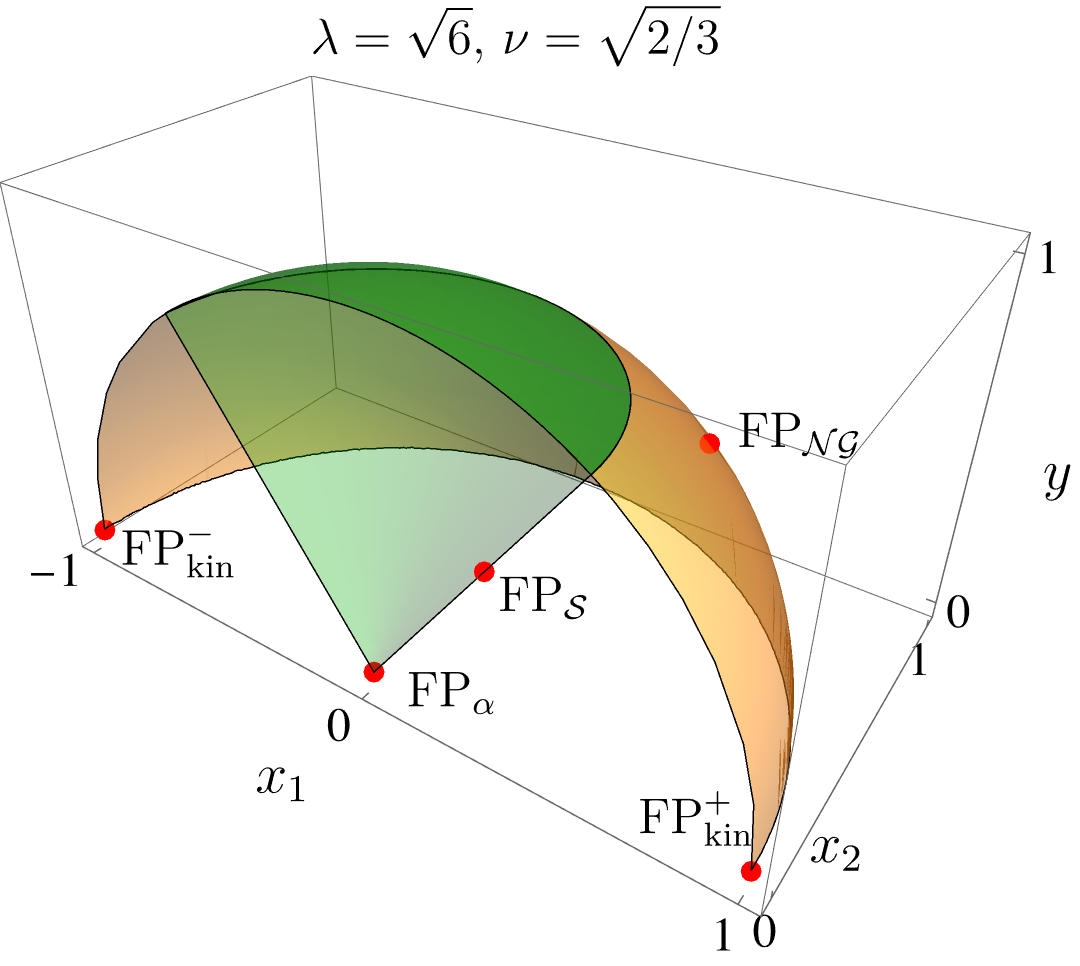}}
\centering 
\rule{\textwidth}{0.5pt} \\ 
\textbf{Matter fluid: $\alpha = 1$} \\ 
{\includegraphics[width=0.31\textwidth]{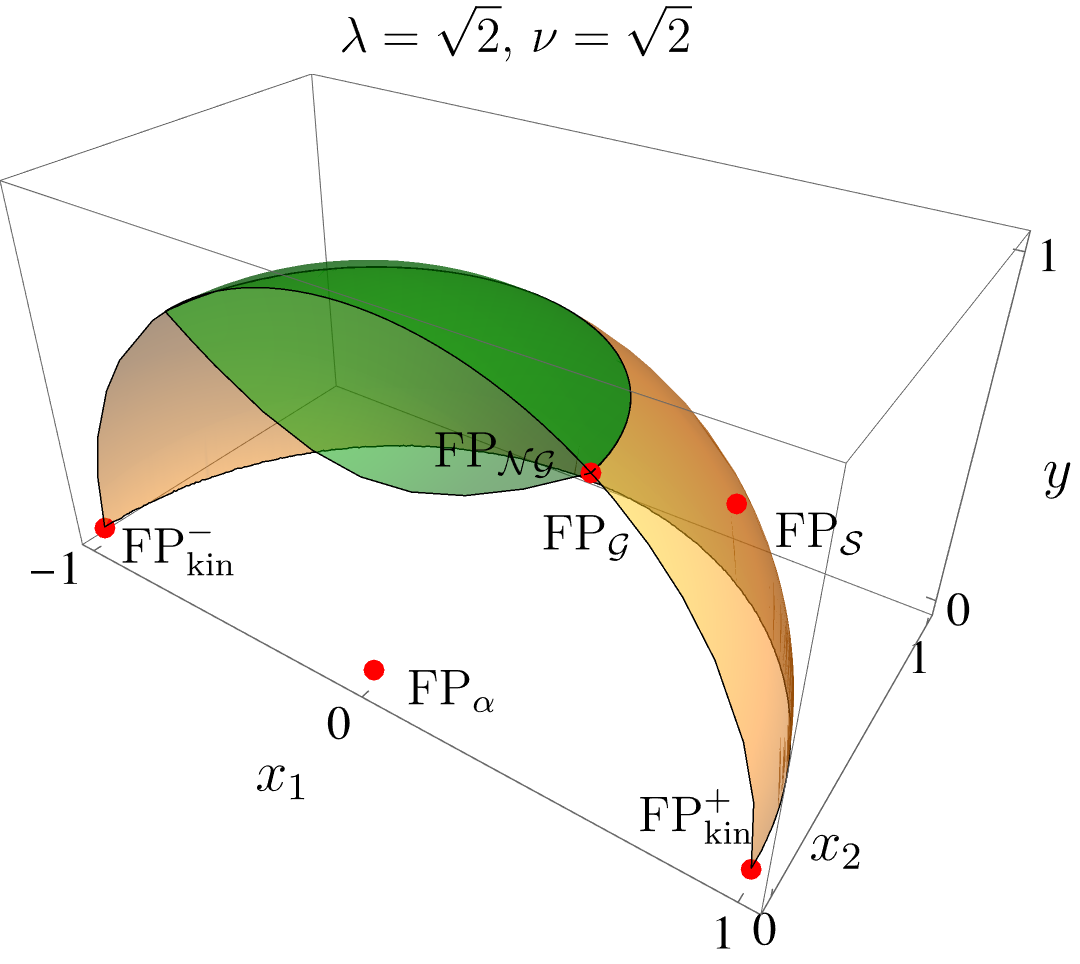} \qquad 
\includegraphics[width=0.31\textwidth]{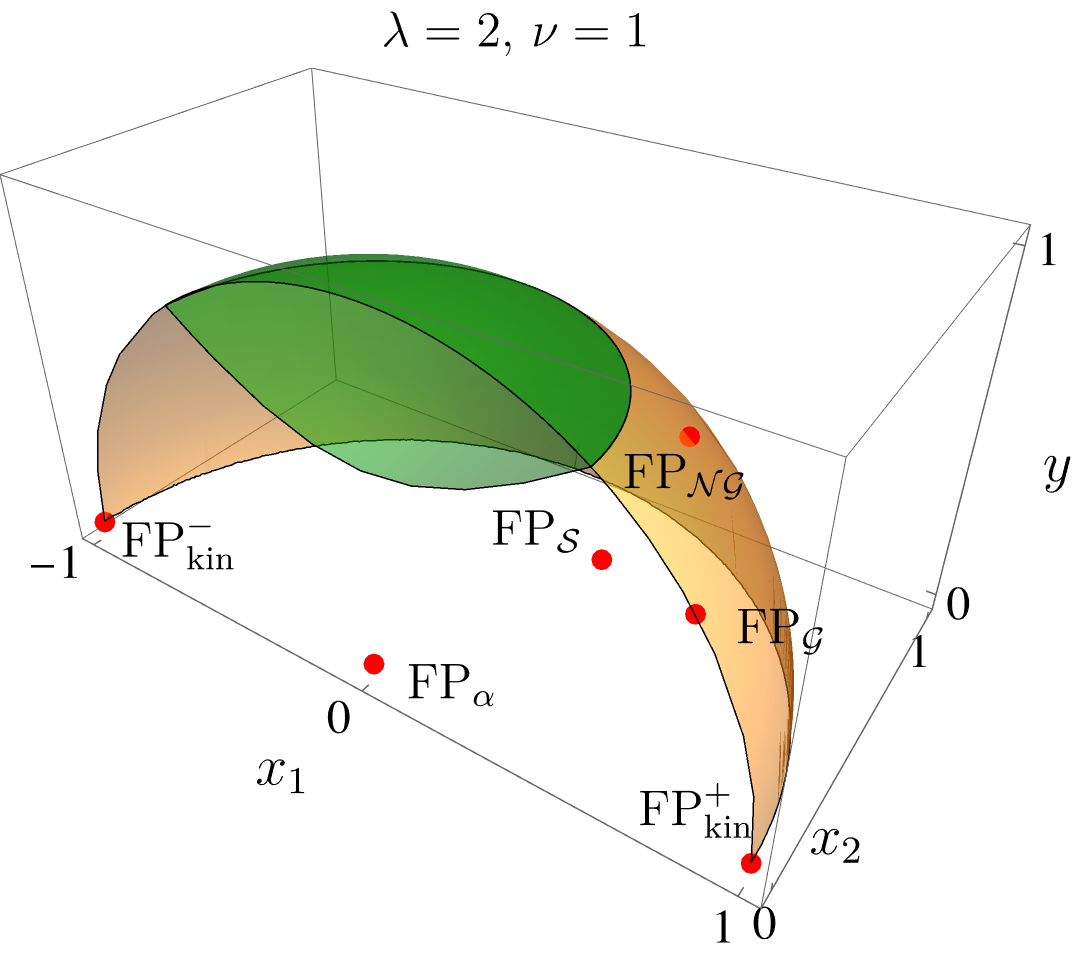} \hfill 
\includegraphics[width=0.31\textwidth]{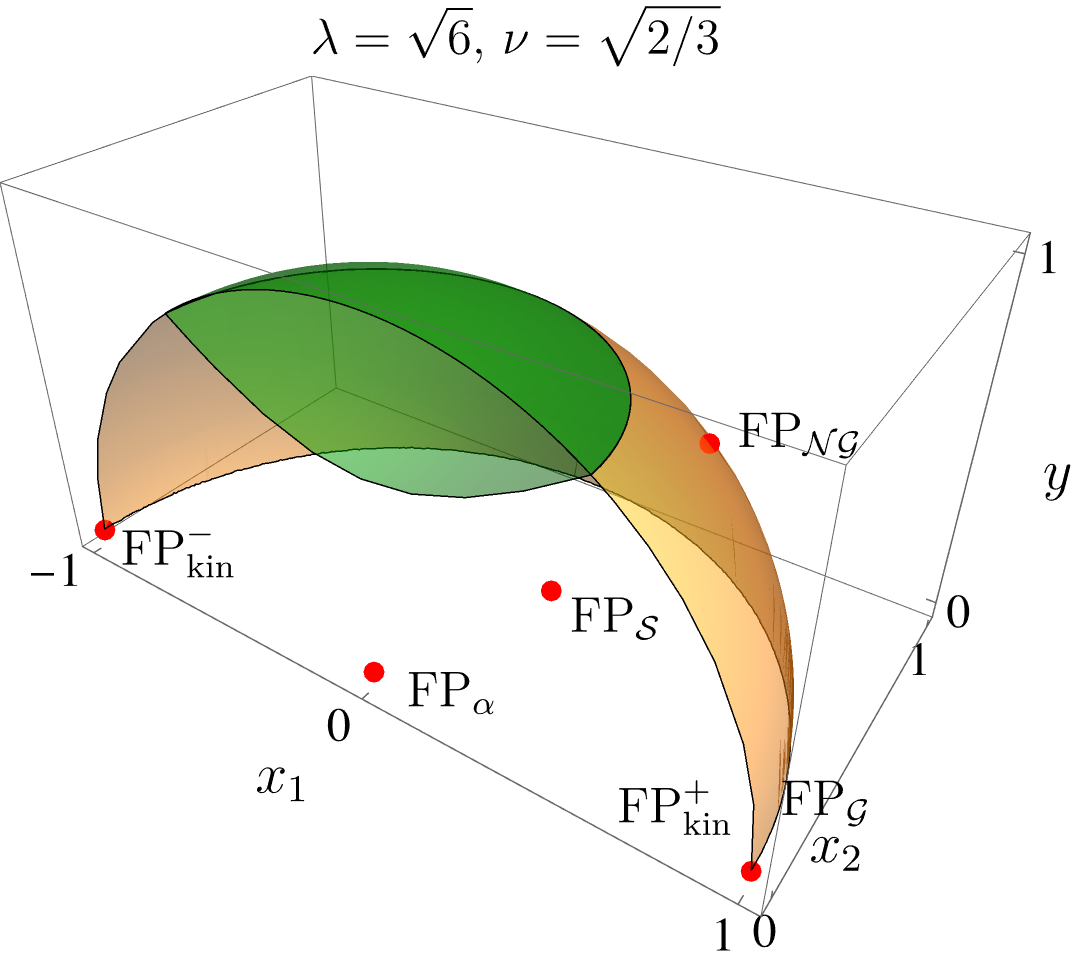}} 
\centering 
\rule{\textwidth}{0.5pt} \\ 
\textbf{Radiation fluid: $\alpha = 4/3$} \\
{\includegraphics[width=0.31\textwidth]{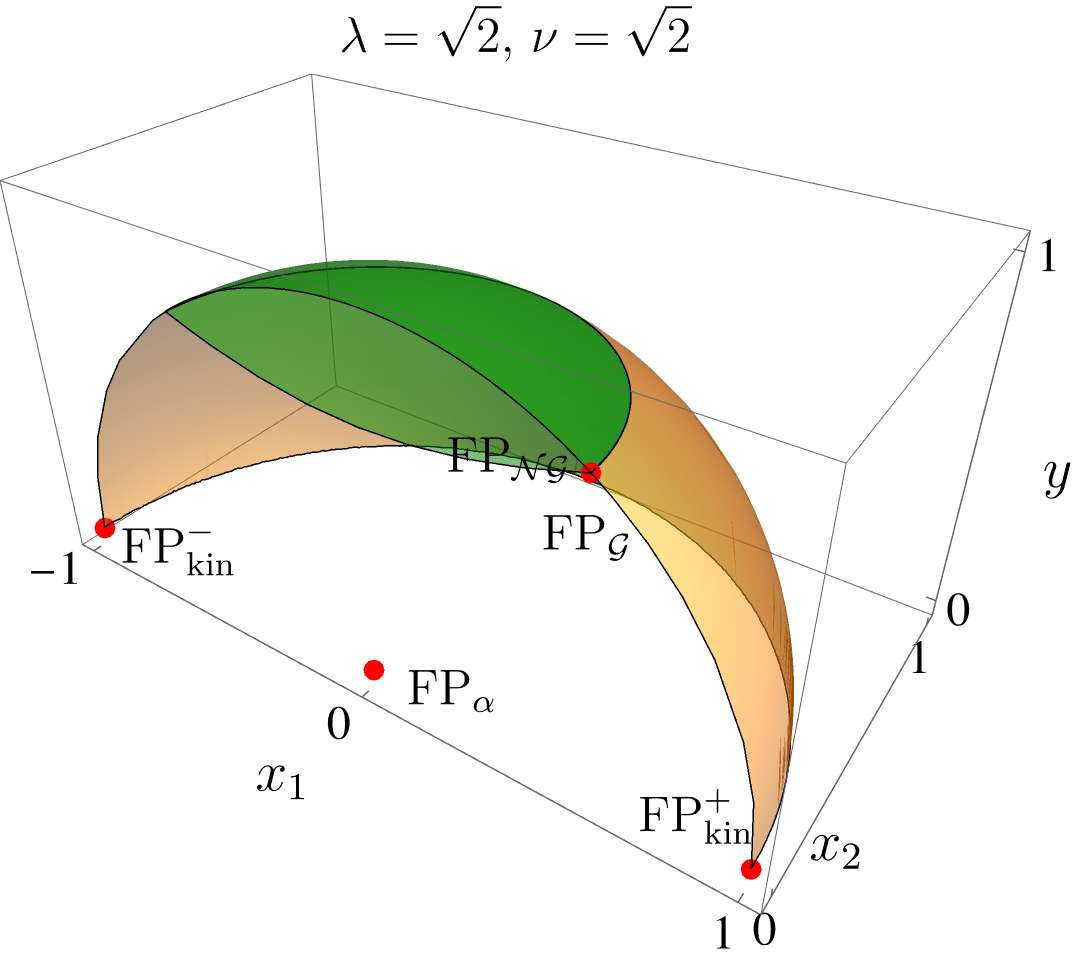} \hfill 
\includegraphics[width=0.31\textwidth]{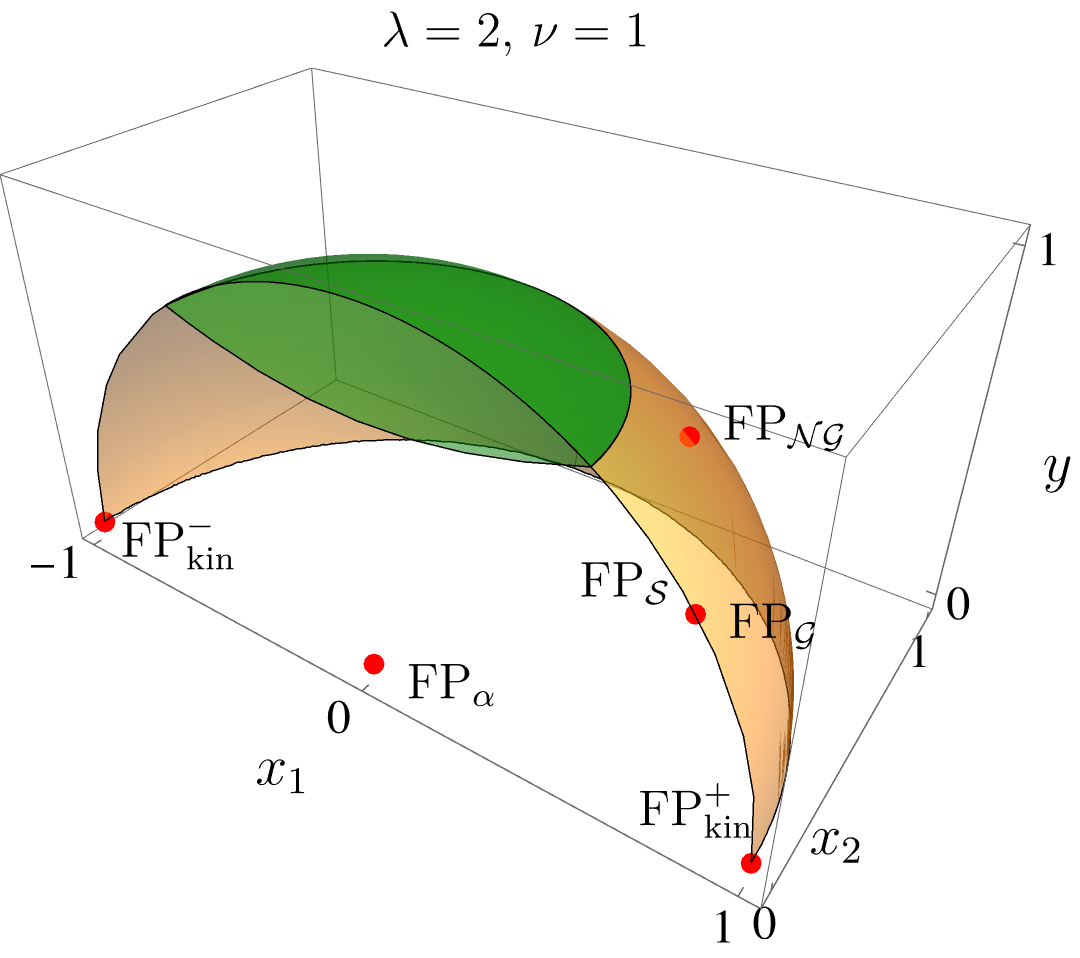} \hfill 
\includegraphics[width=0.31\textwidth]{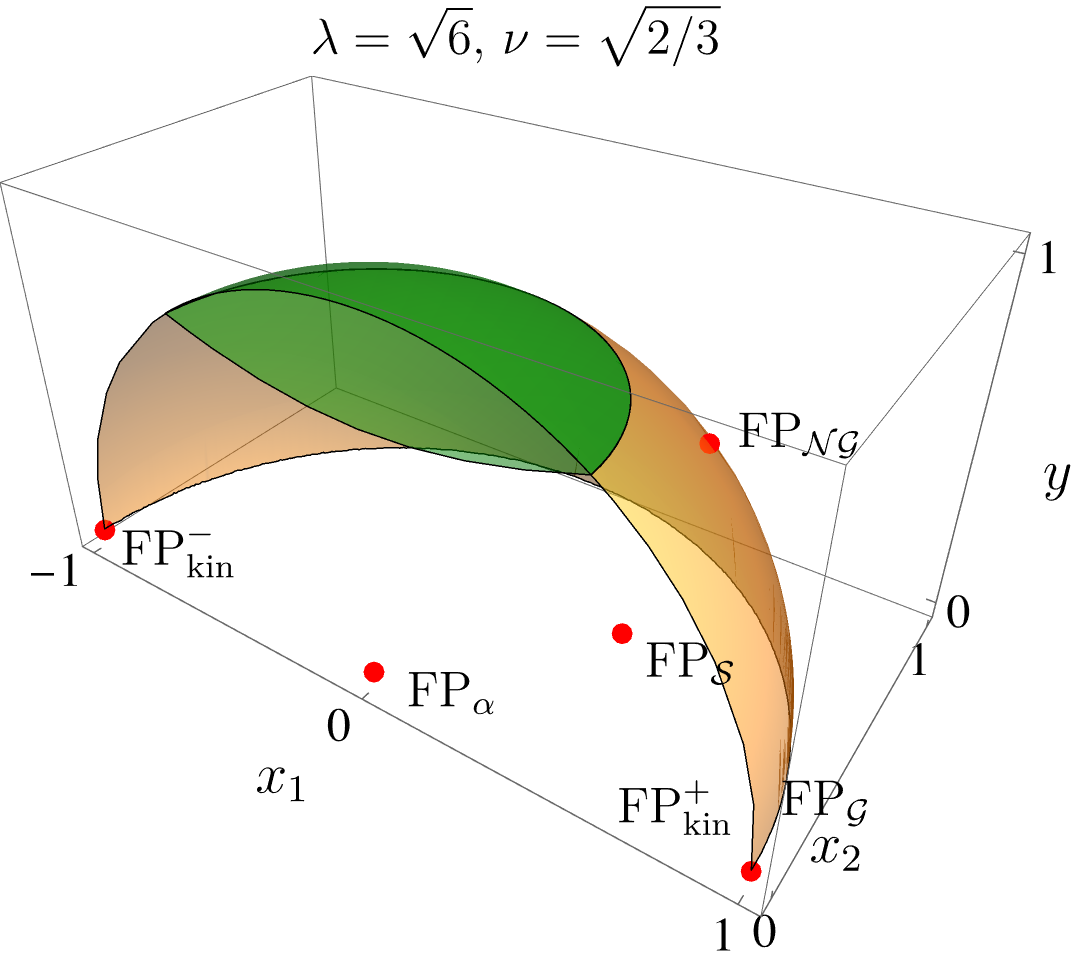}}
\rule{\textwidth}{0.5pt} 
\caption{Fixed point configuration for $\alpha = 2/3, 1$ and $4/3$, corresponding to curvature (top), matter (middle), and radiation (bottom) fluids. Each case is evaluated for $(\lambda,\nu)=\{(\sqrt{2},\sqrt{2}),~(2,1),~(\sqrt{6},\sqrt{2/3})\}$. The green region corresponds to the acceleration phase. The interior of the shell corresponds to $\Omega_\alpha > 0$, whilst the shell itself is the $\Omega_\alpha = 0$ invariant manifold.}
\label{fig:FPaxiodilandbaro}	
\end{figure}

An interesting situation revealed in the these figures is the coexistence of the points $\fp_\mc{S}$, $\fp_\mc{G}$, and $\fp_\mc{NG}$, in all cases, at a ``triple point'' always occurring at the coordinates:
\begin{equation}
    (\lambda,\nu)_\text{tp} = \left(\sqrt{3\alpha}, \frac{\sqrt{3} (2-\alpha)}{2 \sqrt{\alpha}} \right).
\end{equation}

For $\nu > \nu_{\rm tp}$ and $\lambda \gtrsim 1$, the nongeodesic attractor $\mathcal{NG}$ dominates a larger portion of the parameter space, allowing for interesting dynamics where the field turns strongly. As $\alpha$ increases (e.g., going from curvature $\alpha=2/3$ to radiation $\alpha=4/3$), the triple point shifts, enlarging the domain of the  $\mathcal{NG}$ solution. From a physical standpoint, this can be understood by noting that a larger $\alpha$ implies a faster dilution of the barotropic fluid energy contribution.

The rich dynamical structure of the system is visualized in the tomographic phase space portraits of Fig.~\ref{fig:FPaxiodilandbaro}. These plots illustrate the evolution of the fixed points and the acceleration volume across different background fluids (rows) and varying potential/coupling parameters (columns). The phase space is physically bounded by the invariant manifold $\Omega_\alpha = 0$, depicted as the orange spherical shell, which corresponds to the limit where the scalar sector dominates completely. Consequently, the interior of the shell represents trajectories with a nonnegligible contribution from the background fluid—be it curvature, matter, or radiation—while the green volume highlights the region of phase space compatible with accelerated expansion ($q<0$).

A detailed inspection of these snapshots reveals how the background EoS parameter $\alpha$ drastically alters the topology of the acceleration region. For a curvature-dominated background ($\alpha=2/3$, top row), the acceleration condition is relaxed, forming a wide cone that easily encompasses the fixed points. However, as the background fluid becomes stiffer —moving to matter ($\alpha=1$) and radiation ($\alpha=4/3$)— this volume shrinks significantly, retreating towards the pure potential limit ($y=1$). Crucially, this topological change forces the scaling point $\fp_\mc{S}$, which resides in the interior, to migrate towards the kinetic boundary, often expelling it from the viable acceleration zone in high-$\alpha$ scenarios.

Simultaneously, the scalar fixed points $\fp_\mc{G}$ and $\fp_\mc{NG}$ lie permanently on the outer shell, but their positions are governed by the interplay between the potential slope $\lambda$ and the coupling $\nu$. As $\lambda$ increases (left to right), the geodesic point $\fp_\mc{G}$ slides downwards towards the kinetic plane, reflecting the loss of potential dominance. In contrast, a sufficiently large coupling $\nu$ pulls the nongeodesic point $\fp_\mc{NG}$ laterally towards the axionic kinetic $x_2$ axis. This separation is physically significant: it places $\fp_\mc{NG}$ inside the green acceleration region even when the standard geodesic solution has exited it, visually confirming that the axionic turning mechanism provides a robust geometric pathway to acceleration, distinct from simple potential dominance.

\subsection{String theory preferred values}

\begin{figure}[t!]
\centering
\includegraphics[width=0.75\textwidth]{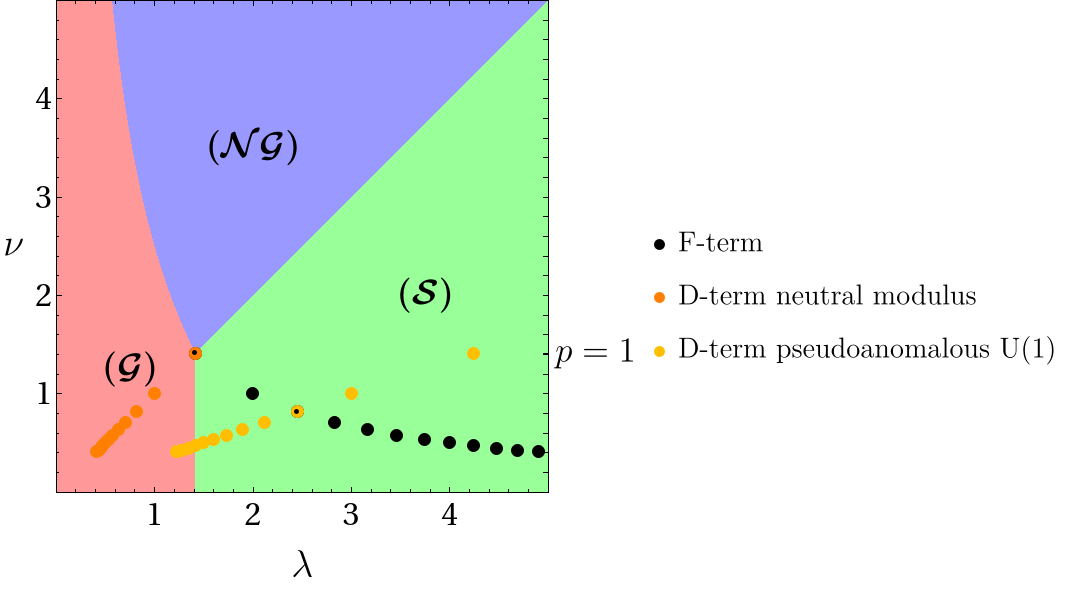}
\caption{Stability phase diagram for the multifield system in a curved universe ($\alpha = 2/3$). The shaded regions indicate the parameter space where the geodesic ($\fp_\mc{G}$, blue), nongeodesic ($\fp_\mc{NG}$, red), and scaling ($\fp_\mc{S}$, green) points are stable. The discrete data points represent string theory realizations for varying integer $p \in [1, 12]$. The F-term scenario (black dots) drives the system towards the scaling solution, while D-term scenarios (dark yellow and orange dots) may fall into the geodesic and/or scaling regions. Notably, none of the standard string realizations naturally populate the nongeodesic region, suggesting that the turning mechanism is disfavored in these minimal setups.}   
\label{fig:axiodilatoncurvatureStabdiag}
\end{figure}

Beyond their phenomenological implications, the dynamical structures derived above acquire particular significance when confronted with explicit UV-motivated constructions. In order to assess whether the regions of parameter space supporting geodesic, nongeodesic, or scaling attractors are naturally populated in string compactifications, we now compare our phase-space classification with representative values emerging from type IIB effective field theories. As detailed in Ref.~\cite{Gallego:2024gay}, the kinetic coupling and potential slopes in these setups are typically quantized in terms of an integer parameter $p$. Specifically, the axionic coupling is generally governed by the parameter $\nu=\sqrt{2/p}$. Regarding the scalar potential, the slope $\lambda$ depends on the specific mechanism generating the potential energy: contributions originating from F-terms typically yield steep slopes, $\lambda=\sqrt{2p}$, whereas D-term contributions can support shallower potentials with $\lambda = \sqrt{2/p}$ or $\lambda = 3\sqrt{2/p}$, for a neutral modulus and a pseudo anomalous $U(1)$ scenario, respectively.\footnote{It is worth noting that the D-term scenarios yielding small $\lambda$ values are of particular interest as they might evade the strictest versions of the de Sitter Swampland conjecture ($\lambda \gtrsim \mathcal{O}(1)$), although their consistency in a full quantum gravity setup remains a subject of active debate.}

Figure~\ref{fig:axiodilatoncurvatureStabdiag} summarizes the interplay between these theoretical priors and the dynamical phase space for a curvature-dominated background ($\alpha = 2/3$). The discrete sequences of points correspond to varying the integer $p$ from 1 to 12 for the three aforementioned scenarios. 

This comparison reveals a structurally nontrivial outcome. Within the minimal modulus–axion constructions considered here, the discrete loci populated by standard F-term and D-term scenarios do not generically overlap with the region where the nongeodesic fixed point $\fp_\mc{NG}$ is both stable and accelerating. Instead, F-term realizations typically fall inside the attraction basin of the scaling solution $\fp_\mc{S}$, while D-term constructions with shallower slopes tend to reside in the geodesic region $\fp_\mc{G}$, reintroducing the familiar requirement of sufficiently flat potentials for acceleration.

Therefore, although the turning mechanism is dynamically viable and enlarges the phase space of accelerating solutions at the level of the autonomous system, its realization within minimal string-motivated parameter regimes appears nongeneric. This observation anticipates the phenomenological analysis of Sec.~\ref{Sec: Constraints}, where the combined dynamical and observational constraints further restrict the viable parameter space.

\subsection{Invariant manifolds}

A complementary and structurally informative approach to analyse the global phase space structure is to focus on invariant submanifolds ---subsystems that admit an independent evolution within a reduced framework. We can identify three primary loci preserved by the flow: the single-field limit ($x_2 = 0$), the scalar-dominated limit ($\Omega_\alpha = 0$), and their intersection ($x_2 = \Omega_\alpha = 0$). While the scalar-dominated case corresponds to the spherical shell discussed in the previous section, here we first concentrate on the $x_2=0$ manifold. This limit describes a single scalar field interacting with a barotropic fluid, reducing the problem to the classic system studied by Copeland \textit{et al.} in Ref.~\cite{Copeland:1997et}, which extends the seminal work of Peebles and Ratra~\cite{Peebles:1987ek}. Subsequently, we consider the $\Omega_\alpha = 0$ locus.

\subsubsection{The $x_2=0$ invariant manifold}

\begin{figure}[hbt!]
\centering
\includegraphics[width=0.45\textwidth]{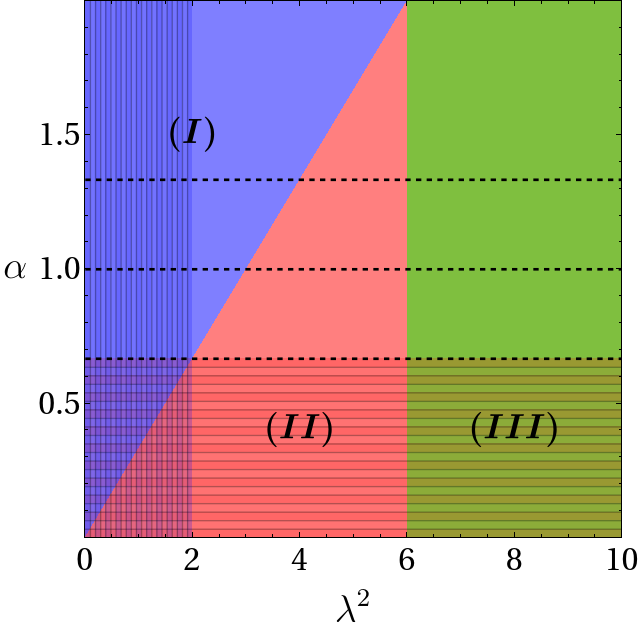}
\caption{Parametric phase diagram for the single scalar field limit ($x_2=0$, or $\nu=0$). The geodesic point $\fp_{\cal G}$ is the attractor in region ($I$), while the scaling point $\fp_\mc{S}$ dominates in regions ($II$) and ($III$). The shaded areas denote the parameter combinations allowing for acceleration in the respective attractors (vertical shading for $\fp_{\cal G}$ and horizontal shading for $\fp_{\cal S}$). Dashed lines indicate the specific cuts for radiation, matter, and curvature fluids.}
\label{fig:parphasespacesinglescalarandbaro}
\end{figure}

Restricting the dynamics to the plane $x_2=0$ allows us to map the stability regions in terms of the potential slope $\lambda$ and the background fluid parameter $\alpha$. As illustrated in the parametric diagram of Fig.~\ref{fig:parphasespacesinglescalarandbaro}, the parameter space is divided into three distinct dynamical regions. In region ($I$), defined by $\alpha > \lambda^2/3$ and $\lambda^2 < 6$, the scaling solution requires $\Omega_\alpha < 0$ and the system invariably evolves towards the field-dominated solution $\fp_{\cal G}$, which serves as the unique attractor. Conversely, in region ($II$) ($\alpha \leq \lambda^2/3$ and $\lambda^2 \leq 6$), the scaling point $\fp_\mc{S}$ enters the space $\Omega_\alpha \ge 0$ and becomes the stable attractor. This stability persists into region ($III$) ($\lambda^2 > 6$), where the potential is too steep to sustain the $\fp_{\cal G}$ solution, leaving $\fp_\mc{S}$ as the sole late-time destination.

Crucially, the condition for accelerated expansion constrains the parameters to low values within these regions: the geodesic point $\fp_\mc{G}$ accelerates only for shallow potentials ($\lambda^2 < 2$), while the scaling point $\fp_\mc{S}$ requires a very soft background EoS ($\alpha < 2/3$). These viability zones are highlighted as shaded areas in Fig.~\ref{fig:parphasespacesinglescalarandbaro}.

\begin{figure}[t!]
\centering
{\includegraphics[width=0.49\textwidth]{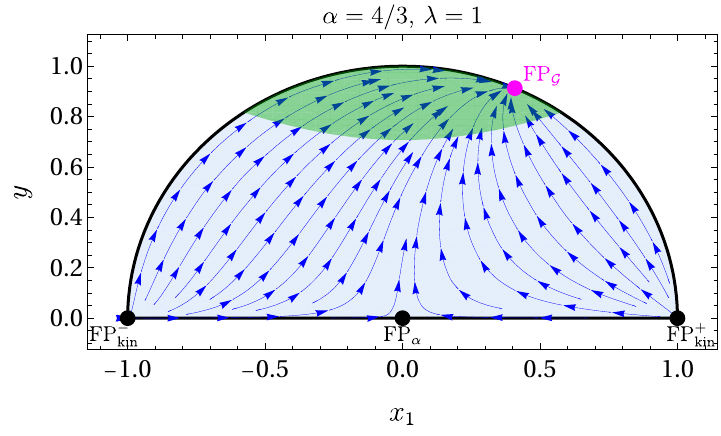} \hfill 
\includegraphics[width=0.49\textwidth]{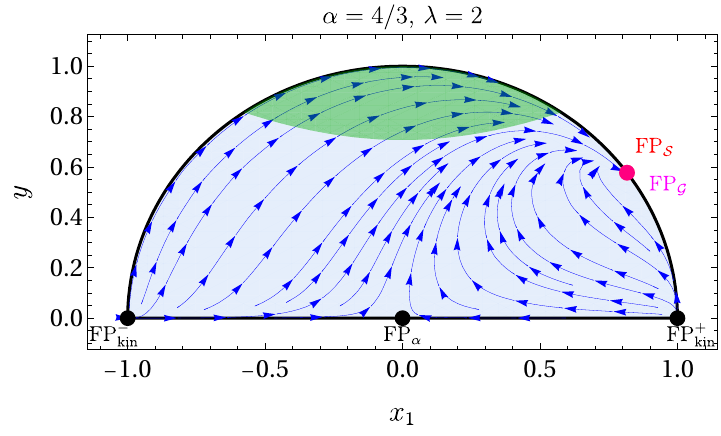}} 
\centering	\\
{\includegraphics[width=0.49\textwidth]{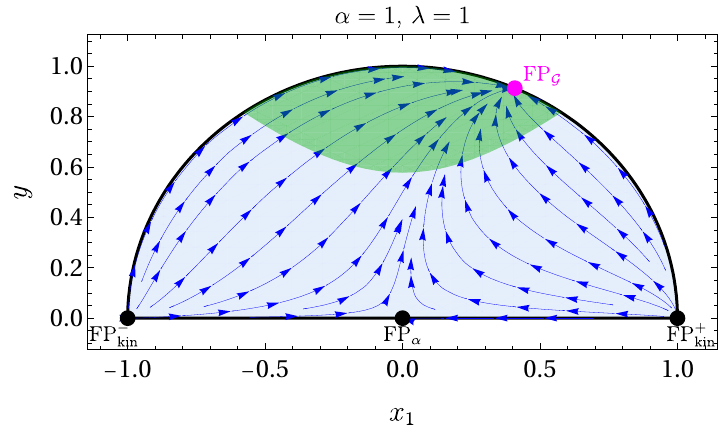} \hfill 
\includegraphics[width=0.49\textwidth]{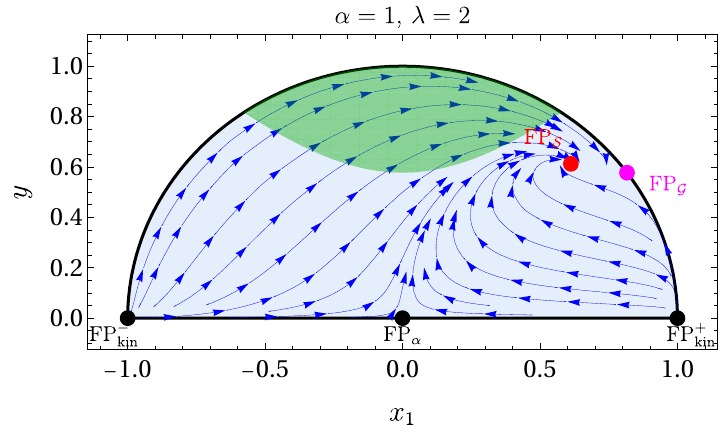}} 
\centering	\\
{\includegraphics[width=0.49\textwidth]{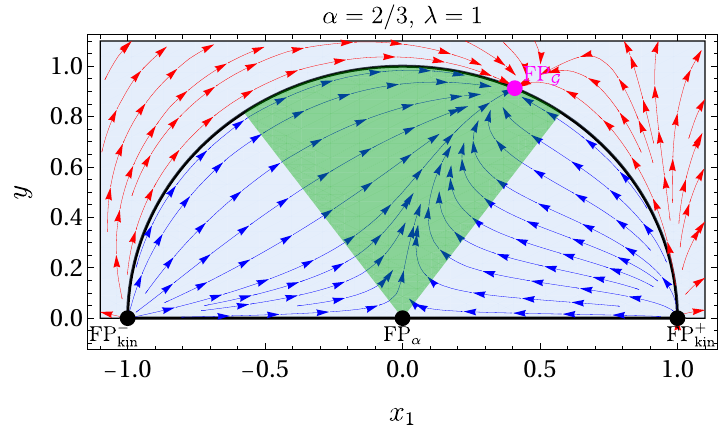} \hfill 
\includegraphics[width=0.49\textwidth]{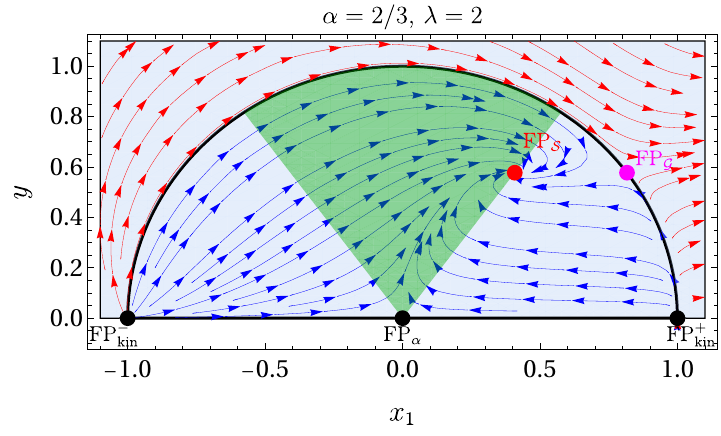}}
\caption{2D phase portraits for the single-field subsystem ($x_2=0$). From top to bottom, the rows correspond to backgrounds with increasing EoS: $\alpha=4/3$ (radiation, top), $\alpha=1$ (matter, middle), and $\alpha=2/3$ (curvature, bottom). The green area represents the acceleration region ($q<0$). Red dots indicate the scaling point $\fp_\mc{S}$, while magenta dots denote the field-dominated point $\fp_\mc{G}$. Note how the acceleration region shrinks and the scaling attractor moves out of it as $\alpha$ increases. In all cases shown for $\lambda = 1$ (left column), the scaling point lies in the $\Omega_\alpha<0$, not allowed for the matter and radiation cases. For $\alpha = 2/3$, the inner region of the circle represents a closed universe, while the outer region represents an open universe.}
\label{fig:phsespacesinglescalarandbaroflat}
\end{figure}

To visualize the resulting cosmological histories, Fig.~\ref{fig:phsespacesinglescalarandbaroflat} presents phase portraits for representative values of $\alpha$, effectively taking $x_2=0$ slices of the full 3D phase space shown previously. The overall conclusion from this reduced analysis is that whenever the scaling solution $\fp_\mc{S}$ is theoretically available (i.e., $\Omega_\alpha > 0$), it typically prevents acceleration unless $\alpha$ is remarkably small. Specifically, for $\alpha > 2/3$, the system exhibits either no acceleration or a single transient phase before tracking the background fluid. However, in the specific case of interest where curvature acts as the barotropic fluid ($\alpha = 2/3$), the dynamics become richer: trajectories may cross the acceleration boundary multiple times, leading to a ``roller-coaster'' expansion history~\cite{DAmico:2020euu}. Finally, only for $\alpha < 2/3$ can the scaling solution itself support an eternal accelerated epoch.

This standard picture changes fundamentally once a nontrivial axionic velocity ($x_2 \neq 0$) is allowed. However, we first analyze the $\Omega_\alpha = 0$ locus. 

\subsubsection{The $\Omega_\alpha=0$ invariant manifold}

This invariant submanifold isolates the dynamics of the two scalar fields, decoupling them from the background fluid. Within the context of string-inspired cosmologies, this system was first investigated in the string frame in Refs.~\cite{Billyard:1999ct, Billyard:2000cz}, and later generalized to arbitrary spacetime dimensions and frames in Ref.~\cite{Sonner:2006yn}.

In this reduced phase space, the dynamics are governed by three types of fixed points: the kination solutions $\fp_\kin^\pm$, the geodesic solution $\fp_{\cal G}$, and the nongeodesic solution $\fp_{\cal NG}$. The latter two play the most significant role, acting as late-time attractors that become degenerate when the coupling satisfies $\nu = (6-\lambda^2)/(2 \lambda)$. Nongeodesic fixed points have garnered considerable attention because they offer a mechanism to reconcile phenomenological requirements with quantum gravity constraints. Specifically, they allow for a slow-roll-like regime ---and consequently accelerated expansion--- even in the presence of the steep potentials ($\lambda \gtrsim \mathcal{O}(1)$) typically predicted by the Swampland conjectures~\cite{Achucarro:2018vey} (for some explicit examples, see Refs.~\cite{Cicoli:2020cfj, Cicoli:2020noz, Akrami:2020zfz, Anguelova:2021jxu, Brinkmann:2022oxy, Gallego:2024gay}).

\begin{figure}[t!]
\centering
\includegraphics[width=0.375\textwidth]{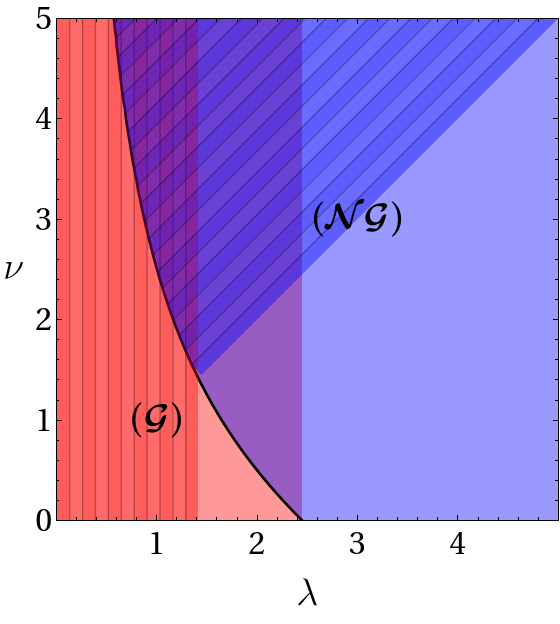}
\caption{Parametric phase space for the pure scalar system ($\Omega_\alpha=0$). The diagram highlights the existence (solid lines) and acceleration (shaded areas) regions for the geodesic ($\cal G$, red) and nongeodesic ($\cal NG$, blue) fixed points. Note that the existence condition for $\fp_\mc{NG}$ ($\nu > (6-\lambda^2)/2\lambda$) defines its attractor region, which completely overlaps with the coexistence domain.}
\label{fig:existenceFPaxiodilaton}
\end{figure}

Physically, this behavior arises because the field space curvature, parametrized by $\nu$, generates a centrifugal force that counteracts the potential gradient. If the turning rate is sufficiently high (specifically, when $\nu > \lambda$), this effect sustains an accelerated phase despite the steepness of the potential. This mechanism is visualized in the parametric phase space of Fig.~\ref{fig:existenceFPaxiodilaton}, where the shaded diagonal region $\nu > \lambda$ marks the domain of nongeodesic acceleration. Furthermore, the basin of attraction for $\fp_\mc{NG}$ is intrinsically larger than that of $\fp_\mc{G}$; indeed, whenever the nongeodesic point exists, it effectively supersedes the geodesic solution as the stable attractor. This dominance stems from the effective EoS: $w_\eff$ for the nongeodesic solution decreases as $\nu$ increases, causing its associated energy density to dilute slower than that of the geodesic mode. It is important to note, however, that this dominance is fragile against the inclusion of a background fluid. As discussed in the previous section, the scaling point $\fp_\mc{S}$ can destabilize $\fp_\mc{NG}$ over significant portions of the parameter space (see Fig.~\ref{fig:phasespacemultiandbaro}).

\begin{figure}[t!]
\centering 
{\includegraphics[width=0.49\textwidth]{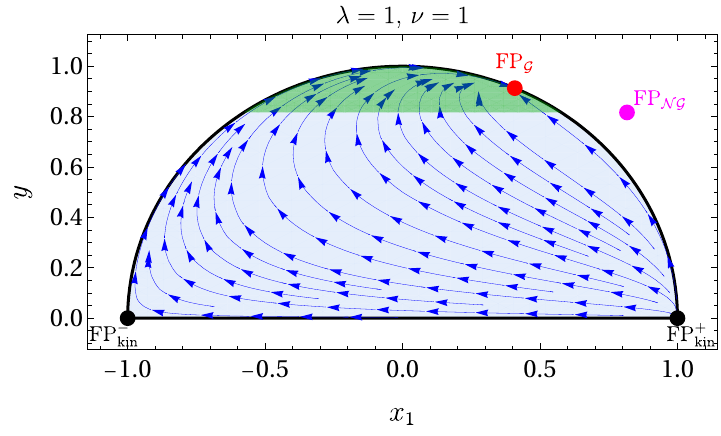} \hfill
\includegraphics[width=0.49\textwidth]{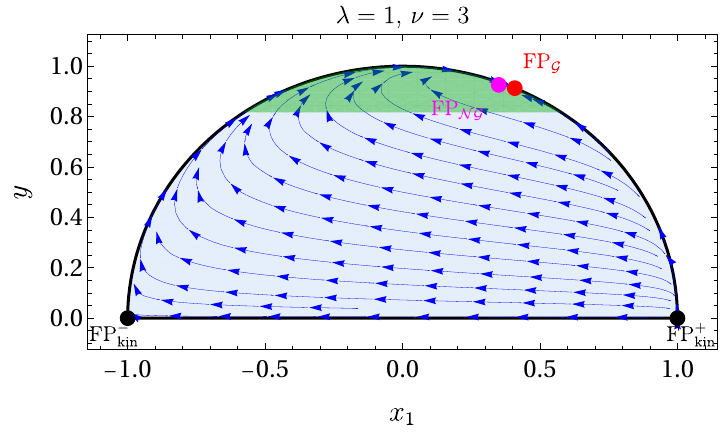}}
\centering
{\includegraphics[width=0.49\textwidth]{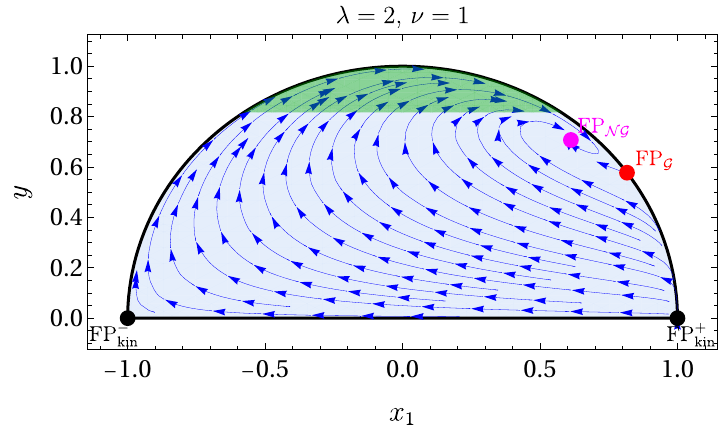} \hfill
\includegraphics[width=0.49\textwidth]{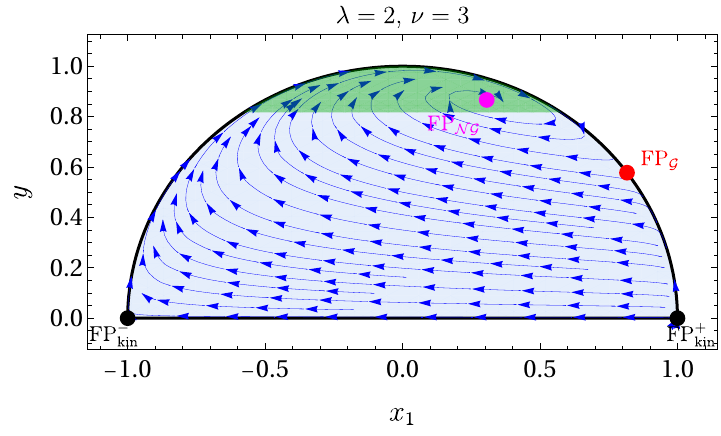}}
\centering
{\includegraphics[width=0.49\textwidth]{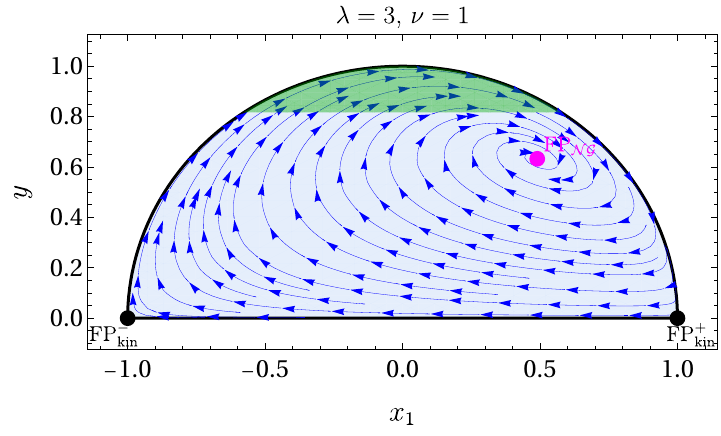} \hfill
\includegraphics[width=0.49\textwidth]{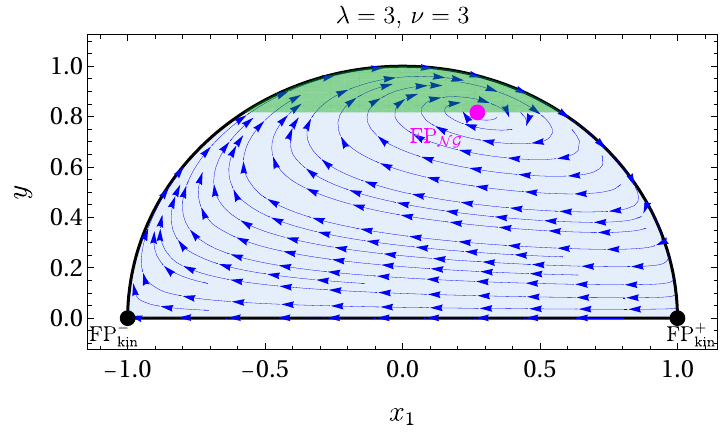}}
\caption{Phase flow on the $\Omega_\alpha = 0$ invariant manifold projected onto the $(x_1, y)$ plane. The boundary of the unit disk represents the single-field limit ($x_2=0$), while the interior corresponds to nontrivial turning trajectories ($x_2 \neq 0$). The green shaded area indicates accelerated expansion ($q<0$). Columns correspond to increasing $\nu$ and rows to increasing $\lambda$.}
\label{fig:twofieldphasespace}
\end{figure}

The dynamical richness of this submanifold is further illustrated in Fig.~\ref{fig:twofieldphasespace}, which displays the phase flow projected onto the $(x_1, y)$ disk (since $x_2^2 = 1 - x_1^2 - y^2$). The topology of the flow confirms that $\fp_\mc{G}$ can support eternal acceleration only for shallow potentials ($\lambda < \sqrt{2}$), whereas the nongeodesic attractor $\fp_\mc{NG}$ unlocks eternal acceleration for steep potentials provided $\lambda < \nu$. Interestingly, the approach to these attractors is not always monotonic. As observed in the phase portraits near the transition line $\nu \approx \lambda$, the system can undergo ``recurrent acceleration''~\cite{Sonner:2006yn}, characterized by spiral trajectories that enter and exit the accelerated region multiple times before settling into the fixed point.

Finally, it is important to caution that the existence of an accelerated fixed point does not automatically guarantee consistency with observational data. Current constraints from DESI~\cite{DESI:2025hao} on the dark energy EoS, parametrized as $w_\phi(a) = w_0 + w_a(1-a)$, yield $w_0 = -0.24^{+0.17}_{-0.64}$ and $w_a = -2.5^{+1.9}$. If we assume the current Universe is tracked by $\fp_\mc{NG}$, the EoS is constant, $w_{\phi,{\cal NG}} = (\lambda-2\nu)/(\lambda+2\nu)$. Inverting this relation yields the constraint:
\begin{equation}
    \lambda = 2\nu \left( \frac{1+w_{\phi,{\cal NG}}}{1-w_{\phi,{\cal NG}}} \right).
\end{equation}
The central value $w_0 = -0.24$ implies a nonaccelerating Universe at $z=0$, which would trivially exclude any accelerating fixed point. However, even if we consider the lower bound of the confidence interval to allow for acceleration (e.g., $w_0 \approx -0.88$), the constraint becomes $\lambda \approx 0.13 \nu$. This is significantly more restrictive than the theoretical condition for acceleration ($\lambda < \nu$). Consequently, to maintain the viability of $\fp_\mc{NG}$ as a description of late-time cosmology, the model parameters must deviate significantly from the central observational values, pushing towards the lower bounds of the confidence interval. Although this result should be regarded only as a mild indication ---since $\fp_\mc{NG}$ cannot account for a Universe with a substantial matter component--- it nevertheless highlights the tension between the turning mechanism and current observational preferences.

\subsubsection{Dynamical features and cosmological histories}

\begin{figure}[t!]
\centering 
{\includegraphics[width=0.49\textwidth]{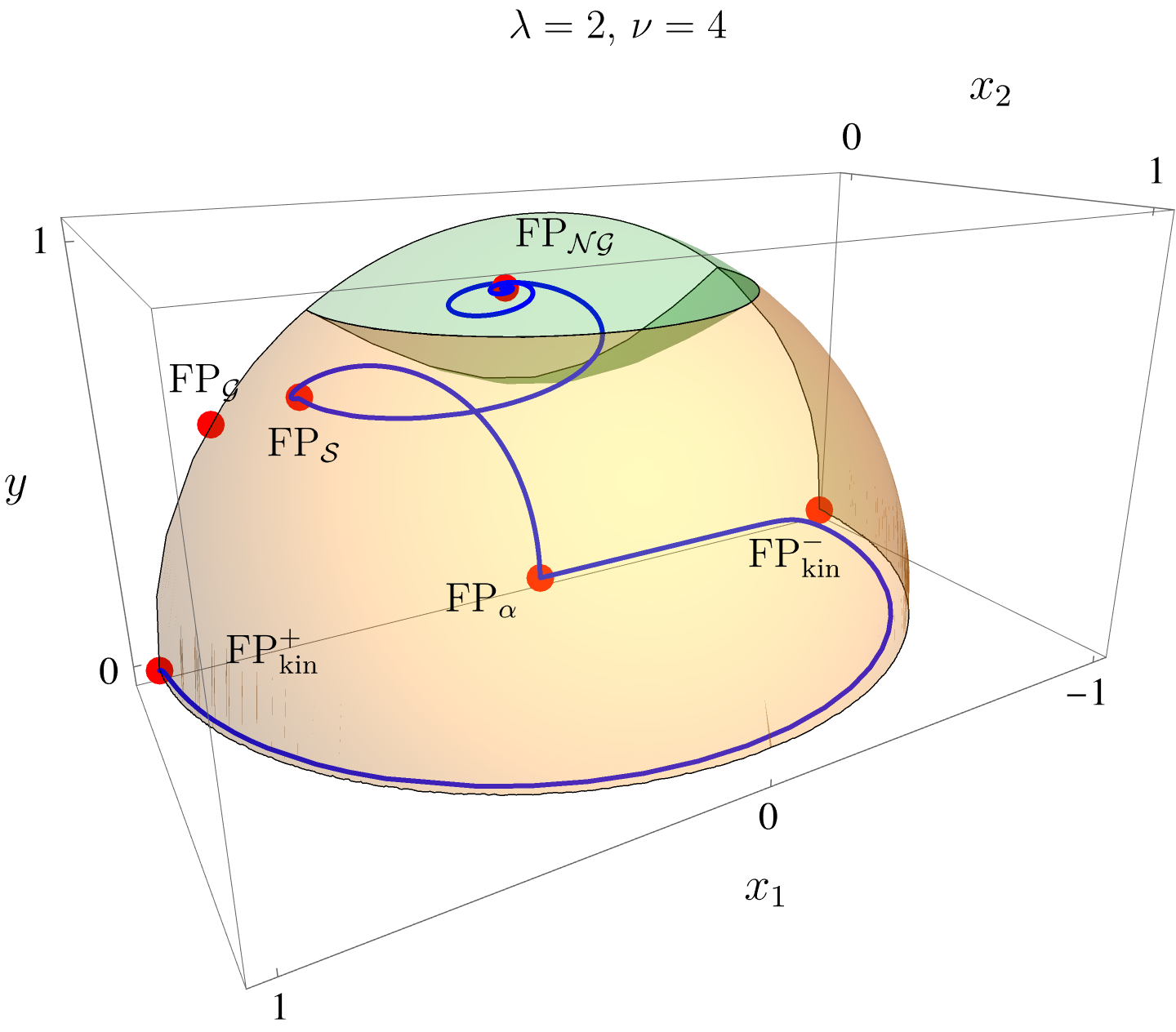} \hfill 
\includegraphics[width=0.45\textwidth]{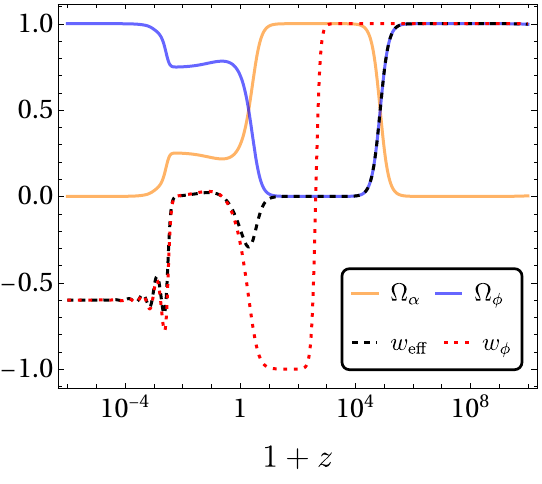}} 
\centering
{\includegraphics[width=0.49\textwidth]{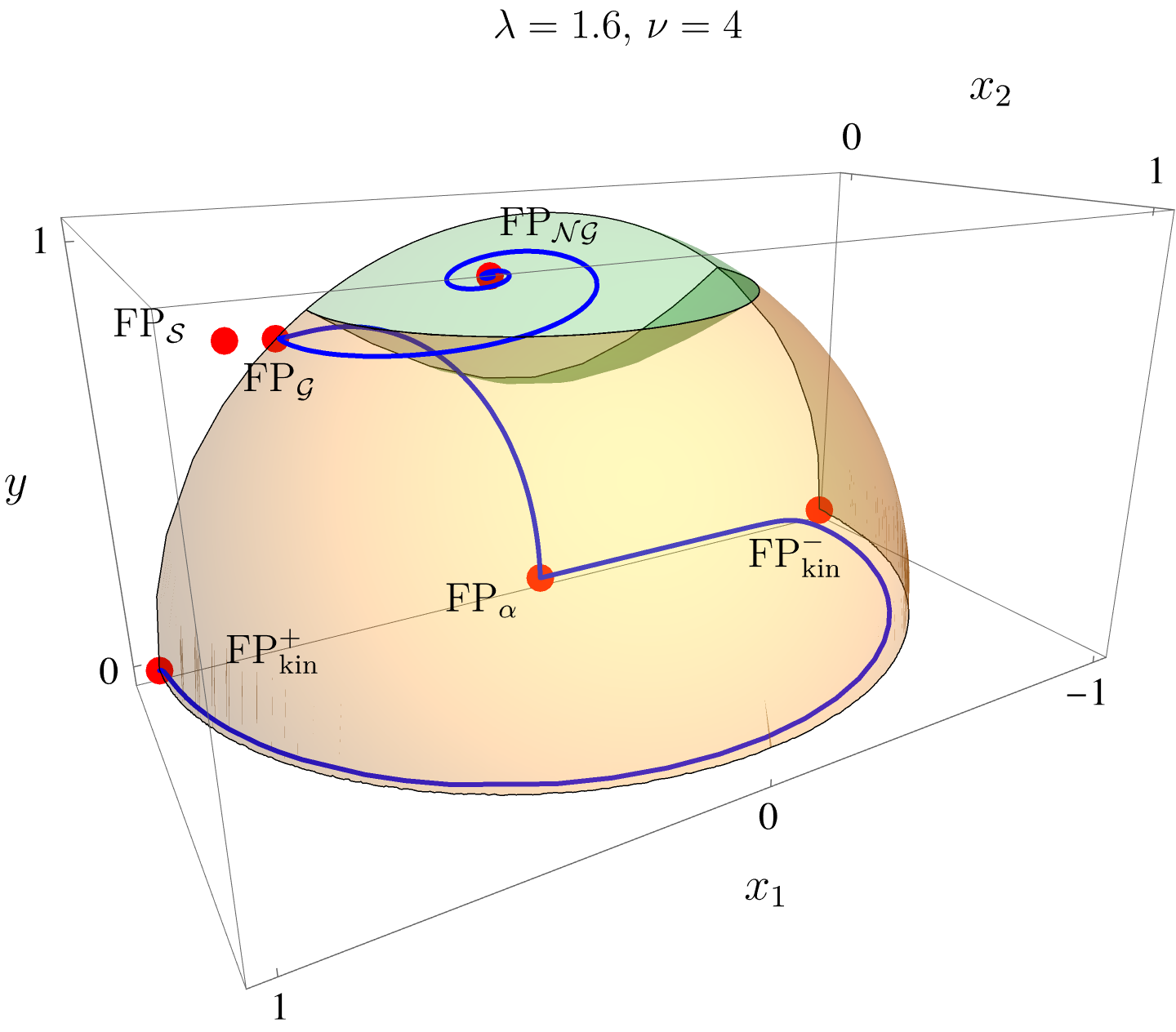} \hfill 
\includegraphics[width=0.45\textwidth]{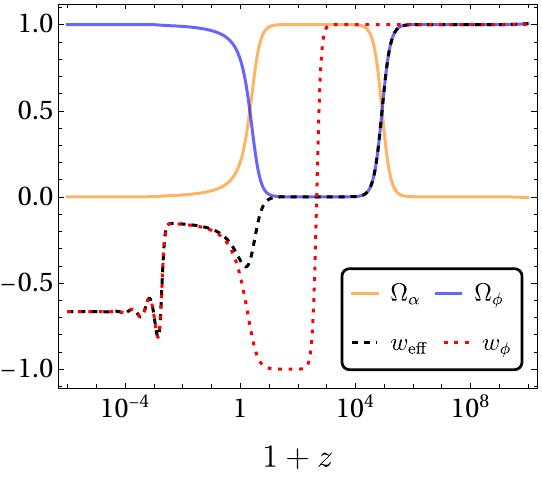}}
\caption{Representative cosmological histories in a matter-dominated background ($\alpha=1$), both culminating in a nongeodesic accelerated attractor $\fp_\mc{NG}$. Top row ($\lambda=2, \nu=4$): The transition is mediated by the scaling fixed point $\fp_\mc{S}$ (red dot on the boundary), which acts as a saddle, temporarily delaying the final approach to acceleration. Bottom row ($\lambda=1.6, \nu=4$): The transition is mediated by the geodesic fixed point $\fp_\mc{G}$. The trajectory exhibits oscillatory behavior, briefly entering and exiting the acceleration region (green) around the saddle before settling into the final attractor. The right panels display the energy densities and the EoS, showing $w_\eff$ exhibiting oscillations which stabilize around $w_\eff \approx -0.6$ in both cases.}
\label{fig:twoscalarsandbaroNGattractor}
\end{figure}

Before addressing the full curved multifield system, it is instructive to illustrate how the stability analysis derived above governs the actual cosmological evolution. Figure~\ref{fig:twoscalarsandbaroNGattractor} presents two representative histories for a matter-dominated background ($\alpha=1$), initialized near the kination regime $\fp_\kin^+$.

In both scenarios, the system eventually evolves towards $\fp_\mc{NG}$, securing a late-time phase of accelerated expansion with $w_\eff \approx -0.6$. However, the trajectory towards this final attractor is nontrivial and heavily influenced by the presence of intermediate saddle points, which dictate the transient behavior of the Universe.

\begin{itemize}
    \item \textbf{Scaling-mediated transition (Top Row, $\lambda=2$):} In this scenario, the trajectory departing from the matter point $\fp_\alpha$ is initially drawn towards the scaling fixed point $\fp_\mc{S}$. Here, $\fp_\mc{S}$ acts as a saddle point: it temporarily captures the flow, forcing the effective EoS to hover near the scaling value (tracking the background) and delaying the onset of full acceleration. Eventually, the system escapes this saddle region and moves into the interior, settling at the nongeodesic attractor $\fp_\mc{NG}$. This case illustrates how the presence of the scaling solution can ``frustrate'' or delay the convergence to the dark energy domination era.

    \item \textbf{Geodesic-mediated transition (Bottom Row, $\lambda=1.6$):} For a shallower potential, the dynamics change. The system bypasses the scaling region and interacts primarily with the geodesic fixed point $\fp_\mc{G}$, which now acts as the dominant saddle. The trajectory exhibits a characteristic spiral behavior (associated with the complex eigenvalues of the system), entering and momentarily exiting the acceleration region as it loops around the attractor. After this transient oscillation, the system finally converges to $\fp_\mc{NG}$, establishing a stable eternal acceleration.
\end{itemize}

These examples highlight a crucial feature of the model: while the late-time attractor is determined by the global stability conditions ($\nu > \lambda$), the intermediate cosmological history —and specifically the timing and nature of the transition to dark energy— depends sensitively on the interplay with the unstable (saddle) fixed points $\fp_\mc{S}$ and $\fp_\mc{G}$.

A second crucial feature, exclusive to the multifield dynamics, is the oscillatory behavior of the EoS observed in the bottom panels. Unlike single-field attractors where the approach is typically monotonic, $\fp_\mc{NG}$ is often a stable spiral (possessing complex eigenvalues). This leads to a ringing effect in $w_\eff$ as the trajectory circles into the attractor, alternating between deep acceleration and stiffer regimes. While this behavior is mathematically robust, our analysis suggests it typically pertains to the asymptotic future rather than the immediate past or present expansion history.

\section{Multi scalar system in a curved Universe}
\label{Sec: Curved Multifield}

In this section we investigate whether spatial curvature and nongeodesic motion can jointly enlarge the viable parameter space of exponential quintessence in a realistic cosmological setting including radiation and matter. 

The dynamical classification of Sec.~\ref{Sec: Single Barotropic} revealed that curvature-induced scaling solutions and nongeodesic attractors arise in distinct invariant manifolds of the full system. However, the existence of these mechanisms in isolation does not guarantee that their effects cannot cooperate during intermediate stages of a realistic cosmological history.  In particular, it remains to be established whether the combined effects of spatial curvature and turning trajectories in field space can relax the restrictions on the slope parameter $\lambda$ required for late-time accelerated expansion, thereby allowing phenomenologically viable solutions even for steep exponential potentials ($\lambda \gtrsim \mathcal{O}(1)$).

We address this question progressively. First, we examine the purely kinematical impact of spatial curvature on the acceleration condition. We then analyze the dynamical evolution in the thawing regime relevant for the present epoch and derive an analytical upper bound on the potential slope. Finally, we confirm these results through a full numerical scan of the multifield system. As we shall see, curvature and turning not only modify the phase-space structure in distinct regimes, but also fail to combine into a viable late-time accelerating solution for steep exponential potentials.

\subsection{Spatial curvature and the kinematic condition for acceleration}
\label{Sec: Insights from Curvature}

We begin by assessing the direct impact of spatial curvature on the present-day acceleration condition. The primary motivation for including curvature, as explored in Refs.~\cite{Andriot:2024jsh, Bhattacharya:2024hep, Alestas:2024gxe}, is the possibility of obtaining viable cosmological histories even with steep exponential potentials ($\lambda \gtrsim \mathcal{O}(1)$). Intuitively, allowing for a nonvanishing spatial curvature contribution might be expected to relax the restrictive constraints on the scalar
field dynamics.

This expectation can be illustrated by examining the effective EoS parameter. Using the relation,
\begin{equation} 
\label{eq:wEffandwPart}
    w_\eff = \sum_i w_i \Omega_i = w_\phi \Omega_\phi + \frac13 (\Omega_r - \Omega_k)\,,
\end{equation}
the requirement of a present-day accelerated phase ($w_{\eff, 0} < -1/3$) translates into a bound for the scalar EoS,
\begin{equation}
    w_{\phi,0} < - \frac{1 + \Omega_{r,0} - \Omega_{k,0}}{3\Omega_{\phi,0}}\,.     
\end{equation}

It is algebraically clear that for an open universe ($\Omega_k > 0$), the numerator decreases, thereby relaxing the upper bound on $w_{\phi,0}$ (i.e., allowing it to be less negative). However, quantitatively, this effect is marginal. Adopting fiducial values consistent with \textit{Planck} data~\cite{Planck:2018vyg},
\begin{equation}
\label{eq:fiducialvaluesflatsinglescalar}
    \Omega_{\phi, 0} \approx 0.695\,, \quad \Omega_{m,0} \approx 0.305\,, \quad \Omega_{r,0} \approx 10^{-4}\,, \quad \Omega_{k,0} = 0\,,
\end{equation}
we find the standard flatness requirement,
\begin{equation}
\label{Eq: bound wphi0}
    w_{\phi,0} \lesssim -0.479.
\end{equation}

If we saturate the current observational bounds with an open universe ($\Omega_{k,0} \sim +0.003$), this bound shifts only slightly to $w_{\phi, 0} \lesssim -0.478$. Thus, while curvature provides a qualitative relaxation of the acceleration condition, it does not by itself dramatically alter the requirements on the dark energy EoS.

A more stringent constraint on the potential slope $\lambda$ arises from the dynamical evolution of the EoS. By differentiating $w_\eff$, one can invert the dynamics to express the potential slope as a function of the cosmological parameters, as done in Ref.~\cite{Andriot:2024jsh} for the invariant manifold $x_2 = 0$.
For the present system, this relation reads
\begin{equation}
\lambda=\frac{9 \Omega_{\phi}+\Omega_k+\Omega_r-9 w_{\eff}^2+3 w_{\eff}'}{\sqrt{6} x_1 \left(3 \Omega_{\phi }+\Omega_r-\Omega_k-3 w_{\eff}\right)}\,.
\end{equation}

To gain physical insight, we can approximate this relation by neglecting the subdominant contributions from radiation and spatial curvature at late times. This is justified observationally since $|\Omega_{k,0}| \lesssim 10^{-3}$,
so curvature corrections enter only at the subpercent level in the expression for $\lambda$. Using the approximation $w_\eff \approx w_\phi \Omega_\phi$ and the scalar continuity equation, the expression simplifies to
\begin{equation}
    \lambda \approx \frac{3(1 - w_\phi^2) + w_\phi'}{\sqrt{6}\, (1 - w_\phi)\, x_1}\,.
\end{equation}

Adopting the CPL parametrization~\cite{Chevallier:2000qy, Linder:2002et}, $w_\phi(a) = w_0 + w_a (1 - a)$, we have $w_\phi' = -w_a a$ (differentiating with respect to $N$). Evaluated at the present time ($a=1$), this yields
\begin{equation}
    \lambda \approx \frac{3(1 - w_0^2) - w_a}{\sqrt{6}\, (1 - w_0)\, x_{1,0}}\,.
\end{equation}

This analytical result connects the fundamental parameter $\lambda$ directly to the observational reconstruction of dark energy. While the pure BAO DESI results suggest a nonaccelerating central value, joint analyses (e.g., DESI + CMB) often favor an accelerated EoS. Taking, for instance, the constraints reported in Ref.~\cite{DESI:2025fii} compatible with acceleration,
\begin{equation} 
\label{eq:desiconstraints} w_0 \approx -0.48, \qquad w_a \approx -1.34\,, \end{equation} 
we obtain the phenomenological bound,
\begin{equation} 
\label{eq:boundlambdafromwprime} 
    \lambda \lesssim \frac{0.58}{x_{1,0}}\,. 
\end{equation}
This relation is illuminating: it demonstrates that satisfying Swampland conditions (large $\lambda$) requires a suppression of the scalar kinetic energy, $x_{1,0}$. Specifically, for the conjecture $\lambda > \sqrt{2}$ to hold, the current kinetic fraction of the field must satisfy
\begin{equation}
    x_{1,0} \lesssim 0.41\,,
\end{equation}
or, neglecting $x_{2,0}^2$, $\Omega_{\phi,\text{kin}}^{(0)} \sim x_{1,0}^2
\lesssim 0.17$.

For comparison, the Friedman constraint,
\begin{equation}
x_{1,0}^2 + x_{2,0}^2 + y_0^2 = \Omega_{\phi,0} \simeq 0.695
\end{equation}
implies the trivial geometric bound $x_{1,0} \le \sqrt{\Omega_{\phi,0}} \approx 0.83$. The constraint derived above is therefore significantly stronger, indicating that steep potentials can only be compatible with acceleration if the scalar field is already strongly frozen today.

More generally, the relation (\ref{eq:boundlambdafromwprime}) reveals a direct connection between the present scalar kinetic energy and the slope of the potential: larger values of $\lambda$ require a progressively smaller scalar velocity today. The dynamical implications of this relation for the late-time evolution will be examined in the
following subsection.

\subsection{Thawing dynamics and the effective single-field limit}
\label{sub:approximate_analytical}

The previous subsection showed that compatibility between accelerated expansion and moderately steep potentials requires the scalar field to be strongly frozen today, with a highly suppressed kinetic energy. A natural framework to describe such configurations is provided by the \emph{thawing} regime in the sense of Ref.~\cite{Caldwell:2005tm}.

In thawing models the scalar field remains nearly frozen during matter domination due to Hubble friction, and only begins to roll significantly at late times. The present epoch therefore corresponds to the early stages of the field evolution away from this frozen configuration rather than to an approach toward a fixed point of the
autonomous system.

Under these conditions the scalar velocity remains small throughout the matter–dark energy transition. As a consequence, both the axionic kinetic component and spatial curvature remain subdominant, and the multifield dynamics effectively reduce to the geodesic submanifold $x_2 \simeq 0$ during the epoch relevant for observations. The consistency of neglecting the axionic component, radiation and curvature within the thawing approximation is verified analytically in the Appendix. The full dynamical behavior of the axionic component and the possibility of significant non–geodesic motion, including the existence of scaling regimes, will be analyzed in detail in Sec.~\ref{Sec: nongedesic and scaling}.


We, therefore, derive an approximate analytical solution describing this thawing phase, with the goal of identifying a dynamical upper bound on the potential slope $\lambda$. In this regime, potential dominance is expected ($y^2 \gg x_1^2$), since the field begins its evolution in a nearly frozen state. The evolution equation for $y$, Eq.~\eqref{Eq: y Eq}, then simplifies to
\begin{equation}
    y' \approx \frac{3}{2}y\,,
\end{equation}
yielding
\begin{equation}
    y(N) \approx y_0 \, e^{3N/2}\,,
\end{equation}
where $y_0$ denotes the value of $y$ at $N=0$, corresponding to the present epoch ($z=0$).

Substituting this into Eq.~\eqref{Eq: x1 Eq} under the same approximations and integrating from a frozen state ($x_1 \to 0$ as $N \to -\infty$), we obtain
\begin{equation}
\label{eq:x1_approx}
    x_1(N) \approx \frac{1}{\sqrt{6}} \lambda y_0^2 e^{3N}
    \quad \implies \quad
    x_{1,0} \approx \frac{\lambda}{\sqrt{6}} y_0^2\,.
\end{equation}

Using the closure relation $\Omega_{\phi 0} = x_{1,0}^2 + y_0^2$, we solve for the present-day values,
\begin{equation}
\label{Eq: x1_0 analytical}
    y_0 = \frac{1}{\lambda}\left(\sqrt{9 + 6\lambda^2 \Omega_{\phi,0}} - 3\right)^{1/2}\,, 
    \quad
    x_{1,0} = \frac{1}{\sqrt{6}\lambda}\left(\sqrt{9 + 6\lambda^2 \Omega_{\phi,0}} - 3\right)\,.
\end{equation}

The instantaneous scalar EoS is $w_\phi \approx -1 + 2x_1^2/y^2$. Evaluated at $z=0$, this yields
\begin{equation}\label{eq:wphi0thawing}
    w_{\phi, 0} = -2 + \sqrt{1 + \frac{2}{3} \lambda^2 \Omega_{\phi, 0}}\,.
\end{equation}

Solving this expression for $\lambda$ yields the slope corresponding to
a given present-day EoS
\begin{equation}
\label{eq:lambdaOmegawanalitical}
    \lambda = \sqrt{ \frac{3}{2\Omega_{\phi, 0}}} 
    \sqrt{(w_{\phi,0} + 1)(w_{\phi,0} + 3)}\,\,.
\end{equation}

Since the derivation is performed in the thawing regime, where matter dominates the expansion history and both curvature and scalar contributions remain subdominant, one expects $w_{\rm eff}\approx0$ throughout most of the evolution. Nevertheless, Eq.~(\ref{eq:wphi0thawing}) provides a useful estimate of the present-day scalar equation of state $w_{\phi,0}$. This can then be translated into a constraint on $\lambda$ by approximating $w_{\rm eff,0}\simeq \Omega_{\phi,0} w_{\phi,0}$ and imposing the acceleration condition $w_{\rm eff,0}<-1/3$. Applying this criterion yields
\begin{equation}
\label{eq:maxanalyticallambdaacceleration}
    \lambda < 1.682 
    \quad 
    (\text{for } \Omega_{\phi,0}=0.7)\,.
\end{equation}

\begin{figure}[t!]
\centering
{\includegraphics[width=0.48\linewidth]{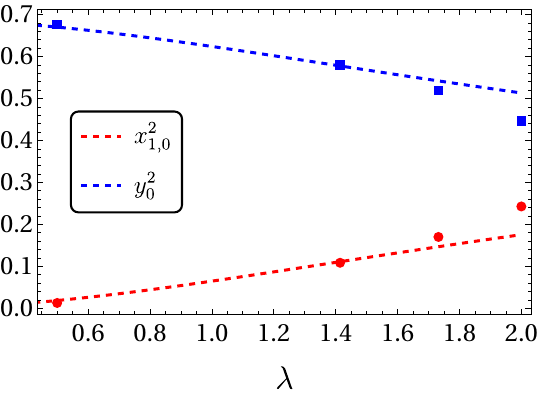} \hfill
\includegraphics[width=0.48\linewidth]{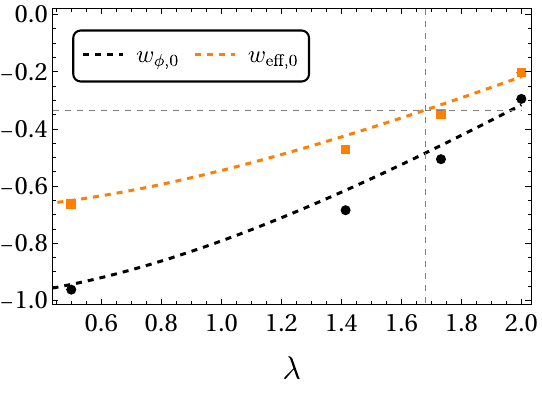}}
\caption{Comparison between the approximate analytical prediction (dashed lines) and the exact numerical solution (points) for the present-day values of some variables as a function of the slope $\lambda$. Left: $x_{1,0}^2$ (red) and $y_0^2$ (blue). Right: $w_{\phi, 0}$ (black) and $w_{\eff, 0}$ (orange). Gray dashed line shows the intersection between $w_{\eff, 0} = -1/3$ and $\lambda = 1.682$.}
\label{fig:x10y10}
\end{figure}

The validity of this approximation is confirmed in Fig.~\ref{fig:x10y10}, where the analytical prediction (dashed lines) shows excellent agreement with the exact numerical solution (points), deviating only for very large $\lambda$ where kinetic dominance sets in.

Care must be taken when interpreting the detailed time evolution, since the analytical approximation assumes a fixed matter background. In the full numerical solution, the onset of acceleration reduces Hubble friction and slightly modifies the field growth. Nevertheless, the relation between $x_{1,0}$ and $y_0$ remains robust.

Substituting the approximation (\ref{eq:x1_approx}) into the
phenomenological bound obtained previously,
$x_{1,0} \lesssim 0.58/\lambda$, and using $y_0^2 \approx \Omega_{\phi,0}$
gives
\begin{equation}
    \frac{\lambda}{\sqrt{6}} \Omega_{\phi,0}
    \lesssim
    \frac{0.58}{\lambda}
    \quad \implies \quad
    \lambda \lesssim
    \sqrt{\frac{1.42}{\Omega_{\phi,0}}}
    \approx 1.42 .
\end{equation}

Using instead the exact expression (\ref{Eq: x1_0 analytical}) yields
\begin{equation}
\label{eq:fullyanaliticalbound}
    \lambda < 1.58
    \quad
    (\text{for } \Omega_{\phi,0}=0.7).
\end{equation}

We therefore arrive at the following dynamical implication: for $\lambda \gtrsim 1.6$, the thawing dynamics generate a kinetic contribution large enough to prevent the onset of acceleration at the matter–dark energy transition. In this regime, the potential slope $\lambda$ becomes the dominant control parameter of the background evolution.

The above derivation relies on the fact that, during the thawing phase, both the axionic kinetic component and spatial curvature remain subdominant. Whether nongeodesic motion can qualitatively modify this bound must be assessed within the full multifield dynamics, which we analyze numerically below.

\subsection{$x_2=0$ invariant manifold: Revisiting the geodesic limit}
\label{sec:x20locus}

Having established that spatial curvature alone does not significantly relax the acceleration condition within the geodesic submanifold, we now revisit explicitly the $x_2=0$ limit as a phenomenological baseline. This case connects directly with the analyses of Refs.~\cite{Andriot:2024jsh, Bhattacharya:2024hep, Alestas:2024gxe} and allows us to isolate the interplay between the exponential slope $\lambda$ and spatial curvature before activating the full multifield dynamics.

As established in the dynamical analysis of Sec.~\ref{Sec: Single Barotropic}, for steep potentials with $\lambda > \sqrt{2}$ —the regime favored by Swampland conjectures~\cite{Obied:2018sgi, Agrawal:2018own}— the geodesic fixed point $\fp_{\cal G}$ ceases to be accelerated. Since this is the only fixed point in the geodesic invariant manifold capable of supporting accelerated expansion, any accelerated phase occurring in this regime cannot correspond to a late-time attractor.

Instead, the system ultimately evolves toward the scaling solution $\fp_\mc{S}$ (when $\alpha<2/3$), which is nonaccelerating. As pointed out originally in Refs.~\cite{Halliwell:1986ja, Copeland:1997et}, and confirmed in recent realistic scans~\cite{Andriot:2024jsh}, this structure naturally allows for a transient period of accelerated expansion while the system evolves toward the scaling attractor.

\begin{figure}[t!]
\centering 
{\includegraphics[width=0.48\textwidth]{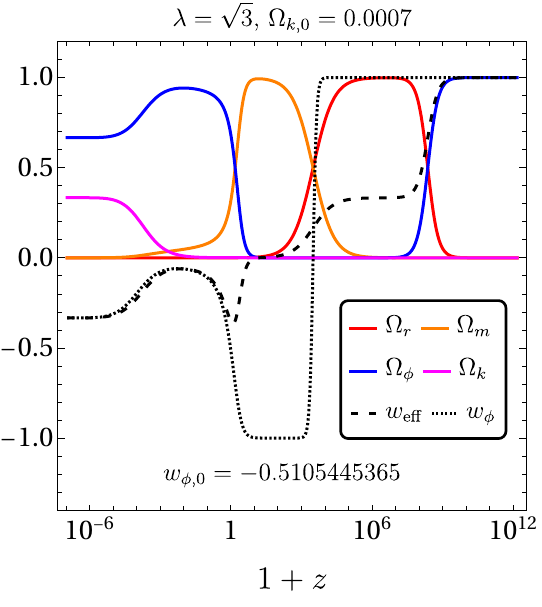} \hfill
\includegraphics[width=0.48\textwidth]{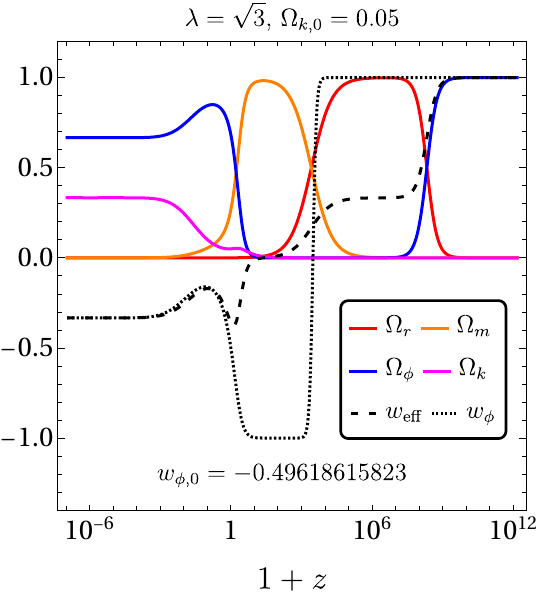}}
\caption{Cosmological evolution of the fractional abundances $\Omega_i$ (solid lines) and the effective EoS $w_\eff$ (dashed) for the single-field limit. Left: A standard curvature scenario ($\Omega_{k, 0} = 0.0007$). Right: An open universe ($\Omega_{k, 0} = 0.05$). In both cases, $w_{\phi, 0}$ is fine-tuned to ensure a consistent radiation-dominated era ($\Delta N_\text{rad} \approx 10$ or $\Delta z_\text{rad} \approx 10^5$) prior to the matter era.}
\label{fig:singlescalarandcurvatureabundances}
\end{figure}

To explore this phenomenology quantitatively, we perform numerical integrations backward from the present day, adopting the fiducial values,
\begin{equation}
\label{eq:fiducialforcasessinglefield}    
    \Omega_{r, 0} = 10^{-4}, \quad \Omega_{\phi, 0} = 0.685.
\end{equation}
Figure~\ref{fig:singlescalarandcurvatureabundances} displays the evolution of the energy abundances and the effective equation of state for two distinct curvature scenarios: a concordancelike value ($\Omega_{k, 0} = 0.0007$) and an open universe with substantial curvature ($\Omega_{k, 0} = 0.05$).

In both cases, it is possible to reproduce the standard sequence of cosmological epochs —radiation domination, followed by matter domination and a late-time accelerated phase— provided that the present-day value of the scalar equation of state $w_{\phi,0}$ is carefully chosen. However, the figure also reveals a crucial technical feature: the extreme sensitivity of the backward evolution to this present-day input parameter.

In the examples shown, achieving a radiation-dominated epoch of sufficient duration ($\Delta N_\text{rad} \gtrsim 10$) requires tuning $w_{\phi, 0}$ up to the tenth significant digit. Conservative estimates, such as those in Ref.~\cite{Alvarez:2019ues}, would impose even stricter tuning requirements. Small deviations from the tuned value lead, when integrating backward, to a rapid growth of the kinetic energy, driving the system into a kination-dominated regime that spoils the standard thermal history.

However, this apparent fine-tuning should not be interpreted as a fundamental pathology of the model. In the standard $\Lambda$CDM or tracking quintessence scenarios, the radiation-dominated fixed point acts as a past attractor, naturally setting the initial conditions. In the present exponential model, by contrast, the dynamical system formally points backwards to a kination-dominated state.

Physically, however, it is natural to assume that the Universe emerges from postinflationary reheating directly into a radiation-dominated phase, with the scalar field frozen or slow-rolling due to Hubble friction. Under this assumption, the early-time configuration is not generic in phase space but dynamically selected. The numerical sensitivity observed in backward integration therefore reflects the requirement of matching this specific early-time configuration when evolving from present-day boundary conditions, rather than signaling an obstruction to the model's cosmological viability. 

The geodesic limit thus provides an instructive baseline. Viable cosmological histories can indeed be constructed, even for moderately steep potentials and in the presence of spatial curvature, but they occupy a restricted region of phase space when expressed in terms of present-day boundary data. In this formulation the accelerated solutions appear only for specific choices of present-day parameters, suggesting a degree of backward fine-tuning.

To assess whether such configurations arise naturally in the full cosmological evolution, it is therefore useful to adopt a complementary perspective and study the forward dynamical evolution of the system. In the next subsection we explore this question through numerical integrations of the multifield dynamics.

\subsection{Full multifield numerical scan}
\label{sec:numericalscan}

The analytical arguments presented in
Sec.~\ref{sub:approximate_analytical} indicate that sustaining
present-day acceleration requires the potential slope to satisfy
$\lambda \lesssim 1.6$–$1.7$. This bound was obtained under the
physically motivated assumption that both spatial curvature and the
axionic kinetic component remain subdominant during the matter–dark
energy transition.

The backward integrations explored in the previous subsection
demonstrate that viable cosmological histories can indeed be
constructed, but they are highly sensitive to the choice of
present-day boundary conditions, especially $w_{\phi,0}$.

To assess whether the accelerated solutions arise naturally in the
cosmological evolution, we now adopt a complementary strategy and
integrate the full multifield system forward in time, without
restricting the dynamics to the geodesic submanifold.

The system is initialized at the matter–radiation equality epoch
($z_{\rm eq} \sim 3200$, corresponding to $N \sim -8$), where the
background evolution is well understood and physically motivated
initial conditions can be imposed.

At equality we require a negligible dark energy contribution,
\begin{equation}
    \Omega_{\DE,\eq} \equiv \Omega_{\phi,\eq} + \Omega_{k,\eq} \approx 10^{-9}\,,
\end{equation}
ensuring that the Universe undergoes the standard radiation and matter dominated epochs before the onset of acceleration. To explore the robustness of the dynamics, we distribute this small energy budget between the scalar and curvature components using a parameter $\kappa \in [0,1]$, such that
\begin{equation}
    \Omega_{\phi,\eq} \approx \kappa\, \Omega_{\DE,\eq},
    \qquad
    \Omega_{k,\eq} \approx (1-\kappa)\, \Omega_{\DE,\eq}.
\end{equation}

We consider three representative cases:
\begin{itemize}
    \item Curvature-dominated initial condition ($\kappa \approx 0.05$),
    \item Equipartition ($\kappa \approx 0.5$),
    \item Scalar-dominated initial condition ($\kappa \approx 0.95$).
\end{itemize}

For each configuration, a grid of initial kinetic values $(x_{1,\eq}, x_{2,\eq})$ satisfying $x_{1,\eq}^2 + x_{2,\eq}^2 \lesssim \Omega_{\phi,\eq}$ is scanned, though extreme values close to the boundary $x_{1,\eq}^2 + x_{2,\eq}^2 \approx \Omega_{\phi, \eq}$ were excluded to avoid numerical singularities. For every trajectory, we define the present epoch ($N_{\text{today}}$) as the moment when the scalar energy fraction reaches the observed value $\Omega_{\phi,0} \approx 0.689$, thereby ensuring a consistent late-time normalization.

\begin{figure}[t]
\centering\includegraphics[width=1\textwidth]{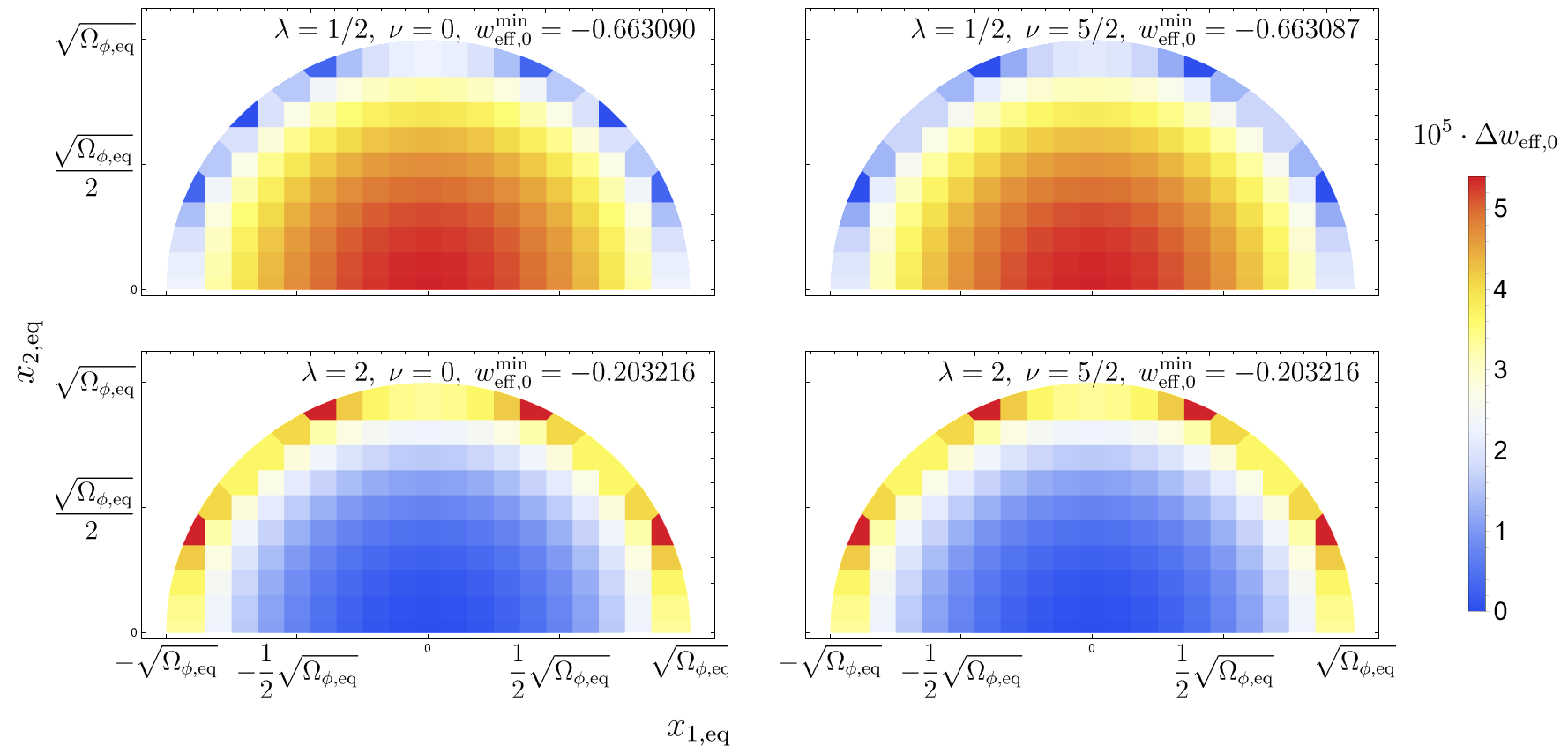}
\caption{Sample of numerical results for the effective EoS parameter today, evaluated at the epoch defined by $\Omega_{\phi,0} = 0.689$. The figure shows the deviation from the minimum value attained in each case, $10^{5}(w_{\mathrm{eff},0}-w_{\mathrm{eff},0}^{\mathrm{min}})$, for $\lambda = 1/2$ (top panels) and $\lambda = 2$ (bottom panels). Left panels correspond to $\nu=0$, right panels to $\nu=5/2$. All cases correspond to the equipartition case, $\Omega_{\phi,\eq} = \Omega_{k,\eq} = 5\times 10^{-10}$.}
\label{fig:sampledataforWeff2}
\end{figure}

Figure~\ref{fig:sampledataforWeff2} shows representative results of the scan for the values of $w_{\mathrm{eff},0}$ for two values of the slope, $\lambda = 1/2$ and $\lambda = 2$, and for two values of the nongeodesic parameter, $\nu = 0$ and $\nu = 5/2$. The plotted quantity corresponds to the deviation of the present-day effective equation-of-state parameter from its minimum value in each case.

Two important features emerge. First, for $\lambda \lesssim 2$, the dispersion in $w_{\eff,0}$ induced by varying the initial kinetic configuration is extremely small, typically of order $10^{-4}$. This confirms that the late-time solution is largely insensitive to the precise initial partition of kinetic energy at equality, characteristic of thawing dynamics, even though sustained acceleration itself only occurs below the analytic threshold $\lambda \simeq 1.7$ derived in the previous subsection. Second, comparing the left ($\nu=0$) and right ($\nu=5/2$) panels reveals that the nongeodesic parameter has a negligible impact on $w_{\eff,0}$ at the background level. While turning trajectories alter the asymptotic structure of the phase space, they do not significantly affect the cosmological evolution between equality and the present epoch.

To provide a more quantitative summary of the scan, we bin the trajectories by the initial kinetic value $x_{1,\eq}$ and compute the mean and standard deviation of the late-time observables for each parameter triplet $(\lambda, \nu, \Omega_{\phi,\eq})$. This procedure allows us to isolate the dominant physical effects while marginalizing over the residual sensitivity to initial conditions.

\begin{figure}[t!]
\centering
{\includegraphics[width=0.49\textwidth]{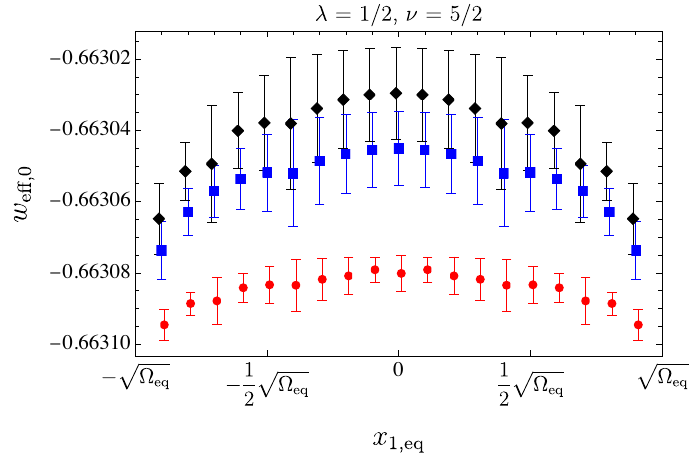} \hfill
\includegraphics[width=0.49\textwidth]{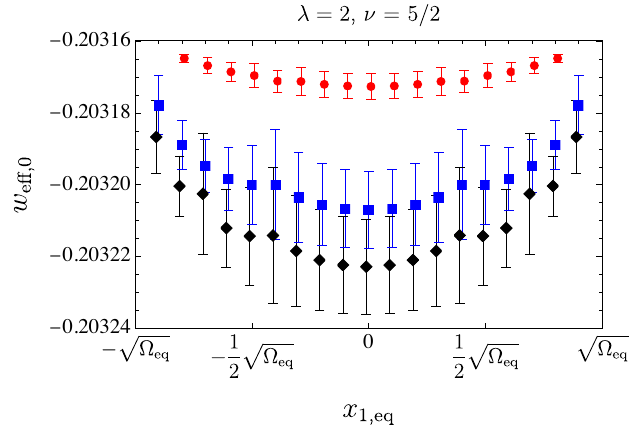}}
\caption{Present effective EoS parameter for different values of $\Omega_{k,\eq}$: $5 \times 10^{-11}$ (red dots), $5\times 10^{-10}$ (blue squares) and $9.5\times 10^{-11}$ (black diamonds). Error bars denote standard deviation due to the $x_{2, \eq}$ induced dispersion. Two values of $\lambda$, $1/2$ (left) and $2$ (right), are shown for $\nu=5/2$.}   
\label{fig:sampledataforWeff}
\end{figure}

An example is shown in Fig.~\ref{fig:sampledataforWeff}, which displays the resulting effective equation of state $w_{\rm eff,0}$ today for a range of initial velocities. This explicitly confirms that for $\lambda \lesssim 2$, the dispersion in $w_{\rm eff}$ due to variations in $(x_{1,\eq}, x_{2,\eq})$ remains minimal (of order $10^{-4}$), indicating that the system effectively tracks a common thawing trajectory regardless of the exact initial state. The small dispersion reflects the dynamical robustness of the thawing evolution, indicating that the analytic threshold $\lambda \simeq 1.7$ derived earlier for sustained acceleration today is robust against variations in the initial conditions.

\begin{figure}[t!]
    \centering
    {\includegraphics[width=0.49\textwidth]{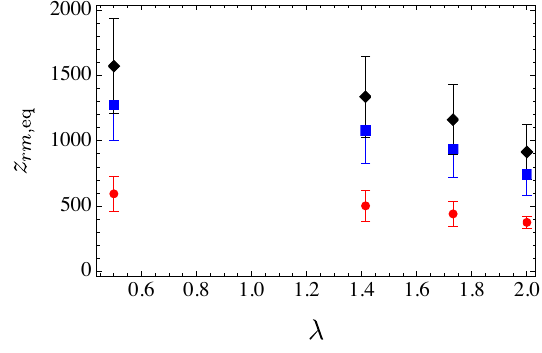} \hfill
    \includegraphics[width=0.475\textwidth]{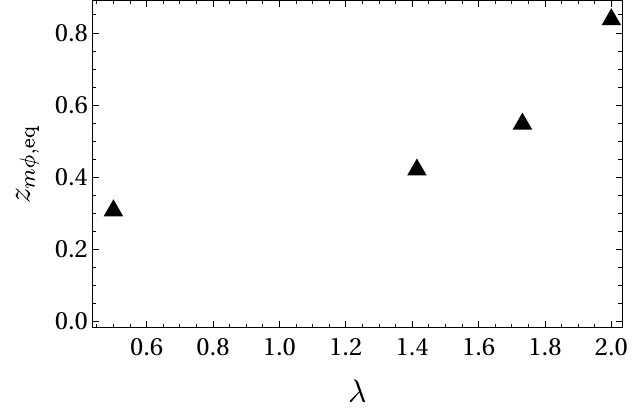}}
    \caption{\textbf{Left:} Redshift of matter-radiation equality. \textbf{Right:} Redshift of matter-scalar equality. Both are plotted as a function of $\lambda$. The dependence on $\lambda$ illustrates how the steepness of the potential affects the global expansion history, shifting the timing of key cosmological epochs. The color code is the same as in Fig.~\ref{fig:sampledataforWeff}.}
    \label{fig:zrmeq}
\end{figure}

A complementary summary of the same binning procedure is presented in Fig.~\ref{fig:zrmeq}, which displays the redshift of matter–radiation equality and matter–scalar equality as functions of $\lambda$.
The right panel (matter–scalar equality) exhibits the expected behavior: increasing $\lambda$ leads to an earlier onset of scalar domination. This reflects the enhanced kinetic contribution associated with steeper potentials, which accelerates the departure from matter domination and modifies the timing of the dark energy transition.

More remarkably, however, the left panel reveals that the redshift of matter–radiation equality, $z_{rm,\mathrm{eq}}$, also undergoes nonnegligible shifts as $\lambda$ varies. Although the dark energy fraction at equality is fixed to be extremely small, the integrated expansion history between equality and the present depends on the scalar dynamics. Consequently, different values of $\lambda$ lead to distinct late-time normalizations of the Hubble rate and therefore to measurable shifts in the epoch at which $\rho_m = \rho_r$.

Even more striking is the fact that the dispersion induced by varying the initial energy partition between the two scalar components and curvature is of the same order as the shift produced by changing $\lambda$ itself. The error bars and the separation among the three curves clearly show that redistributing the small dark-energy fraction at equality —whether between the radial and axionic components or between scalar energy and curvature— produces variations in $z_{rm,\mathrm{eq}}$ comparable to those obtained by modifying the potential slope.

This observation highlights a genuine degeneracy at the level of background observables: within the present framework, variations in scalar dynamics and variations in curvature can imprint effects of similar magnitude on early-Universe milestones. At the same time, it identifies $z_{rm, \mathrm{eq}}$ as an especially sensitive probe of the underlying field space dynamics. Since this epoch is tightly constrained by CMB measurements and directly affects the sound horizon and the subsequent growth of structure, even moderate shifts may provide strong indirect bounds on the combined scalar–curvature sector. This interplay between sensitivity and degeneracy motivates a closer examination of the relative dynamical roles of the fundamental parameters, to which we now turn.

\subsubsection{Parameter hierarchy: $\lambda$ vs\ $\nu$}

The statistical analysis of the trajectories confirms, at the fully numerical level, the hierarchy of physical effects anticipated from the analytical thawing treatment. This hierarchy is crucial for interpreting the degeneracies discussed above and for guiding subsequent parameter inference.

\paragraph{1. The decoupling of $\nu$.}
A central result of the numerical scan is the effective decoupling of the nongeodesic parameter $\nu$ at the background level. Comparing the left ($\nu=0$) and right ($\nu=2.5$) panels of Fig.~\ref{fig:sampledataforWeff2}, we observe virtually identical distributions for $w_{\rm eff,0}$. While $\nu$ significantly alters the asymptotic structure of the phase space —determining whether the system ultimately approaches the nongeodesic attractor $\fp_{\cal NG}$— its impact on the trajectory from matter-radiation equality to the present epoch is negligible.

Physically, this behavior is understood from the fact that during the thawing phase the turning component remains subdominant, $x_2^2 \ll x_1^2, y^2$. The background evolution is therefore controlled almost entirely by the radial degree of freedom. Consequently, observational constraints on $\nu$ derived from background geometrical probes are expected to be weak or largely uninformative, unless supplemented by perturbative or future asymptotic observables.

\paragraph{2. The control parameter $\lambda$.}
By contrast, the potential slope $\lambda$ clearly emerges as the primary control parameter of late-time cosmology. As shown in Fig.~\ref{fig:sampledataforWeff}, there is a sharp transition in viability around $\lambda \approx 1.7$. For $\lambda \lesssim 1.7$, the mean equation of state lies safely within the acceleration regime ($w_{\rm eff} < -1/3$). As $\lambda$ increases beyond this threshold, $w_{\rm eff}$ rapidly approaches zero, reflecting the growing kinetic contribution $x_{1,0}^2$ and the consequent loss of accelerated expansion.

This numerical threshold is in remarkable agreement with the analytical bound $\lambda < 1.68$ obtained in Eq.~\eqref{eq:maxanalyticallambdaacceleration}, confirming that the thawing approximation captures the essential physics of the matter–dark energy transition.

Moreover, the influence of $\lambda$ extends beyond late-time acceleration. As illustrated in Fig.~\ref{fig:zrmeq}, increasing $\lambda$ systematically shifts the redshifts of both matter–scalar and matter–radiation equality. In particular, the sensitivity of $z_{rm, \mathrm{eq}}$ to $\lambda$ is comparable to the shifts induced by redistributing energy between the scalar components and curvature. This reinforces the conclusion that $\lambda$ governs the integrated expansion history and that early-Universe observables may provide complementary constraints on the slope, albeit subject to degeneracies with curvature and field space energy partition.

\subsection{Structural absence of a nongeodesic scaling solution}\label{Sec: nongedesic and scaling}

The numerical hierarchy identified above admits a deeper dynamical explanation. 

Although the full system is genuinely multifield, the numerical phase-space trajectories indicate that the second field does not participate in the scaling behavior that characterizes its radial partner. As shown in Sec.~\ref{Sec: Single Barotropic}, while the system admits the scaling fixed point $\fp_\mc{S}$ and the nongeodesic scalar-dominated point $\fp_\mc{NG}$, no fixed point simultaneously combines scaling behavior with nongeodesic motion in the presence of a background fluid. This raises a structural question: why does the axion fail to admit a scaling solution analogous to that of $\phi_1$? In what follows, we provide an analytical explanation of this behavior and show that the absence of axion tracking is a generic property of the system rather than a numerical artifact.

To address this, consider a matter-dominated universe where the background density scales as $\rho_m \propto a^{-3} \propto t^{-2}$, with $H = 2/3t$. For the radial field $\phi_1$ to track this background (scaling regime), its energy density must satisfy $\rho_{\phi_1} \propto t^{-2}$. Assuming the dynamics of $\phi_1$ are dominated by its potential $V(\phi_1) = V_0 e^{-\lambda \phi_1/\M}$ and kinetic terms (temporarily neglecting the coupling to $\phi_2$), the scaling ansatz implies
\begin{equation}
    \rho_{\phi_1} \sim \dot{\phi}_1^2 \sim \frac{1}{t^2}  
    \quad \implies \quad 
    \phi_1(t) = \phi_{c} \ln \left(\frac{t}{t_0}\right)\,,
\end{equation}
where $\phi_c$ and $t_0$ are integration constants. Inserting this logarithmic evolution into the Klein-Gordon equation (Eq.~\ref{Eq: phi eom}) yields a consistent relation between the amplitude and the potential slope,
\begin{equation}
    \frac{\phi_c}{\M} = \frac{2}{\lambda}\,, 
    \qquad 
    \frac{V_0}{\M^2} = \frac{2}{\lambda^2}\,.
\end{equation}
Notice that it is precisely the existence of the exponential potential that allows the kinetic and potential contributions to scale proportionally to the Hubble friction term. In other words, the specific exponential form enables a self-consistent balance between potential energy, kinetic energy, and Hubble damping, preventing the field from diluting purely kinetically ($a^{-6}$) and thereby generating the attractor solution $\fp_\mc{S}$.

In stark contrast, such a mechanism is absent for the axion field. Since $\phi_2$ lacks a potential, its equation of motion is purely a conservation law for its momentum. The covariant current equation reads
\begin{equation}
    \nabla_\mu \left( f^2(\phi_1)\nabla^\mu \phi_2 \right) = 0 
    \quad \implies \quad 
    \dot{\phi}_2 = \frac{J}{a^3 f^2(\phi_1)}\,,
\end{equation}
where $J$ is the conserved Noether current. Consequently, the energy density of the axion evolves as
\begin{equation}
    \rho_{\phi_2} 
    = \frac{1}{2} f^2(\phi_1) \dot{\phi}_2^2 
    = \frac{J^2}{2} \frac{1}{a^6 f^2(\phi_1)}\,.
\end{equation}
If the coupling is negligible ($f \approx \text{const}$), the axion behaves as a stiff fluid, $\rho_{\phi_2} \propto a^{-6}$. Since matter scales as $a^{-3}$, the axion density decays extremely rapidly relative to the background, rendering scaling impossible. In this limit, no attractor mechanism exists that could lock $\rho_{\phi_2}$ to $\rho_m$.

Even in the presence of coupling, scaling requires a highly restrictive condition. Substituting the scaling solution for the radial field, $\phi_1 \propto \ln t \propto \ln a^{3/2}$, into the coupling function $f(\phi_1) \propto e^{-\nu \phi_1}$, we find
\begin{equation}
    \rho_{\phi_2} \propto a^{-6} e^{2\nu \phi_1} 
    \propto a^{-6\left( 1 - \frac{\nu}{\lambda} \right)}.
\end{equation}
For the axion to scale with matter ($\rho_{\phi_2} \propto a^{-3}$), one would require the precise tuning
\begin{equation}
    \nu = \frac{\lambda}{2}.
\end{equation}
This condition is highly restrictive, since $\lambda$ and $\nu$ are independent parameters of the theory —typically determined by the underlying UV construction— and are not expected to satisfy such a relation in general. Moreover, as shown in Table~\ref{tab:stabilitymultifieldandbaro}, the relation $\lambda = 2\nu$ coincides exactly with the bifurcation boundary between $\fp_\mc{S}$ and $\fp_\mc{NG}$. Axion scaling therefore occurs only on the codimension-one surface of the full dynamical system separating distinct dynamical regimes, rather than within an open region of parameter space.\footnote{A particular UV-motivated configuration that satisfies this relation and lies on this boundary is the triple point discussed in Sec.~\ref{Sec: Single Barotropic}, as shown in Fig.~\ref{fig:axiodilatoncurvatureStabdiag}.}

We therefore conclude that a genuine scaling solution involving the axion does not arise as a stable attractor of the system. Unlike the radial field, the axion lacks an intrinsic mechanism capable of locking its energy density to that of the background fluid. Except for the finely tuned case $\lambda = 2\nu$, corresponding to the bifurcation boundary of the dynamical system, the axion evolves independently of the matter component and does not exhibit tracking behavior.

In the parameter regime relevant for late-time acceleration ($\nu > \lambda$), the axion energy density redshifts more slowly than matter and can eventually become dynamically relevant as the system evolves toward the nongeodesic fixed point $\fp_\mc{NG}$. However, in the absence of a tracking attractor during the matter era, its contribution remains subdominant throughout the intermediate expansion history. Only when the evolution departs from matter domination does the axion participate in the onset of the dark energy phase. This explains why, despite being present in the full multifield system, the axion plays a negligible role in the background dynamics during the epochs probed by current observations, in agreement with our numerical results.

Analogous considerations apply to a generic barotropic fluid with EoS $w=\alpha-1$. For illustrative purposes, and in direct connection with the present epoch, we have focused on the matter case $\alpha=1$. However, the same reasoning extends straightforwardly to arbitrary $\alpha$, leading to the condition,
\begin{equation}
    \lambda = \frac{2\alpha\,\nu}{2-\alpha}\,,
\end{equation}
for the axion to scale with the background fluid. This requirement again coincides precisely with the coexistence boundary between the nongeodesic and the scaling fixed points.

More generally, this demonstrates that axion scaling ---in the absence of a scalar potential and driven purely by nongeodesic effects--- does not correspond to a robust attractor within the physical parameter space. Instead, it can only occur on the bifurcation surface separating distinct dynamical regimes. Away from this boundary, no tracking solution exists that allows the axion to follow the background fluid.

\subsection{Curvature evolution and late-time degeneracies}

Collecting the analytical and numerical results of this section, a coherent dynamical hierarchy emerges.

Spatial curvature alone does not substantially relax the acceleration condition. The thawing dynamics impose a robust upper bound $\lambda \lesssim 1.6$--$1.7$, confirmed by the full multifield scan. The nongeodesic parameter $\nu$ remains largely decoupled from the background evolution during the matter and early dark-energy eras, reflecting the structural absence of a nongeodesic scaling attractor in the presence of a background fluid.

At the same time, the integrated expansion history exhibits nontrivial degeneracies. Variations in the potential slope, in the initial scalar–curvature energy partition, and in the curvature abundance can induce comparable shifts in early-Universe milestones, particularly in the redshift of matter–radiation equality. While late-time observables such as $w_{\rm eff,0}$ are remarkably robust, equality epochs retain a residual sensitivity to the underlying multifield dynamics.

These results therefore suggest a clear observational expectation: background data should primarily constrain the potential slope $\lambda$ through the requirement of late-time acceleration, while the parameter $\nu$ is unlikely to be tightly bounded at this level. At the same time, precision measurements of early-Universe physics may indirectly probe the scalar–curvature sector through its impact on the integrated expansion history.

We now turn to a quantitative confrontation with observational data to assess how these dynamical expectations are realized in practice.

\section{Observational Constraints} 
\label{Sec: Constraints} 

In this section, we confront the curved multifield model with current cosmological observations. In particular, we test whether the dynamical expectations derived in the previous section —most notably the requirement of a sufficiently small potential slope to sustain late- time acceleration—are consistent with current data.

While analogous analyses including spatial curvature have recently been performed in the single-field exponential scenario \cite{Bhattacharya:2024hep, Alestas:2024gxe}, a systematic confrontation of the minimal multifield setup with data is still lacking. Our primary goal is therefore to determine the observational constraints on the potential slope $\lambda$, which directly controls the onset of acceleration and provides the link to UV-motivated expectations. In this sense, bounding $\lambda$ observationally amounts to testing the extent to which steep exponential potentials, as suggested by string-inspired constructions and the Swampland conjectures, remain compatible with late-time cosmology.

To perform the parameter inference, we compute the background evolution of the model using a modified version of the Boltzmann solver \texttt{CLASS}~\cite{Lesgourgues:2011re, Blas:2011rf}. The cosmological parameter space is explored through the Metropolis–Hastings algorithm using the Monte Carlo Markov Chain (MCMC) code \texttt{MontePython}~\cite{Brinckmann:2018cvx}, allowing us to propagate the dynamical scalar-field evolution consistently into the cosmological observables.

\subsection{Datasets}

To constrain the background dynamics, we employ a robust combination of geometric probes. Specifically, we utilize:

\begin{itemize} 
    \item \textbf{CMB:} Compressed likelihoods based on the acoustic scale shift parameters ($\ell_a, R_a$) and the baryon density $\omega_b$ from the \textit{Planck} 2018 release~\cite{Planck:2018vyg}, which effectively capture the geometric information of the last scattering surface. 
    \item \textbf{SNe Ia:} Luminosity distance measurements from the Pantheon+ sample~\cite{Brout:2022vxf}, constraining the expansion history in the late universe ($0.01 < z < 2.3$). 
    \item \textbf{BAO:} Baryon acoustic oscillation distance measurements. We utilize low-redshift anchors from 6dFGS~\cite{2011MNRAS.416.3017B} and SDSS DR7 MGS~\cite{Ross:2014qpa} to anchor the low-redshift distance ladder. 
    \item \textbf{CC:} Cosmic Chronometers measurements of the Hubble parameter $H(z)$~\cite{Guo:2015gpa, Moresco:2016mzx}, serving as a model-independent metric of the expansion rate.
\end{itemize}

Since the observational probes employed in this analysis enter primarily through geometric constraints on the expansion history, we restrict the implementation to the homogeneous background evolution of the scalar sector, and therefore do not include linear perturbations of the scalar fields in the Boltzmann hierarchy.

\subsection{Boltzmann solver implementation} 

To fully capture the interplay between the geometry and the field sector, we extend the dynamical system to include the curvature as an evolving degree of freedom. We define the dimensionless variable $\tilde z$ associated with the curvature energy density:
\begin{equation}
    \tilde z \equiv \frac{\sqrt{|k|}}{a},
\end{equation}
such that the Friedman constraint includes the curvature contribution dynamically:
\begin{equation} 
    H^2 = \frac{1}{3 \M^2}(\rho_m + \rho_r) + H^2\left(\tilde{x}_1^2 + \tilde{x}_2^2 + \tilde{y}^2 + \tilde z^2\right).
\end{equation} 
In our modified \texttt{CLASS} implementation, we solve the coupled system for the vector $\mathbf{v} = \{\tilde x_1, \tilde x_2, \tilde y, \tilde z\}$. These variables are initialized deep in the radiation era ($z=10^{14}$) and evolved self-consistently. We employ the \textit{shooting method} already implemented in \texttt{CLASS} to ensure that the cosmological evolution from these high-redshift initial conditions converges to a Universe consistent with the properties specified by the input cosmological parameters.

Crucially, rather than treating the present-day curvature parameter $\Omega_{k,0}$ as a free parameter, we promote the curvature contribution to a dynamical variable determined by its initial conditions. Specifically, we initialize $\tilde z$ with a trace abundance at equality, consistent with inflationary expectations of near flatness, and evolve it self-consistently alongside the scalar fields. The shooting procedure is implemented to ensure that the total energy budget satisfies the Friedman constraint at $z=0$. In this way, the curvature fraction $\Omega_k(z) \propto \tilde z^2/H^2$ emerges dynamically from the integrated expansion history, rather than being imposed as an external parameter. This procedure effectively replaces the usual flat prior on $\Omega_{k,0}$ by a prior on the initial curvature amplitude deep in the radiation era.

\subsection{Results and discussion}

\begin{figure}[t]
    \centering
    \includegraphics[width=\linewidth]{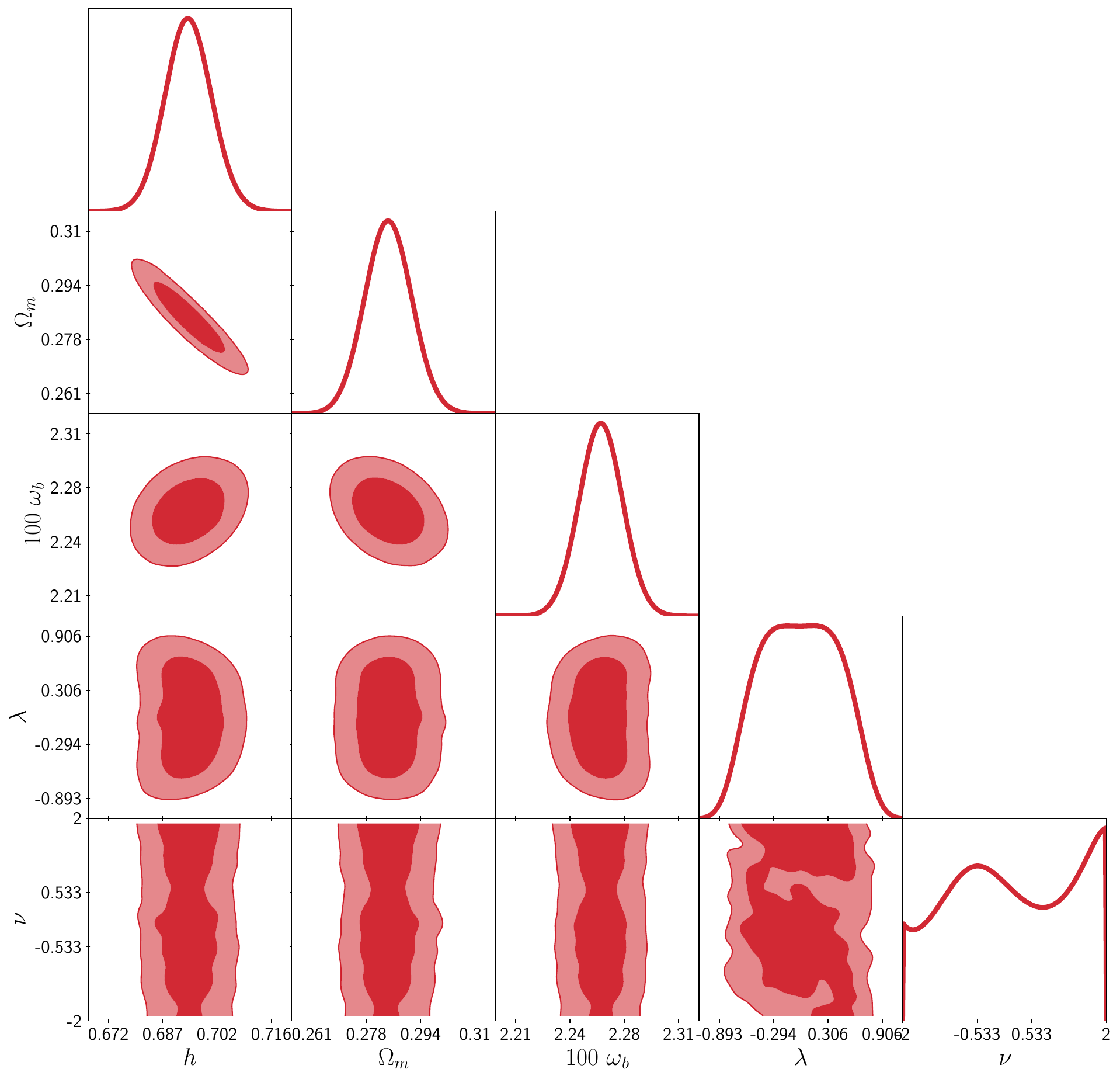} 
    \caption{Posterior distributions for the model parameters obtained from the combined analysis (CMB + SNe Ia + BAO + CC). The analysis incorporates the full dynamical evolution of the curvature field $\tilde{z}$. Despite this, the potential slope $\lambda$ remains constrained to small values.}
    \label{fig:triangle_plot}
\end{figure}

We performed a statistical inference on the parameter vector $\mathcal{P} \equiv \{h, \Omega_m, \omega_b, \lambda, \nu\}$. The curvature density $\Omega_k(z)$ is not treated as a free parameter, but is dynamically reconstructed from the evolution of the variable $\tilde{z}$ and the Hubble flow. The resulting marginalized posterior distributions are displayed in Fig.~\ref{fig:triangle_plot}.

The analysis yields two crucial physical conclusions:
\begin{enumerate}
    \item \textbf{Suppression of the potential slope:} The posterior for $\lambda$ is strongly peaked at small values, with a $95\%$ C.L. upper bound $\lambda \lesssim 0.75$. This result is particularly noteworthy because the analysis incorporates the full dynamical evolution of spatial curvature. Even with this additional degree of freedom, the data favor a nearly flat potential to reproduce the observed late-time acceleration. Physically, this bound arises because sufficiently steep exponential potentials fail to produce a negative enough equation of state during the late-time thawing phase. In practice, this bound leaves little room for steep exponential potentials ($\lambda \gtrsim 1$), indicating that the inclusion of dynamical curvature does not significantly relax the tension between accelerated expansion and UV-motivated expectations.
    
    \item \textbf{Decoupling of the axion:} The posterior distribution for the coupling parameter $\nu$ is effectively flat, indicating that current background data provide no significant constraint on the axionic turning rate. This observational insensitivity directly reflects the numerical evolution described in Sec.~\ref{sec:numericalscan}: during the redshift interval relevant for present-day observables, the axion remains energetically subdominant ($x_2 \approx 0$) and has a negligible impact on the expansion history. As a result, the nongeodesic mechanism remains phenomenologically dormant at the background level. No significant degeneracy between $\lambda$ and $\nu$ is observed in the posterior distribution, confirming that the turning rate remains dynamically suppressed during the observable epoch.
\end{enumerate}

These results are summarized in Table~\ref{tab:parameter_constraints}.

\begin{table}[htb!]
    \centering
    \caption{Mean values and $1\sigma$ errors (or $95\%$ upper limits) for the model parameters obtained from the combined analysis.}
    \label{tab:parameter_constraints}
    \begin{tabular}{l c} 
        \toprule
        Parameter & Constraint (68\% C.L.) \\
        \midrule
        $h$ & $0.694 \pm 0.006$ \\
        $\Omega_m$ & $0.284 \pm 0.007$ \\
        $100\omega_b$ & $2.26 \pm 0.014$ \\
        \midrule
        $\lambda$ & $< 0.75$ \quad ($95\%$ C.L.) \\
        $\nu$ & Unconstrained \\
        \bottomrule
    \end{tabular}
\end{table}

Now we compare our findings with those reported in Refs.~\cite{Bhattacharya:2024hep} and \cite{Alestas:2024gxe}, where the single-field limit of the model (corresponding to our $x_2 = 0$ case) is investigated.

In Ref.~\cite{Bhattacharya:2024hep}, the authors assume negligible initial field velocity and amplitude, while the overall normalization of the potential is fixed through the Friedman constraint. Under these assumptions, the potential slope $\lambda$ becomes the only independent model parameter. Our approach is conceptually similar in that we also fix the initial conditions and determine the potential amplitude via a shooting procedure that enforces the Friedman constraint at the present time. However, in our case both $\lambda$ and $\nu$ remain as free parameters, although we find that $\nu$ is observationally unconstrained at the background level.

A key difference lies in the datasets and methodology employed. Reference~\cite{Bhattacharya:2024hep} performs a full perturbation-level analysis and uses more constraining datasets, including the full CMB likelihood and DESI BAO measurements, whereas our analysis is restricted to background geometric probes (CMB distance priors, SNe Ia, BAO, and cosmic chronometers). Despite these methodological differences, both analyses agree in excluding steep exponential potentials ($\lambda \gtrsim 1$) at high statistical significance. Interestingly, Ref.~\cite{Bhattacharya:2024hep} also reports that $\lambda = 0$ is disfavored at the $2\sigma$ level, a result that we do not observe in our background-only analysis. This difference likely reflects the additional constraining power of perturbation observables.

On the other hand, Ref.~\cite{Alestas:2024gxe} revisits the background dynamics while allowing both the initial field velocity $\dot{\phi}_i$ and the potential amplitude $V_0$ to vary as free parameters during inference. Their resulting bound on $\lambda$ is qualitatively similar to ours. However, when both $V_0$ and the initial kinetic energy are treated as independent parameters, special care must be taken to ensure consistency with the Friedman constraint. In particular, one must specify which cosmological parameter is derived from the closure condition, since not all background quantities can be freely varied simultaneously. The treatment of this consistency condition is not entirely explicit in Ref.~\cite{Alestas:2024gxe}, which may partly explain minor quantitative differences with respect to our results.

Overall, the agreement across these independent analyses strengthens the conclusion that current cosmological data favor very small values of the potential slope, and that allowing for dynamical curvature or additional field space structure does not significantly relax this constraint at the background level.

\section{Conclusions}
\label{Sec: Conclusions}

In this work we have carried out a systematic dynamical and phenomenological analysis of the minimal multifield exponential dark energy model, consisting of a modulus--axion system evolving on a curved field space and embedded in a curved FLRW spacetime with radiation, matter, and spatial curvature components. Our goal was to determine how the interplay between field space curvature and spacetime curvature constrains the late-time cosmological dynamics and whether their combined effect enlarges the viable parameter space of exponential quintessence.

\subsection*{Global dynamical structure}

We provided a complete phase-space classification of the autonomous system in the presence of a generic barotropic fluid. All critical points were identified, their stability properties determined, and the invariant submanifolds organizing the global flow characterized. The system admits three possible late-time attractors: the geodesic scalar-dominated solution $\fp_{\mc G}$, the nongeodesic turning solution $\fp_{\mc NG}$, and the scaling solution $\fp_{\mc S}$ associated with the background fluid. Their stability regions are separated by well-defined bifurcation curves in the $(\lambda,\nu)$ parameter space.

When confronted with parameter values motivated by effective supergravity constructions from string compactifications, we find that the preferred regions typically lie within the basin of attraction of the scaling solution, where the scalar sector tracks the background fluid and the system does not exhibit accelerated expansion. In particular, the discrete loci corresponding to standard $F$-term and $D$-term realizations do not generically populate the region of parameter space where the nongeodesic accelerated fixed point is both stable and viable. This suggests that sustained turning-induced acceleration is not generically realized within minimal string-motivated constructions.

More precisely, this conclusion applies within the asymptotic regime of moduli spaces ($\phi \to \infty$), where the exponential potentials and curved field space geometries considered here provide the relevant effective description. Outside this regime, more intricate dynamics could in principle arise.

Within this framework, a central structural result of our analysis is the absence of a stable nongeodesic scaling fixed point in the presence of a background fluid within an open region of parameter space. While the radial field admits a scaling solution due to the exponential form of its potential, the axion lacks a potential term and obeys a conserved current equation. Its energy density therefore scales as $a^{-6}$ (modulo coupling effects), preventing it from tracking the background in a manner analogous to the radial field. This analytical mechanism explains why no fixed point simultaneously combines scaling behavior and sustained turning dynamics once matter is included. The absence of axion tracking is thus a structural property of the system rather than a numerical artifact.

Although $\fp_{\mc NG}$ may be a stable spiral in phase space, inducing damped oscillatory convergence of the effective equation of state, this behavior is restricted to the asymptotic future and is not relevant for the observable era.

\subsection*{Numerical evolution and parameter hierarchy}

We complemented the analytical classification with an extensive numerical scan of model parameters and initial conditions around radiation--matter equality, allowing the reconstruction of complete cosmological histories.

The numerical analysis confirms a clear hierarchy among the fundamental parameters. While the nongeodesic coupling $\nu$ determines the asymptotic structure of the phase space and whether the system ultimately approaches $\fp_{\mc NG}$, the potential slope $\lambda$ emerges as the primary control parameter of the observable background evolution. Moreover, the full dynamical evolution reproduces the thawing prediction of a critical value near $\lambda \simeq 1.7$ and shows that this estimate remains largely unaffected by the presence of the nongeodesic component $x_2$.


Most background observables are robust under variations of initial conditions within the viable region of parameter space. A notable exception is the redshift of radiation--matter equality, $z_{rm, \rm eq}$, which exhibits enhanced sensitivity to the model parameters. Variations in $\lambda$ shift $z_{rm, \rm eq}$ in a way that mimics a redistribution of energy between the scalar sector and spatial curvature. As a result, $z_{rm, \rm eq}$ acts as a sensitive but degenerate probe of the dark sector: while it constrains the total nonstandard energy contribution, it does not disentangle the internal energy partition of the multifield system nor clearly distinguish it from curvature effects.

Within the region consistent with present-day cosmological abundances, both the axionic degree of freedom and spatial curvature remain dynamically subdominant during the observable epochs. The late-time universe is therefore generically described by an effective thawing regime close to the geodesic submanifold.

\subsection*{Constraints and ultraviolet implications}

Using the effective late-time description, we derived a dynamical upper bound on the potential slope from the requirement of a viable matter--dark energy transition,
\[
\lambda < 1.6 ,
\]
in agreement with the numerical scan. 

A subsequent MCMC analysis, which employed background observations from Planck 2018, Pantheon+, BAO, and Cosmic Chronometers, further constrains the model to
\[
\lambda \lesssim 0.75 \qquad (95\% \ \mathrm{C.L.}),
\]
while leaving the nongeodesic coupling parameter $\nu$ essentially unconstrained. This observational decoupling of $\nu$ is consistent with the dynamical hierarchy identified above and reflects the fact that the axionic degree of freedom remains energetically subdominant during the observable epoch.

Taken together, these results indicate that the combined effects of spatial curvature and nongeodesic motion do not enlarge the viable region of parameter space relative to the effectively single-field case. Although field space curvature enriches the global dynamical structure and permits qualitatively distinct asymptotic behaviors, realistic cosmological evolution drives the system toward an effectively geodesic thawing regime during the observable era.

These conclusions have direct implications for ultraviolet-motivated scenarios. Exponential potentials with $\lambda = \mathcal{O}(1)$ arise naturally in many string-inspired effective theories and are compatible with Swampland-type slope bounds. Within the minimal curved multifield exponential setup analyzed here, however, neither spatial curvature nor turning effects reconcile steep exponential slopes with sustained late-time acceleration. Current observations instead select significantly smaller values of $\lambda$, reinforcing the tension between simple ultraviolet-motivated constructions and the requirements of dark energy phenomenology.

In this sense, the minimal two-field exponential model in curved spacetime provides a dynamically consistent and structurally well-controlled framework in which this tension can be clearly identified. Any successful reconciliation of steep ultraviolet-motivated potentials with the observed acceleration would therefore require departures from the minimal structure considered here, such as extended field sectors, modified potentials, or additional interactions.

A natural extension of the minimal framework considered here is to relax the assumption of an exactly flat axionic direction. The absence of an explicit potential for $\phi_2$ plays an important role in determining the phase-space structure of the model and in the absence of nongeodesic scaling solutions. More general axion--dilaton scenarios, in which the axion acquires a potential through symmetry-breaking effects, have recently been explored in a variety of cosmological contexts~\cite{Alexander:2022own,Burgess:2021obw,Smith:2024ibv,Berera:2025jbj,Smith:2025uaq}. The additional dynamical structure introduced by such potentials can modify the fixed-point structure and potentially alter some of the conclusions obtained within the minimal model. Extending the present dynamical-systems analysis to these scenarios constitutes an interesting direction for future work.

As a final remark, it is worth comparing our background constraints with recent results in the literature. Although steep potentials are heavily restricted, Ref.~\cite{Bhattacharya:2024hep} has shown that a single scalar field with a very flat potential is also disfavored by a robust combination of recent datasets (Planck, DESI, Pantheon$+$, Union3, and DES-Y5), yielding $\lambda = 0.77^{+0.18}_{-0.15}$ at $68\%$ C.L. Therefore, the phenomenological richness of multifield models at the perturbative level might offer a mechanism to relax the bound on $\lambda$. The presence of a second field inherently generates isocurvature modes that source the cosmological perturbations differently than adiabatic initial conditions. These modes can alter the decay of gravitational potentials, introducing nontrivial corrections to the Integrated Sachs-Wolfe (ISW) effect. Additionally, unlike canonical single-field models where the sound speed equals the speed of light, multifield interactions can reduce the effective sound speed, allowing dark energy to cluster and impact structure formation. Whether these perturbative effects can compensate for a steeper potential and push the upper bound of $\lambda$ closer to order-one values is a highly relevant question that we leave for future work.

\acknowledgments

D.G. acknowledges support from a sabbatical leave granted by the Universidad Pedagógica y Tecnológica de Colombia and gratefully acknowledges the hospitality of the Universidad del Valle during a research visit in which part of this work was developed. J.B.O.Q. acknowledges support from Vicerrectoría de Investigaciones-Universidad del Valle, Grant No. 71394.

We thank the anonymous referee for a careful and detailed assessment of the manuscript. Their comments and suggestions have significantly improved the clarity and presentation of this work.

D.G. dedicates this work to the victims of war and violence worldwide, and in particular to those suffering in Sudan and Palestine.

\appendix


\section{Consistency of the thawing approximation}
\label{app:subdominant}

In Sec.~\ref{sub:approximate_analytical} we derived an analytical
description of the thawing regime assuming that the axionic kinetic
component and spatial curvature remain subdominant,
so that the dynamics effectively reduce to the geodesic
submanifold $x_2 \simeq 0$. In this appendix we verify the
self-consistency of this approximation by studying the evolution of
the remaining variables within the analytical solution obtained
in that section.

\subsection*{Radiation and curvature}

Under the thawing assumption, potential domination, the evolution equations for the radiation abundance and curvature variable become
\begin{equation}
    \Omega_r' = -(1+3y^2)\Omega_r\,, \qquad 
    z' = \frac12(1-3y^2)\Omega_r\, .
\end{equation}
Thus, using the thawing solution $y(N) \simeq y_0 e^{3N/2}$ obtained in Sec.~\ref{sub:approximate_analytical}, the radiation abundance therefore follows
\begin{equation}
\Omega_r(N) = \Omega_{r0} \exp\!\left[-N - y_0^2\!\left(e^{3N}-1\right)\right]\,,
\end{equation}
which shows the expected dilution behavior during the matter and
dark-energy dominated epochs.

For the curvature variable one finds the approximate solution,
\begin{equation}
z(N) \simeq \left[ \Omega_{k0} \exp\!\left( N - y_0^2\!\left(e^{3N}-1\right)
\right) \right]^{1/2}\,.
\end{equation}

This function reaches a maximum at
\begin{equation}
N_{\rm max} =\frac13 \ln\!\left(\frac{1}{3y_0^2}\right)\,,
\end{equation}
with amplitude of order $\Omega_{k0}^{1/2}$. For observationally viable cosmologies, where
$\Omega_{k0}\ll1$ and $\Omega_{\phi0}\gtrsim0.1$, the curvature contribution therefore remains small throughout the thawing evolution.
This justifies neglecting curvature in the analytical treatment,
although in principle it could become relevant if the system
approaches the scaling–curvature fixed point discussed in
Sec.~\ref{Sec: nongedesic and scaling}.

\subsection*{Axionic kinetic component}

For the axionic variable we obtain
\begin{align}
x_2'&= x_2 \left( -\frac32 -3y^2 +2\sqrt6\,\nu x_1\right)\,,
\\
&= -x_2
\left(\frac32+ (3-2\lambda\nu)y^2 \right)\, .
\end{align}

Substituting the thawing solution for $y(N)$ gives
\begin{equation}
x_2' = -\frac12\left[ 3 + 2(3-2\lambda\nu)y_0^2 e^{3N} \right]x_2\, .
\end{equation}

For $\lambda\nu < 3/2$, the bracket is always positive and the axionic
component decays monotonically. The solution reads
\begin{equation}
x_2(N)
=
x_{20}
\exp\!\left[
-\frac{3N}{2}
+
\frac16 y_0^2(3-2\lambda\nu)
\left(1-e^{3N}\right)
\right].
\end{equation}

Growth of $x_2$ can only occur when $\lambda\nu>3/2$ and
\begin{equation}
N >
\frac13
\ln
\!\left[
\frac{3}{y_0^2(2\lambda\nu-3)}
\right].
\end{equation}
This region overlaps with the basin of attraction of the
non–geodesic fixed point, where $x_2$ eventually approaches
a non–vanishing value.

However, for initial conditions compatible with the thawing regime
the axionic component is expected to be extremely small,
$x_{2{\rm in}}\ll1$.
Integrating from an initial time $N_0$ gives
\begin{equation}
x_2(N)
=
x_{2{\rm in}}
\exp
\left[
\frac16
\left(
e^{3N}-e^{3N_0}
\right)
y_0^2(2\lambda\nu-3)
-\frac32(N-N_0)
\right].
\end{equation}

Evaluated today ($N=0$), this yields approximately
\begin{equation}
x_{20}
\simeq
x_{2{\rm in}}
\exp
\left[
\frac16 y_0^2(2\lambda\nu-3)
\right].
\end{equation}

For moderate values of $\lambda$ and $\nu$ this correction changes the
initial amplitude by at most a few orders of magnitude. Since
$x_{2{\rm in}}$ is expected to be extremely small (typically
$\lesssim10^{-4}$), the axionic kinetic component remains negligible
throughout the thawing evolution.

We therefore conclude that neglecting $x_2$, $\Omega_r$ and curvature
in the analytical treatment of Sec.~\ref{sub:approximate_analytical}
is self-consistent within the parameter region relevant for the
late-time cosmological evolution considered in this work.
\bibliographystyle{JHEP}
\bibliography{kasistedDE.bib}

@article{Achucarro:2018vey,
  title = {The String Swampland Constraints Require Multi-Field Inflation},
  author = {Ach{\'u}carro, Ana and Palma, Gonzalo A.},
  year = 2019,
  month = feb,
  journal = {JCAP},
  volume = {02},
  number = {02},
  eprint = {1807.04390},
  primaryclass = {hep-th},
  pages = {041},
  issn = {1475-7516},
  doi = {10.1088/1475-7516/2019/02/041},
  urldate = {2023-02-15},
  archiveprefix = {arXiv},
  langid = {english}
}

@article{Agrawal:2018own,
  title = {On the {{Cosmological Implications}} of the {{String Swampland}}},
  author = {Agrawal, Prateek and Obied, Georges and Steinhardt, Paul J. and Vafa, Cumrun},
  year = 2018,
  month = sep,
  journal = {Phys. Lett. B},
  volume = {784},
  eprint = {1806.09718},
  primaryclass = {hep-th},
  pages = {271--276},
  issn = {03702693},
  doi = {10.1016/j.physletb.2018.07.040},
  urldate = {2025-09-15},
  archiveprefix = {arXiv}
}

@article{Akrami:2020zfz,
  title = {Multi-Field Dark Energy: {{Cosmic}} Acceleration on a Steep Potential},
  shorttitle = {Multi-Field Dark Energy},
  author = {Akrami, Yashar and Sasaki, Misao and Solomon, Adam R. and Vardanyan, Valeri},
  year = 2021,
  month = jun,
  journal = {Phys. Lett. B},
  volume = {819},
  eprint = {2008.13660},
  primaryclass = {astro-ph.CO},
  pages = {136427},
  issn = {0370-2693},
  doi = {10.1016/j.physletb.2021.136427},
  urldate = {2023-02-07},
  archiveprefix = {arXiv},
  langid = {english}
}

@article{Alestas:2024gxe,
  title = {Is Curvature-Assisted Quintessence Observationally Viable?},
  author = {Alestas, George and Delgado, Matilda and Ruiz, Ignacio and Akrami, Yashar and Montero, Miguel and Nesseris, Savvas},
  year = 2024,
  month = nov,
  journal = {Phys. Rev. D},
  volume = {110},
  number = {10},
  eprint = {2406.09212},
  primaryclass = {hep-th},
  pages = {106010},
  issn = {2470-0010, 2470-0029},
  doi = {10.1103/PhysRevD.110.106010},
  urldate = {2024-12-30},
  archiveprefix = {arXiv},
  langid = {english}
}

@article{Alexander:2019rsc,
  ids = {alexanderAxiondilatonDestabilizationHubble2019a,alexanderAxiondilatonDestabilizationHubble2019b},
  title = {Axion-Dilaton Destabilization and the {{Hubble}} Tension},
  author = {Alexander, Stephon and McDonough, Evan},
  year = 2019,
  month = oct,
  journal = {Phys. Lett. B},
  volume = {797},
  eprint = {1904.08912},
  primaryclass = {astro-ph.CO},
  pages = {134830},
  issn = {03702693},
  doi = {10.1016/j.physletb.2019.134830},
  urldate = {2022-11-29},
  archiveprefix = {arXiv},
  langid = {english}
}

@article{Amendola:2016saw,
  title = {Cosmology and Fundamental Physics with the {{Euclid}} Satellite},
  author = {Amendola, Luca and Appleby, Stephen and Avgoustidis, Anastasios and others},
  year = 2018,
  month = apr,
  journal = {Living Rev. Rel.},
  volume = {21},
  number = {1},
  eprint = {1606.00180},
  primaryclass = {astro-ph.CO},
  pages = {2},
  issn = {2367-3613, 1433-8351},
  doi = {10.1007/s41114-017-0010-3},
  urldate = {2022-09-15},
  archiveprefix = {arXiv},
  langid = {english}
}

@article{Andriot:2023wvg,
  title = {Accelerated Expansion of an Open Universe and String Theory Realizations},
  author = {Andriot, David and Tsimpis, Dimitrios and Wrase, Timm},
  year = 2023,
  month = dec,
  journal = {Phys. Rev. D},
  volume = {108},
  number = {12},
  eprint = {2309.03938},
  primaryclass = {hep-th},
  pages = {123515},
  publisher = {American Physical Society},
  doi = {10.1103/PhysRevD.108.123515},
  urldate = {2025-03-21},
  archiveprefix = {arXiv}
}

@article{Andriot:2024jsh,
  title = {Exponential Quintessence: Curved, Steep and Stringy?},
  shorttitle = {Exponential Quintessence},
  author = {Andriot, David and Parameswaran, Susha and Tsimpis, Dimitrios and Wrase, Timm and Zavala, Ivonne},
  year = 2024,
  month = aug,
  journal = {JHEP},
  volume = {08},
  number = {8},
  eprint = {2405.09323},
  primaryclass = {hep-th},
  pages = {117},
  issn = {1029-8479},
  doi = {10.1007/JHEP08(2024)117},
  urldate = {2024-12-30},
  archiveprefix = {arXiv},
  langid = {english}
}

@article{Andriot:2024sif,
  title = {Quintessence: An Analytical Study, with Theoretical and Observational Applications},
  shorttitle = {Quintessence},
  author = {Andriot, David},
  year = 2025,
  month = jun,
  journal = {Fortsch. Phys.},
  volume = {73},
  number = {6},
  eprint = {2410.17182},
  primaryclass = {hep-th},
  pages = {e70007},
  doi = {10.1002/prop.70007},
  archiveprefix = {arXiv}
}

@article{Anguelova:2021jxu,
  title = {Dark {{Energy}} from {{Inspiraling}} in {{Field Space}}},
  author = {Anguelova, Lilia and Dumancic, John and Gass, Richard and Wijewardhana, L. C. R.},
  year = 2022,
  month = mar,
  journal = {JCAP},
  volume = {03},
  number = {03},
  eprint = {2111.12136},
  primaryclass = {hep-th},
  pages = {018},
  issn = {1475-7516},
  doi = {10.1088/1475-7516/2022/03/018},
  urldate = {2025-10-26},
  archiveprefix = {arXiv}
}

@article{AtacamaCosmologyTelescope:2025nti,
  title = {The {{Atacama Cosmology Telescope}}: {{DR6 Constraints}} on {{Extended Cosmological Models}}},
  shorttitle = {The {{Atacama Cosmology Telescope}}},
  author = {Calabrese, Erminia and Hill, J. Colin and Jense, Hidde T. and Posta, Adrien La and {Abril-Cabezas}, Irene and Addison, Graeme E. and Ade, Peter A. R. and Aiola, Simone and Alford, Tommy and Alonso, David and Amiri, Mandana and An, Rui and Atkins, Zachary and Austermann, Jason E. and Barbavara, Eleonora and Barbieri, Nicola and Battaglia, Nicholas and Battistelli, Elia Stefano and Beall, James A. and Bean, Rachel and Beheshti, Ali and Beringue, Benjamin and Bhandarkar, Tanay and Biermann, Emily and Bolliet, Boris and Bond, J. Richard and Capalbo, Valentina and Carrero, Felipe and Chen, Stephen and Chesmore, Grace and Cho, Hsiao-mei and Choi, Steve K. and Clark, Susan E. and Cothard, Nicholas F. and Coughlin, Kevin and Coulton, William and Crichton, Devin and Crowley, Kevin T. and Darwish, Omar and Devlin, Mark J. and Dicker, Simon and Duell, Cody J. and Duff, Shannon M. and Duivenvoorden, Adriaan J. and Dunkley, Jo and Dunner, Rolando and Villagra, Carmen Embil and Fankhanel, Max and Farren, Gerrit S. and Ferraro, Simone and Foster, Allen and Freundt, Rodrigo and Fuzia, Brittany and Gallardo, Patricio A. and Garrido, Xavier and Gerbino, Martina and Giardiello, Serena and Gill, Ajay and Givans, Jahmour and Gluscevic, Vera and Goldstein, Samuel and Golec, Joseph E. and Gong, Yulin and Guan, Yilun and Halpern, Mark and Harrison, Ian and Hasselfield, Matthew and He, Adam and Healy, Erin and Henderson, Shawn and Hensley, Brandon and {Herv{\'i}as-Caimapo}, Carlos and Hilton, Gene C. and Hilton, Matt and Hincks, Adam D. and Hlo{\v z}ek, Ren{\'e}e and Ho, Shuay-Pwu Patty and Hood, John and Hornecker, Erika and Huber, Zachary B. and Hubmayr, Johannes and Huffenberger, Kevin M. and Hughes, John P. and Ikape, Margaret and Irwin, Kent and Isopi, Giovanni and Joshi, Neha and Keller, Ben and Kim, Joshua and Knowles, Kenda and Koopman, Brian J. and Kosowsky, Arthur and Kramer, Darby and Kusiak, Aleksandra and Lague, Alex and Lakey, Victoria and Lattanzi, Massimiliano and Lee, Eunseong and Li, Yaqiong and Li, Zack and Limon, Michele and Lokken, Martine and Louis, Thibaut and Lungu, Marius and MacCrann, Niall and MacInnis, Amanda and Madhavacheril, Mathew S. and Maldonado, Diego and Maldonado, Felipe and {Mallaby-Kay}, Maya and Marques, Gabriela A. and van Marrewijk, Joshiwa and McCarthy, Fiona and McMahon, Jeff and Mehta, Yogesh and Menanteau, Felipe and Moodley, Kavilan and Morris, Thomas W. and Mroczkowski, Tony and Naess, Sigurd and Namikawa, Toshiya and Nati, Federico and Nerval, Simran K. and Newburgh, Laura and Nicola, Andrina and Niemack, Michael D. and Nolta, Michael R. and {Orlowski-Scherer}, John and Pagano, Luca and Page, Lyman A. and Pandey, Shivam and Partridge, Bruce and Sarmiento, Karen Perez and Prince, Heather and Puddu, Roberto and Qu, Frank J. and Ragavan, Damien C. and Guachalla, Bernardita Ried and Rogers, Keir K. and Rojas, Felipe and Sakuma, Tai and Schaan, Emmanuel and Schmitt, Benjamin L. and Sehgal, Neelima and Shaikh, Shabbir and Sherwin, Blake D. and Sierra, Carlos and Sievers, Jon and Sif{\'o}n, Crist{\'o}bal and Simon, Sara and Sonka, Rita and Spergel, David N. and Staggs, Suzanne T. and Storer, Emilie and Surrao, Kristen and Switzer, Eric R. and Tampier, Niklas and Thiele, Leander and Thornton, Robert and Trac, Hy and Tucker, Carole and Ullom, Joel and Vale, Leila R. and Engelen, Alexander Van and Lanen, Jeff Van and Vargas, Cristian and Vavagiakis, Eve M. and Wagoner, Kasey and Wang, Yuhan and Wenzl, Lukas and Wollack, Edward J. and Zheng, Kaiwen},
  year = 2025,
  month = nov,
  journal = {JCAP},
  volume = {11},
  eprint = {2503.14454},
  primaryclass = {astro-ph.CO},
  pages = {063},
  doi = {10.1088/1475-7516/2025/11/063},
  urldate = {2025-05-07},
  archiveprefix = {arXiv},
  collaboration = {ACT}
}

@article{Bernardo:2022ztc,
  ids = {bernardoheliudsonDarkSectorModel2022},
  title = {Towards a {{Dark Sector Model}} from {{String Theory}}},
  author = {Bernardo, Heliudson and Brandenberger, Robert and Fr{\"o}hlich, J{\"u}rg},
  year = 2022,
  month = sep,
  journal = {JCAP},
  volume = {09},
  number = {09},
  eprint = {2201.04668},
  primaryclass = {hep-th},
  pages = {040},
  issn = {1475-7516},
  doi = {10.1088/1475-7516/2022/09/040},
  urldate = {2022-12-22},
  archiveprefix = {arXiv}
}

@article{Bhattacharya:2024hep,
  title = {Cosmological Constraints on Curved Quintessence},
  author = {Bhattacharya, Sukannya and Borghetto, Giulia and Malhotra, Ameek and Parameswaran, Susha and Tasinato, Gianmassimo and Zavala, Ivonne},
  year = 2024,
  month = sep,
  journal = {JCAP},
  volume = {09},
  number = {09},
  eprint = {2405.17396},
  primaryclass = {astro-ph.CO},
  pages = {073},
  issn = {1475-7516},
  doi = {10.1088/1475-7516/2024/09/073},
  urldate = {2024-12-30},
  archiveprefix = {arXiv}
}

@article{Billyard:1999ct,
  title = {Qualitative Analysis of Early Universe Cosmologies},
  author = {Billyard, Andrew P. and Coley, Alan A. and Lidsey, James E.},
  year = 1999,
  journal = {J. Math. Phys.},
  volume = {40},
  number = {10},
  eprint = {gr-qc/9907043},
  pages = {5092},
  issn = {0022-2488},
  doi = {10.1063/1.533017},
  urldate = {2025-03-31},
  archiveprefix = {arXiv}
}

@article{Billyard:2000cz,
  title = {Cyclical Behavior in Early Universe Cosmologies},
  author = {Billyard, Andrew P. and Coley, Alan A. and Lidsey, James E.},
  year = 2000,
  journal = {J. Math. Phys.},
  volume = {41},
  number = {9},
  eprint = {gr-qc/0005118},
  pages = {6277--6283},
  issn = {0022-2488},
  doi = {10.1063/1.1286878},
  urldate = {2025-03-31},
  archiveprefix = {arXiv}
}

@article{Blaback:2013fca,
  title = {Accelerated {{Universes}} from Type {{IIA Compactifications}}},
  author = {Bl{\aa}b{\"a}ck, Johan and Danielsson, Ulf and Dibitetto, Giuseppe},
  year = 2014,
  month = mar,
  journal = {JCAP},
  volume = {03},
  eprint = {1310.8300},
  primaryclass = {hep-th},
  pages = {003},
  doi = {10.1088/1475-7516/2014/03/003},
  archiveprefix = {arXiv}
}

@article{Brennan:2017rbf,
  title = {The {{String Landscape}}, the {{Swampland}}, and the {{Missing Corner}}},
  author = {Brennan, T. Daniel and Carta, Federico and Vafa, Cumrun},
  year = 2017,
  month = dec,
  journal = {PoS},
  volume = {TASI2017},
  eprint = {1711.00864},
  primaryclass = {hep-th},
  pages = {015},
  doi = {10.22323/1.305.0015},
  urldate = {2022-11-09},
  archiveprefix = {arXiv}
}

@article{Brinkmann:2022oxy,
  title = {Stringy Multifield Quintessence and the {{Swampland}}},
  author = {Brinkmann, Max and Cicoli, Michele and Dibitetto, Giuseppe and Pedro, Francisco G.},
  year = 2022,
  month = nov,
  journal = {JHEP},
  volume = {11},
  number = {11},
  eprint = {2206.10649},
  primaryclass = {hep-th},
  pages = {044},
  issn = {1029-8479},
  doi = {10.1007/JHEP11(2022)044},
  urldate = {2023-02-04},
  archiveprefix = {arXiv},
  langid = {english}
}

@article{Brown:2017osf,
  title = {Hyperbolic {{Inflation}}},
  author = {Brown, Adam R.},
  year = 2018,
  month = dec,
  journal = {Phys. Rev. Lett.},
  volume = {121},
  number = {25},
  eprint = {1705.03023},
  primaryclass = {hep-th},
  pages = {251601},
  publisher = {American Physical Society},
  doi = {10.1103/PhysRevLett.121.251601},
  urldate = {2023-02-17},
  archiveprefix = {arXiv}
}

@article{Buniy:2006ed,
  title = {Does String Theory Predict an Open Universe?},
  author = {Buniy, Roman V. and Hsu, Stephen D. H. and Zee, A.},
  year = 2008,
  journal = {Phys. Lett. B},
  volume = {660},
  number = {4},
  eprint = {hep-th/0610231},
  pages = {382--385},
  issn = {0370-2693},
  doi = {10.1016/j.physletb.2008.01.007},
  urldate = {2024-12-30},
  archiveprefix = {arXiv}
}

@article{Caldwell:1997ii,
  title = {Cosmological {{Imprint}} of an {{Energy Component}} with {{General Equation}} of {{State}}},
  author = {Caldwell, R. R. and Dave, Rahul and Steinhardt, Paul J.},
  year = 1998,
  journal = {Phys. Rev. Lett.},
  volume = {80},
  number = {8},
  eprint = {astro-ph/9708069},
  pages = {1582--1585},
  publisher = {American Physical Society},
  doi = {10.1103/PhysRevLett.80.1582},
  urldate = {2025-04-25},
  archiveprefix = {arXiv}
}

@article{Catena:2007jf,
  title = {Axion--Dilaton Cosmology and Dark Energy},
  author = {Catena, Riccardo and M{\"o}ller, Jan},
  year = 2008,
  journal = {JCAP},
  volume = {03},
  number = {03},
  eprint = {0709.1931},
  primaryclass = {hep-ph},
  pages = {012},
  issn = {1475-7516},
  doi = {10.1088/1475-7516/2008/03/012},
  urldate = {2023-09-19},
  archiveprefix = {arXiv},
  langid = {english}
}

@article{Cespedes:2020xpn,
  title = {Lorentzian Vacuum Transitions: {{Open}} or Closed Universes?},
  shorttitle = {Lorentzian Vacuum Transitions},
  author = {C{\'e}spedes, Sebasti{\'a}n and {de Alwis}, Senarath P. and Muia, Francesco and Quevedo, Fernando},
  year = 2021,
  month = jul,
  journal = {Phys. Rev. D},
  volume = {104},
  number = {2},
  eprint = {2011.13936},
  primaryclass = {hep-th},
  pages = {026013},
  publisher = {American Physical Society},
  doi = {10.1103/PhysRevD.104.026013},
  urldate = {2025-11-07},
  archiveprefix = {arXiv}
}

@article{Chen:2016eyp,
  title = {{{CONSTRAINTS ON NON-FLAT COSMOLOGIES WITH MASSIVE NEUTRINOS AFTER PLANCK}} 2015},
  author = {Chen, Yun and Ratra, Bharat and Biesiada, Marek and Li, Song and Zhu, Zong-Hong},
  year = 2016,
  month = sep,
  journal = {Astrophys. J.},
  volume = {829},
  number = {2},
  eprint = {1603.07115},
  primaryclass = {astro-ph.CO},
  pages = {61},
  publisher = {The American Astronomical Society},
  issn = {0004-637X},
  doi = {10.3847/0004-637X/829/2/61},
  urldate = {2024-06-19},
  archiveprefix = {arXiv},
  langid = {english}
}

@article{Christodoulidis:2019jsx,
  ids = {christodoulidisScalingAttractorsMultifield2019a},
  title = {Scaling Attractors in Multi-Field Inflation},
  author = {Christodoulidis, Perseas and Roest, Diederik and Sfakianakis, Evangelos I.},
  year = 2019,
  month = dec,
  journal = {JCAP},
  volume = {12},
  number = {12},
  eprint = {1903.06116},
  primaryclass = {hep-th},
  pages = {059},
  publisher = {IOP Publishing},
  issn = {1475-7516},
  doi = {10.1088/1475-7516/2019/12/059},
  urldate = {2023-02-17},
  archiveprefix = {arXiv},
  langid = {english}
}

@article{Cicoli:2012tz,
  title = {Natural Quintessence in String Theory},
  author = {Cicoli, Michele and Pedro, Francisco G and Tasinato, Gianmassimo},
  year = 2012,
  journal = {JCAP},
  volume = {07},
  number = {07},
  eprint = {1203.6655},
  primaryclass = {hep-th},
  pages = {044},
  publisher = {IOP Publishing},
  issn = {1475-7516},
  doi = {10.1088/1475-7516/2012/07/044},
  archiveprefix = {arXiv}
}

@article{Cicoli:2020cfj,
  ids = {cicoliNewAcceleratingSolutions2020,cicoliNewAcceleratingSolutions2020b},
  title = {New Accelerating Solutions in Late-Time Cosmology},
  author = {Cicoli, Michele and Dibitetto, Giuseppe and Pedro, Francisco G.},
  year = 2020,
  month = may,
  journal = {Phys. Rev. D},
  volume = {101},
  number = {10},
  eprint = {2002.02695},
  primaryclass = {gr-qc},
  pages = {103524},
  issn = {2470-0010, 2470-0029},
  doi = {10.1103/PhysRevD.101.103524},
  urldate = {2024-06-04},
  archiveprefix = {arXiv},
  langid = {english}
}

@article{Cicoli:2020noz,
  title = {Out of the Swampland with Multifield Quintessence?},
  author = {Cicoli, Michele and Dibitetto, Giuseppe and Pedro, Francisco G.},
  year = 2020,
  month = oct,
  journal = {JHEP},
  volume = {10},
  number = {10},
  eprint = {2007.11011},
  primaryclass = {hep-th},
  pages = {035},
  issn = {1029-8479},
  doi = {10.1007/JHEP10(2020)035},
  urldate = {2023-02-04},
  archiveprefix = {arXiv},
  langid = {english}
}

@article{Clarkson:2007bc,
  title = {Dynamical Dark Energy or Simply Cosmic Curvature?},
  author = {Clarkson, Chris and Cort{\^e}s, Marina and Bassett, Bruce},
  year = 2007,
  journal = {JCAP},
  volume = {08},
  number = {08},
  eprint = {astro-ph/0702670},
  pages = {011},
  issn = {1475-7516},
  doi = {10.1088/1475-7516/2007/08/011},
  urldate = {2024-12-30},
  archiveprefix = {arXiv},
  langid = {english}
}

@article{Copeland:1997et,
  ids = {copelandExponentialPotentialsCosmological1998a},
  title = {Exponential Potentials and Cosmological Scaling Solutions},
  author = {Copeland, Edmund J. and Liddle, Andrew R. and Wands, David},
  year = 1998,
  journal = {Phys. Rev. D},
  volume = {57},
  number = {8},
  eprint = {gr-qc/9711068},
  pages = {4686--4690},
  publisher = {American Physical Society},
  doi = {10.1103/PhysRevD.57.4686},
  urldate = {2023-02-07},
  archiveprefix = {arXiv}
}

@article{Copeland:2006wr,
  title = {Dynamics of Dark Energy},
  author = {Copeland, Edmund J. and Sami, M. and Tsujikawa, Shinji},
  year = 2006,
  journal = {Int. J. Mod. Phys. D},
  volume = {15},
  number = {11},
  eprint = {hep-th/0603057},
  pages = {1753--1936},
  doi = {10.1142/S021827180600942X},
  archiveprefix = {arXiv}
}

@article{DAmico:2020euu,
  title = {Rollercoaster Cosmology},
  author = {D'Amico, Guido and Kaloper, Nemanja},
  year = 2021,
  month = aug,
  journal = {JCAP},
  volume = {08},
  number = {08},
  eprint = {2011.09489},
  primaryclass = {hep-th},
  pages = {058},
  publisher = {IOP Publishing},
  issn = {1475-7516},
  doi = {10.1088/1475-7516/2021/08/058},
  urldate = {2025-10-26},
  archiveprefix = {arXiv},
  langid = {english}
}

@article{deCruzPerez:2024shj,
  title = {Updated Observational Constraints on Spatially Flat and Nonflat},
  author = {P{\'e}rez, Javier de Cruz and Park, Chan-Gyung and Ratra, Bharat},
  year = 2024,
  month = jul,
  journal = {Phys. Rev. D},
  volume = {110},
  number = {2},
  eprint = {2404.19194},
  primaryclass = {astro-ph.CO},
  pages = {023506},
  publisher = {American Physical Society},
  doi = {10.1103/PhysRevD.110.023506},
  urldate = {2025-11-05},
  archiveprefix = {arXiv}
}

@article{DESI:2016fyo,
  title = {The {{DESI Experiment Part I}}: {{Science}},{{Targeting}}, and {{Survey Design}}},
  shorttitle = {The {{DESI Experiment Part I}}},
  author = {Aghamousa, Amir and Aguilar, Jessica and Ahlen, Steve and others},
  year = 2016,
  month = dec,
  journal = {arXiv:1611.00036 [astro-ph.IM]},
  eprint = {1611.00036},
  primaryclass = {astro-ph.IM},
  doi = {10.48550/arXiv.1611.00036},
  urldate = {2024-06-03},
  archiveprefix = {arXiv},
  collaboration = {DESI}
}

@article{DESI:2024mwx,
  title = {{{DESI}} 2024 {{VI}}: {{Cosmological Constraints}} from the {{Measurements}} of {{Baryon Acoustic Oscillations}}},
  shorttitle = {{{DESI}} 2024 {{VI}}},
  author = {Collaboration, {\relax DESI} and Adame, A. G. and Aguilar, J. and Ahlen, S. and Alam, S. and Alexander, D. M. and Alvarez, M. and Alves, O. and Anand, A. and Andrade, U. and Armengaud, E. and Avila, S. and Aviles, A. and Awan, H. and {Bahr-Kalus}, B. and Bailey, S. and Baltay, C. and Bault, A. and Behera, J. and BenZvi, S. and Bera, A. and Beutler, F. and Bianchi, D. and Blake, C. and Blum, R. and Brieden, S. and Brodzeller, A. and Brooks, D. and {Buckley-Geer}, E. and Burtin, E. and Calderon, R. and Canning, R. and Rosell, A. Carnero and Cereskaite, R. and {Cervantes-Cota}, J. L. and Chabanier, S. and Chaussidon, E. and {Chaves-Montero}, J. and Chen, S. and Chen, X. and Claybaugh, T. and Cole, S. and Cuceu, A. and Davis, T. M. and Dawson, K. and de la Macorra, A. and de Mattia, A. and Deiosso, N. and Dey, A. and Dey, B. and Ding, Z. and Doel, P. and Edelstein, J. and Eftekharzadeh, S. and Eisenstein, D. J. and Elliott, A. and Fagrelius, P. and Fanning, K. and Ferraro, S. and Ereza, J. and Findlay, N. and Flaugher, B. and {Font-Ribera}, A. and {Forero-S{\'a}nchez}, D. and {Forero-Romero}, J. E. and Frenk, C. S. and {Garcia-Quintero}, C. and Gazta{\~n}aga, E. and {Gil-Mar{\'i}n}, H. and Gontcho, S. Gontcho A. and {Gonzalez-Morales}, A. X. and {Gonzalez-Perez}, V. and Gordon, C. and Green, D. and Gruen, D. and Gsponer, R. and Gutierrez, G. and Guy, J. and Hadzhiyska, B. and Hahn, C. and Hanif, M. M. S. and {Herrera-Alcantar}, H. K. and Honscheid, K. and Howlett, C. and Huterer, D. and Ir{\v s}i{\v c}, V. and Ishak, M. and Juneau, S. and Kara{\c c}ayl{\i}, N. G. and Kehoe, R. and Kent, S. and Kirkby, D. and Kremin, A. and Krolewski, A. and Lai, Y. and Lan, T.-W. and Landriau, M. and Lang, D. and Lasker, J. and Goff, J. M. Le and Guillou, L. Le and Leauthaud, A. and Levi, M. E. and Li, T. S. and Linder, E. and Lodha, K. and Magneville, C. and Manera, M. and Margala, D. and Martini, P. and Maus, M. and McDonald, P. and {Medina-Varela}, L. and Meisner, A. and {Mena-Fern{\'a}ndez}, J. and Miquel, R. and Moon, J. and Moore, S. and Moustakas, J. and Mudur, N. and Mueller, E. and {Mu{\~n}oz-Guti{\'e}rrez}, A. and Myers, A. D. and Nadathur, S. and Napolitano, L. and Neveux, R. and Newman, J. A. and Nguyen, N. M. and Nie, J. and Niz, G. and Noriega, H. E. and Padmanabhan, N. and Paillas, E. and {Palanque-Delabrouille}, N. and Pan, J. and Penmetsa, S. and Percival, W. J. and Pieri, M. M. and Pinon, M. and Poppett, C. and Porredon, A. and Prada, F. and {P{\'e}rez-Fern{\'a}ndez}, A. and {P{\'e}rez-R{\`a}fols}, I. and Rabinowitz, D. and Raichoor, A. and {Ram{\'i}rez-P{\'e}rez}, C. and {Ramirez-Solano}, S. and Ravoux, C. and Rashkovetskyi, M. and Rezaie, M. and Rich, J. and Rocher, A. and Rockosi, C. and Roe, N. A. and {Rosado-Marin}, A. and Ross, A. J. and Rossi, G. and Ruggeri, R. and {Ruhlmann-Kleider}, V. and Samushia, L. and Sanchez, E. and Saulder, C. and Schlafly, E. F. and Schlegel, D. and Schubnell, M. and Seo, H. and Shafieloo, A. and Sharples, R. and Silber, J. and Slosar, A. and Smith, A. and Sprayberry, D. and Tan, T. and Tarl{\'e}, G. and Taylor, P. and Trusov, S. and {Ure{\~n}a-L{\'o}pez}, L. A. and Vaisakh, R. and Valcin, D. and Valdes, F. and {Vargas-Maga{\~n}a}, M. and Verde, L. and Walther, M. and Wang, B. and Wang, M. S. and Weaver, B. A. and Weaverdyck, N. and Wechsler, R. H. and Weinberg, D. H. and White, M. and Yu, J. and Yu, Y. and Yuan, S. and Y{\`e}che, C. and Zaborowski, E. A. and Zarrouk, P. and Zhang, H. and Zhao, C. and Zhao, R. and Zhou, R. and Zhuang, T. and Zou, H.},
  year = 2025,
  month = feb,
  journal = {JCAP},
  volume = {02},
  number = {02},
  eprint = {2404.03002},
  primaryclass = {astro-ph.CO},
  pages = {021},
  issn = {1475-7516},
  doi = {10.1088/1475-7516/2025/02/021},
  urldate = {2025-11-06},
  archiveprefix = {arXiv},
  collaboration = {DESI}
}

@article{DESI:2025fii,
  title = {Extended {{Dark Energy}} Analysis Using {{DESI DR2 BAO}} Measurements},
  author = {Lodha, K. and Calderon, R. and Matthewson, W. L. and Shafieloo, A. and Ishak, M. and Pan, J. and {Garcia-Quintero}, C. and Huterer, D. and Valogiannis, G. and {Ure{\~n}a-L{\'o}pez}, L. A. and Kamble, N. V. and Parkinson, D. and Kim, A. G. and Zhao, G. B. and {Cervantes-Cota}, J. L. and Rohlf, J. and {Lozano-Rodr{\'i}guez}, F. and {Rom{\'a}n-Herrera}, J. O. and {Abdul-Karim}, M. and Aguilar, J. and Ahlen, S. and Alves, O. and Andrade, U. and Armengaud, E. and Aviles, A. and BenZvi, S. and Bianchi, D. and Brodzeller, A. and Brooks, D. and Burtin, E. and Canning, R. and Rosell, A. Carnero and Casas, L. and Castander, F. J. and Charles, M. and Chaussidon, E. and {Chaves-Montero}, J. and Chebat, D. and Claybaugh, T. and Cole, S. and Cuceu, A. and Dawson, K. S. and de la Macorra, A. and de Mattia, A. and Deiosso, N. and Demina, R. and Dey, Arjun and Dey, Biprateep and Ding, Z. and Doel, P. and Eisenstein, D. J. and Elbers, W. and Ferraro, S. and {Font-Ribera}, A. and {Forero-Romero}, J. E. and Garrison, Lehman H. and Gazta{\~n}aga, E. and {Gil-Mar{\'i}n}, H. and Gontcho, S. Gontcho A. and {Gonzalez-Morales}, A. X. and Gutierrez, G. and Guy, J. and Hahn, C. and Herbold, M. and {Herrera-Alcantar}, H. K. and Honscheid, K. and Howlett, C. and Juneau, S. and Kehoe, R. and Kirkby, D. and Kisner, T. and Kremin, A. and Lahav, O. and Lamman, C. and Landriau, M. and Guillou, L. Le and Leauthaud, A. and Levi, M. E. and Li, Q. and Magneville, C. and Manera, M. and Martini, P. and Meisner, A. and {Mena-Fern{\'a}ndez}, J. and Miquel, R. and Moustakas, J. and Santos, D. Mu{\~n}oz and {Mu{\~n}oz-Guti{\'e}rrez}, A. and Myers, A. D. and Nadathur, S. and Niz, G. and Noriega, H. E. and Paillas, E. and {Palanque-Delabrouille}, N. and Percival, W. J. and Pieri, Matthew M. and Poppett, C. and Prada, F. and {P{\'e}rez-Fern{\'a}ndez}, A. and {P{\'e}rez-R{\`a}fols}, I. and {Ram{\'i}rez-P{\'e}rez}, C. and Rashkovetskyi, M. and Ravoux, C. and Ross, A. J. and Rossi, G. and {Ruhlmann-Kleider}, V. and Samushia, L. and Sanchez, E. and Schlegel, D. and Schubnell, M. and Seo, H. and Sinigaglia, F. and Sprayberry, D. and Tan, T. and Tarl{\'e}, G. and Taylor, P. and Turner, W. and {Vargas-Maga{\~n}a}, M. and Walther, M. and Weaver, B. A. and Wolfson, M. and Y{\`e}che, C. and Zarrouk, P. and Zhou, R. and Zou, H.},
  year = 2025,
  month = oct,
  journal = {Phys. Rev. D},
  volume = {112},
  number = {8},
  eprint = {2503.14743},
  primaryclass = {astro-ph.CO},
  pages = {083511},
  doi = {10.1103/w4c6-1r5j},
  urldate = {2025-04-11},
  archiveprefix = {arXiv},
  collaboration = {DESI}
}

@article{DESI:2025hao,
  title = {Frequentist {{Cosmological Constraints}} from {{Full-Shape Clustering Measurements}} in {{DESI DR1}}},
  author = {Morawetz, J. and others},
  year = 2025,
  month = aug,
  journal = {arXiv:2508.11811 [astro-ph.CO]},
  eprint = {2508.11811},
  primaryclass = {astro-ph.CO},
  archiveprefix = {arXiv},
  collaboration = {DESI}
}

@article{DESI:2025wyn,
  title = {Dynamical {{Dark Energy}} in Light of the {{DESI DR2 Baryonic Acoustic Oscillations Measurements}}},
  author = {Gu, Gan and others},
  year = 2025,
  month = sep,
  journal = {Nature Astron.},
  eprint = {2504.06118},
  primaryclass = {astro-ph.CO},
  doi = {10.1038/s41550-025-02669-6},
  archiveprefix = {arXiv},
  collaboration = {DESI}
}

@article{DESI:2025zgx,
  title = {{{DESI DR2 Results II}}: {{Measurements}} of {{Baryon Acoustic Oscillations}} and {{Cosmological Constraints}}},
  shorttitle = {{{DESI DR2 Results II}}},
  author = {Abdul Karim, M. and others},
  year = 2025,
  month = oct,
  journal = {Phys. Rev. D},
  volume = {112},
  number = {8},
  eprint = {2503.14738},
  primaryclass = {astro-ph.CO},
  pages = {083515},
  doi = {10.1103/tr6y-kpc6},
  archiveprefix = {arXiv},
  collaboration = {DESI}
}

@article{DiValentino:2019qzk,
  title = {Planck Evidence for a Closed {{Universe}} and a Possible Crisis for Cosmology},
  author = {Di Valentino, Eleonora and Melchiorri, Alessandro and Silk, Joseph},
  year = 2019,
  month = nov,
  journal = {Nature Astron.},
  volume = {4},
  number = {2},
  eprint = {1911.02087},
  primaryclass = {astro-ph.CO},
  pages = {196--203},
  publisher = {Nature Publishing Group},
  issn = {2397-3366},
  doi = {10.1038/s41550-019-0906-9},
  urldate = {2025-11-05},
  archiveprefix = {arXiv},
  copyright = {2019 The Author(s), under exclusive licence to Springer Nature Limited},
  langid = {english}
}

@article{Dossett:2012kd,
  title = {Spatial Curvature and Cosmological Tests of General Relativity},
  author = {Dossett, Jason N. and Ishak, Mustapha},
  year = 2012,
  journal = {Phys. Rev. D},
  volume = {86},
  number = {10},
  eprint = {1205.2422},
  primaryclass = {astro-ph.CO},
  pages = {103008},
  publisher = {American Physical Society},
  doi = {10.1103/PhysRevD.86.103008},
  urldate = {2025-11-05},
  archiveprefix = {arXiv}
}

@article{Gallego:2024gay,
  title = {Anisotropic Dark Energy from String Compactifications},
  author = {Gallego, Diego and {Orjuela-Quintana}, J. Bayron and {Valenzuela-Toledo}, C{\'e}sar A.},
  year = 2024,
  month = apr,
  journal = {JHEP},
  volume = {04},
  number = {4},
  eprint = {2402.09570},
  primaryclass = {hep-th},
  pages = {131},
  issn = {1029-8479},
  doi = {10.1007/JHEP04(2024)131},
  urldate = {2024-06-04},
  archiveprefix = {arXiv},
  langid = {english}
}

@article{Gebhardt:2021vfo,
  title = {The {{Hobby}}--{{Eberly Telescope Dark Energy Experiment}} ({{HETDEX}}) {{Survey Design}}, {{Reductions}}, and {{Detections}}*},
  author = {Gebhardt, Karl and Cooper, Erin Mentuch and Ciardullo, Robin and Acquaviva, Viviana and Bender, Ralf and Bowman, William P. and Castanheira, Barbara G. and Dalton, Gavin and Davis, Dustin and de Jong, Roelof S. and DePoy, D. L. and Devarakonda, Yaswant and Dongsheng, Sun and Drory, Niv and Fabricius, Maximilian and Farrow, Daniel J. and Feldmeier, John and Finkelstein, Steven L. and Froning, Cynthia S. and Gawiser, Eric and Gronwall, Caryl and Herold, Laura and Hill, Gary J. and Hopp, Ulrich and House, Lindsay R. and Janowiecki, Steven and Jarvis, Matthew and Jeong, Donghui and Jogee, Shardha and Kakuma, Ryota and Kelz, Andreas and Kollatschny, W. and Komatsu, Eiichiro and Krumpe, Mirko and Landriau, Martin and Liu, Chenxu and Niemeyer, Maja Lujan and MacQueen, Phillip and Marshall, Jennifer and Mawatari, Ken and McLinden, Emily M. and Mukae, Shiro and Nagaraj, Gautam and Ono, Yoshiaki and Ouchi, Masami and Papovich, Casey and Sakai, Nao and Saito, Shun and Schneider, Donald P. and Schulze, Andreas and Shanmugasundararaj, Khavvia and Shetrone, Matthew and Sneden, Chris and Snigula, Jan and Steinmetz, Matthias and Thomas, Benjamin P. and Thomas, Brianna and Tuttle, Sarah and Urrutia, Tanya and Wisotzki, Lutz and Wold, Isak and Zeimann, Gregory and Zhang, Yechi},
  year = 2021,
  month = dec,
  journal = {Astrophys. J.},
  volume = {923},
  number = {2},
  eprint = {2110.04298},
  primaryclass = {astro-ph.IM},
  pages = {217},
  publisher = {The American Astronomical Society},
  issn = {0004-637X},
  doi = {10.3847/1538-4357/ac2e03},
  urldate = {2022-11-09},
  archiveprefix = {arXiv},
  langid = {english}
}

@article{Gong:2007wx,
  title = {Dark {{Energy}} and {{Cosmic Curvature}}: {{Monte Carlo Markov Chain Approach}}},
  author = {Gong, Yungui and Wu, Qiang and {Anzhong Wang} and Wang, Anzhong},
  year = 2008,
  journal = {Astrophys. J.},
  volume = {681},
  number = {1},
  eprint = {0708.1817},
  primaryclass = {astro-ph},
  pages = {27--39},
  doi = {10.1086/588598},
  archiveprefix = {arXiv}
}

@article{Gonzalez:2021ojp,
  title = {Testing the Consistency between Cosmological Data: The Impact of Spatial Curvature and the Dark Energy {{EoS}}},
  shorttitle = {Testing the Consistency between Cosmological Data},
  author = {Gonzalez, Javier E. and Benetti, Micol and {von Marttens}, Rodrigo and Alcaniz, Jailson},
  year = 2021,
  month = nov,
  journal = {JCAP},
  volume = {11},
  number = {11},
  eprint = {2104.13455},
  primaryclass = {astro-ph.CO},
  pages = {060},
  publisher = {IOP Publishing},
  issn = {1475-7516},
  doi = {10.1088/1475-7516/2021/11/060},
  urldate = {2025-11-05},
  archiveprefix = {arXiv},
  langid = {english}
}

@article{Gosenca:2015qha,
  title = {Dynamical {{Analysis}} of {{Scalar Field Cosmologies}} with {{Spatial Curvature}}},
  author = {Gosenca, Mateja and Coles, Peter},
  year = 2016,
  journal = {Open J. Astrophys.},
  volume = {1},
  number = {1},
  eprint = {1502.04020},
  primaryclass = {gr-qc},
  pages = {1},
  publisher = {Maynooth Academic Publishing},
  doi = {10.21105/astro.1502.04020},
  urldate = {2024-06-19},
  archiveprefix = {arXiv},
  langid = {english}
}

@article{Guth:2012ww,
  title = {What Can the Observation of Nonzero Curvature Tell Us?},
  author = {Guth, Alan H. and Nomura, Yasunori},
  year = 2012,
  journal = {Phys. Rev. D},
  volume = {86},
  number = {2},
  eprint = {1203.6876},
  primaryclass = {hep-th},
  pages = {023534},
  publisher = {American Physical Society},
  doi = {10.1103/PhysRevD.86.023534},
  urldate = {2025-11-07},
  archiveprefix = {arXiv}
}

@article{Halliwell:1986ja,
  title = {Scalar Fields in Cosmology with an Exponential Potential},
  author = {Halliwell, J. J.},
  year = 1987,
  journal = {Phys. Lett. B},
  volume = {185},
  number = {3},
  pages = {341},
  issn = {0370-2693},
  doi = {10.1016/0370-2693(87)91011-2},
  urldate = {2025-01-01}
}

@article{Handley:2019tkm,
  title = {Curvature Tension: {{Evidence}} for a Closed Universe},
  shorttitle = {Curvature Tension},
  author = {Handley, Will},
  year = 2021,
  month = feb,
  journal = {Phys. Rev. D},
  volume = {103},
  number = {4},
  eprint = {1908.09139},
  primaryclass = {astro-ph.CO},
  pages = {L041301},
  publisher = {American Physical Society},
  doi = {10.1103/PhysRevD.103.L041301},
  urldate = {2025-11-05},
  archiveprefix = {arXiv}
}

@article{Horn:2017kmv,
  title = {Positive Curvature and Scalar Field Tunneling in the Landscape},
  author = {Horn, Bart},
  year = 2019,
  month = jan,
  journal = {Phys. Rev. D},
  volume = {99},
  number = {2},
  eprint = {1707.03851},
  primaryclass = {hep-th},
  pages = {025010},
  publisher = {American Physical Society},
  doi = {10.1103/PhysRevD.99.025010},
  urldate = {2025-01-17},
  archiveprefix = {arXiv}
}

@article{Kaloper:2005aj,
  title = {Of Pngb Quintessence},
  author = {Kaloper, Nemanja and Sorbo, Lorenzo},
  year = 2006,
  journal = {JCAP},
  volume = {04},
  eprint = {astro-ph/0511543},
  pages = {007},
  doi = {10.1088/1475-7516/2006/04/007},
  archiveprefix = {arXiv}
}

@article{Kaloper:2008qs,
  title = {Where in the {{String Landscape}} Is {{Quintessence}}},
  author = {Kaloper, Nemanja and Sorbo, Lorenzo},
  year = 2009,
  journal = {Phys. Rev. D},
  volume = {79},
  eprint = {0810.5346},
  primaryclass = {hep-th},
  pages = {043528},
  doi = {10.1103/PhysRevD.79.043528},
  archiveprefix = {arXiv}
}

@article{kamionkowskiDarkEnergyString2014b,
  ids = {kamionkowskiDarkEnergyString2014},
  title = {Dark {{Energy}} from the {{String Axiverse}}},
  author = {Kamionkowski, Marc and Pradler, Josef and Walker, Devin G. E.},
  year = 2014,
  month = dec,
  journal = {Phys. Rev. Lett.},
  volume = {113},
  number = {25},
  eprint = {1409.0549},
  primaryclass = {hep-ph},
  pages = {251302},
  publisher = {American Physical Society},
  doi = {10.1103/PhysRevLett.113.251302},
  urldate = {2022-10-12},
  archiveprefix = {arXiv}
}

@article{Kleban:2012ph,
  title = {Spatial Curvature Falsifies Eternal Inflation},
  author = {Kleban, Matthew and Schillo, Marjorie},
  year = 2012,
  journal = {JCAP},
  volume = {06},
  number = {06},
  eprint = {1202.5037},
  primaryclass = {astro-ph.CO},
  pages = {029},
  issn = {1475-7516},
  doi = {10.1088/1475-7516/2012/06/029},
  urldate = {2025-11-07},
  archiveprefix = {arXiv},
  langid = {english}
}

@article{Komatsu:2014ioa,
  title = {Results from the {{Wilkinson Microwave Anisotropy Probe}}},
  author = {Komatsu, Eiichiro and Bennett, Charles L. and Barnes, C. and Bean, R. and Bennett, C. L. and Dor{\'e}, O. and Dunkley, J. and Gold, B. and Greason, M. R. and Halpern, M. and Hill, R. S. and Hinshaw, G. and Jarosik, N. and Kogut, A. and Komatsu, E. and Larson, D. and Limon, M. and Meyer, S. S. and Nolta, M. R. and Odegard, N. and Page, L. and Peiris, H. V. and Smith, K. M. and Spergel, D. N. and Tucker, G. S. and Verde, L. and Weiland, J. L. and Wollack, E. and Wright, E. L. and {(on behalf of the WMAP science team)}},
  year = 2014,
  journal = {PTEP},
  volume = {2014},
  number = {6},
  eprint = {1404.5415},
  primaryclass = {astro-ph.CO},
  pages = {06B102},
  issn = {2050-3911},
  doi = {10.1093/ptep/ptu083},
  urldate = {2024-06-03},
  archiveprefix = {arXiv},
  collaboration = {WMAP Science Team}
}

@article{Leonard:2016evk,
  title = {Spatial Curvature Endgame: {{Reaching}} the Limit of Curvature Determination},
  shorttitle = {Spatial Curvature Endgame},
  author = {Leonard, C. Danielle and Bull, Philip and Allison, Rupert},
  year = 2016,
  month = jul,
  journal = {Phys. Rev. D},
  volume = {94},
  number = {2},
  eprint = {1604.01410},
  primaryclass = {astro-ph.CO},
  pages = {023502},
  issn = {2470-0010, 2470-0029},
  doi = {10.1103/PhysRevD.94.023502},
  urldate = {2025-01-02},
  archiveprefix = {arXiv},
  copyright = {http://link.aps.org/licenses/aps-default-license},
  langid = {english}
}

@article{Martin:2012bt,
  title = {{Everything you always wanted to know about the cosmological constant problem (but were afraid to ask)}},
  author = {Martin, J{\'e}r{\^o}me},
  year = 2012,
  journal = {Comptes Rendus Physique},
  volume = {13},
  number = {6-7},
  eprint = {1205.3365},
  primaryclass = {astro-ph.CO},
  pages = {566--665},
  publisher = {arXiv},
  issn = {1878-1535},
  doi = {10.1016/j.crhy.2012.04.008},
  urldate = {2024-06-03},
  archiveprefix = {arXiv},
  langid = {french}
}

@article{McDonough:2022pku,
  title = {Towards {{Early Dark Energy}} in {{String Theory}}},
  author = {McDonough, Evan and Scalisi, Marco},
  year = 2023,
  month = oct,
  journal = {JHEP},
  volume = {10},
  eprint = {2209.00011},
  primaryclass = {hep-th},
  pages = {118},
  doi = {10.1007/JHEP10(2023)118},
  urldate = {2022-10-12},
  archiveprefix = {arXiv}
}

@article{Obied:2018sgi,
  title = {De {{Sitter Space}} and the {{Swampland}}},
  author = {Obied, Georges and Ooguri, Hirosi and Spodyneiko, Lev and Vafa, Cumrun},
  year = 2018,
  month = jul,
  journal = {arXiv:1806.08362 [hep-th]},
  eprint = {1806.08362},
  primaryclass = {hep-th},
  doi = {10.48550/arXiv.1806.08362},
  urldate = {2025-04-26},
  archiveprefix = {arXiv}
}

@article{Ooguri:2006in,
  ids = {ooguriGeometryStringLandscape2007b},
  title = {On the Geometry of the String Landscape and the Swampland},
  author = {Ooguri, Hirosi and Vafa, Cumrun},
  year = 2007,
  journal = {Nucl. Phys. B},
  volume = {766},
  number = {1},
  eprint = {hep-th/0605264},
  pages = {21--33},
  issn = {0550-3213},
  doi = {10.1016/j.nuclphysb.2006.10.033},
  urldate = {2022-11-09},
  archiveprefix = {arXiv},
  langid = {english}
}

@article{Ooguri:2016pdq,
  title = {Non-Supersymmetric {{AdS}} and the {{Swampland}}},
  author = {Ooguri, Hirosi and Vafa, Cumrun},
  year = 2017,
  journal = {Adv. Theor. Math. Phys.},
  volume = {21},
  eprint = {1610.01533},
  primaryclass = {hep-th},
  pages = {1787--1801},
  doi = {10.4310/ATMP.2017.v21.n7.a8},
  urldate = {2022-11-09},
  archiveprefix = {arXiv}
}

@article{Padmanabhan:2002ji,
  ids = {padmanabhanCosmologicalConstantWeight2003},
  title = {Cosmological Constant---the Weight of the Vacuum},
  author = {Padmanabhan, T.},
  year = 2003,
  journal = {Phys. Rept.},
  volume = {380},
  number = {5},
  eprint = {hep-th/0212290},
  pages = {235--320},
  issn = {0370-1573},
  doi = {10.1016/S0370-1573(03)00120-0},
  urldate = {2024-06-03},
  archiveprefix = {arXiv}
}

@article{Palti:2019pca,
  title = {The {{Swampland}}: {{Introduction}} and {{Review}}},
  shorttitle = {The {{Swampland}}},
  author = {Palti, Eran},
  year = 2019,
  month = jun,
  journal = {Fortsch. Phys.},
  volume = {67},
  number = {6},
  eprint = {1903.06239},
  primaryclass = {hep-th},
  pages = {1900037},
  issn = {1521-3978},
  doi = {10.1002/prop.201900037},
  urldate = {2022-11-16},
  archiveprefix = {arXiv},
  langid = {english}
}

@article{Patil:2024mno,
  title = {Role of {{Spatial Curvature}} in a {{Dark Energy Interacting Model}}},
  author = {Patil, Trupti and Panda, Sukanta},
  year = 2025,
  month = jan,
  journal = {Eur. Phys. J. Plus},
  volume = {140},
  number = {1},
  eprint = {2405.20170},
  primaryclass = {astro-ph.CO},
  pages = {48},
  doi = {10.1140/epjp/s13360-025-05977-y},
  urldate = {2024-12-30},
  archiveprefix = {arXiv}
}

@article{Peebles:1987ek,
  title = {Cosmology with a Time-Variable Cosmological 'Constant'},
  author = {Peebles, P. J. E. and Ratra, Bharat},
  year = 1988,
  journal = {Astrophys. J. Lett.},
  volume = {325},
  pages = {L17},
  issn = {0004-637X, 1538-4357},
  doi = {10.1086/185100},
  urldate = {2024-06-04},
  langid = {english}
}

@article{Peebles:2002gy,
  ids = {peeblesCosmologicalConstantDark2003},
  title = {The Cosmological Constant and Dark Energy},
  author = {Peebles, P. J. E. and Ratra, Bharat},
  year = 2003,
  journal = {Rev. Mod. Phys.},
  volume = {75},
  number = {2},
  eprint = {astro-ph/0207347},
  pages = {559--606},
  publisher = {American Physical Society},
  issn = {0034-6861, 1539-0756},
  doi = {10.1103/RevModPhys.75.559},
  urldate = {2024-06-04},
  archiveprefix = {arXiv},
  copyright = {http://link.aps.org/licenses/aps-default-license},
  langid = {english}
}

@article{Perivolaropoulos:2021jda,
  title = {Challenges for {{$\Lambda$CDM}}: {{An}} Update},
  shorttitle = {Challenges for {{$\Lambda$CDM}}},
  author = {Perivolaropoulos, L. and Skara, F.},
  year = 2022,
  month = dec,
  journal = {New Astron. Rev.},
  volume = {95},
  eprint = {2105.05208},
  primaryclass = {astro-ph.CO},
  pages = {101659},
  issn = {1387-6473},
  doi = {10.1016/j.newar.2022.101659},
  urldate = {2022-12-14},
  archiveprefix = {arXiv},
  langid = {english}
}

@article{Planck:2018vyg,
  title = {{\emph{Planck}} 2018 Results: {{VI}}. {{Cosmological}} Parameters},
  shorttitle = {{\emph{Planck}} 2018 Results},
  author = {Aghanim, N. and Akrami, Y. and Ashdown, M. and others},
  year = 2020,
  month = sep,
  journal = {Astron. Astrophys.},
  volume = {641},
  eprint = {1807.06209},
  primaryclass = {astro-ph.CO},
  pages = {A6},
  issn = {0004-6361, 1432-0746},
  doi = {10.1051/0004-6361/201833910},
  urldate = {2024-06-03},
  archiveprefix = {arXiv},
  collaboration = {Planck},
  copyright = {https://www.edpsciences.org/en/authors/copyright-and-licensing}
}

@article{Ratra:1987rm,
  title = {Cosmological Consequences of a Rolling Homogeneous Scalar Field},
  author = {Ratra, Bharat and Peebles, P. J. E.},
  year = 1988,
  journal = {Phys. Rev. D},
  volume = {37},
  number = {12},
  pages = {3406},
  publisher = {American Physical Society},
  doi = {10.1103/PhysRevD.37.3406},
  urldate = {2025-01-01}
}

@article{Revello:2023hro,
  title = {Attractive (s)Axions: Cosmological Trackers at the Boundary of Moduli Space},
  shorttitle = {Attractive (s)Axions},
  author = {Revello, Filippo},
  year = 2024,
  month = may,
  journal = {JHEP},
  volume = {05},
  number = {5},
  eprint = {2311.12429},
  primaryclass = {hep-th},
  pages = {037},
  issn = {1029-8479},
  doi = {10.1007/JHEP05(2024)037},
  urldate = {2026-01-16},
  archiveprefix = {arXiv},
  langid = {english}
}

@article{Riess:2016jrr,
  title = {A 2.4\% {{DETERMINATION OF THE LOCAL VALUE OF THE HUBBLE CONSTANT}}},
  author = {Riess, Adam G. and Macri, Lucas M. and Hoffmann, Samantha L. and Scolnic, Dan and Casertano, Stefano and Filippenko, Alexei V. and Tucker, Brad E. and Reid, Mark J. and Jones, David O. and Silverman, Jeffrey M. and Chornock, Ryan and Challis, Peter and Yuan, Wenlong and Brown, Peter J. and Foley, Ryan J.},
  year = 2016,
  month = jul,
  journal = {Astrophys. J.},
  volume = {826},
  number = {1},
  eprint = {1604.01424},
  primaryclass = {astro-ph.CO},
  pages = {56},
  issn = {1538-4357},
  doi = {10.3847/0004-637X/826/1/56},
  urldate = {2022-11-29},
  archiveprefix = {arXiv}
}

@article{Riess:2020fzl,
  title = {Cosmic {{Distances Calibrated}} to 1\% {{Precision}} with {{Gaia EDR3 Parallaxes}} and {{Hubble Space Telescope Photometry}} of 75 {{Milky Way Cepheids Confirm Tension}} with {{$\Lambda$CDM}}},
  author = {Riess, Adam G. and Casertano, Stefano and Yuan, Wenlong and Bowers, J. Bradley and Macri, Lucas and Zinn, Joel C. and Scolnic, Dan},
  year = 2021,
  month = feb,
  journal = {Astrophys. J. Lett.},
  volume = {908},
  number = {1},
  eprint = {2012.08534},
  primaryclass = {astro-ph.CO},
  pages = {L6},
  publisher = {The American Astronomical Society},
  issn = {2041-8205},
  doi = {10.3847/2041-8213/abdbaf},
  urldate = {2022-12-06},
  archiveprefix = {arXiv},
  langid = {english}
}

@article{Schoneberg:2021qvd,
  title = {The {{H0 Olympics}}: {{A}} Fair Ranking of Proposed Models},
  shorttitle = {The {{H0 Olympics}}},
  author = {Sch{\"o}neberg, Nils and Abell{\'a}n, Guillermo Franco and S{\'a}nchez, Andrea P{\'e}rez and Witte, Samuel J. and Poulin, Vivian and Lesgourgues, Julien},
  year = 2022,
  month = oct,
  journal = {Phys. Rept.},
  series = {The {{H}}\_0 {{Olympics}}: {{A}} Fair Ranking of Proposed Models},
  volume = {984},
  eprint = {2107.10291},
  primaryclass = {astro-ph.CO},
  pages = {2228},
  issn = {0370-1573},
  doi = {10.1016/j.physrep.2022.07.001},
  urldate = {2022-11-30},
  archiveprefix = {arXiv},
  langid = {english}
}

@article{Sonner:2006yn,
  title = {Recurrent Acceleration in Dilaton-Axion Cosmology},
  author = {Sonner, Julian and Townsend, Paul K.},
  year = 2006,
  journal = {Phys. Rev. D},
  volume = {74},
  number = {10},
  eprint = {hep-th/0608068},
  pages = {103508},
  publisher = {American Physical Society},
  doi = {10.1103/PhysRevD.74.103508},
  urldate = {2023-02-07},
  archiveprefix = {arXiv}
}

@article{SPT-3G:2025bzu,
  title = {{{SPT-3G D1}}: {{CMB}} Temperature and Polarization Power Spectra and Cosmology from 2019 and 2020 Observations of the {{SPT-3G Main}} Field},
  shorttitle = {{{SPT-3G D1}}},
  author = {Camphuis, E. and others},
  year = 2025,
  month = jun,
  journal = {arXiv:2506.20707 [astro-ph.CO]},
  eprint = {2506.20707},
  primaryclass = {astro-ph.CO},
  archiveprefix = {arXiv},
  collaboration = {SPT-3G}
}

@article{SPT:2019fqo,
  title = {Constraints on {{Cosmological Parameters}} from the 500 Deg2 {{SPTPOL Lensing Power Spectrum}}},
  author = {Bianchini, F. and Wu, W. L. K. and Ade, P. A. R. and Anderson, A. J. and Austermann, J. E. and Avva, J. S. and Beall, J. A. and Bender, A. N. and Benson, B. A. and Bleem, L. E. and Carlstrom, J. E. and Chang, C. L. and Chaubal, P. and Chiang, H. C. and Citron, R. and Moran, C. Corbett and Crawford, T. M. and Crites, A. T. and {de Haan}, T. and Dobbs, M. A. and Everett, W. and Gallicchio, J. and George, E. M. and Gilbert, A. and Gupta, N. and Halverson, N. W. and Harrington, N. and Henning, J. W. and Hilton, G. C. and Holder, G. P. and Holzapfel, W. L. and Hrubes, J. D. and Huang, N. and Hubmayr, J. and Irwin, K. D. and Knox, L. and Lee, A. T. and Li, D. and Lowitz, A. and Manzotti, A. and McMahon, J. J. and Meyer, S. S. and Millea, M. and Mocanu, L. M. and Montgomery, J. and Nadolski, A. and Natoli, T. and Nibarger, J. P. and Noble, G. and Novosad, V. and Omori, Y. and Padin, S. and Patil, S. and Pryke, C. and Reichardt, C. L. and Ruhl, J. E. and Saliwanchik, B. R. and Sayre, J. T. and Schaffer, K. K. and Sievers, C. and Simard, G. and Smecher, G. and Stark, A. A. and Story, K. T. and Tucker, C. and Vanderlinde, K. and Veach, T. and Vieira, J. D. and Wang, G. and Whitehorn, N. and Yefremenko, V.},
  year = 2020,
  month = jan,
  journal = {Astrophys. J.},
  volume = {888},
  number = {2},
  eprint = {1910.07157},
  primaryclass = {astro-ph.CO},
  pages = {119},
  publisher = {The American Astronomical Society},
  issn = {0004-637X},
  doi = {10.3847/1538-4357/ab6082},
  urldate = {2025-11-05},
  archiveprefix = {arXiv},
  collaboration = {SPT},
  langid = {english}
}

@article{SupernovaCosmologyProject:1998vns,
  title = {Measurements of {{\textohm}} and {{$\Lambda$}} from 42 {{High-Redshift Supernovae}}},
  author = {Perlmutter, S. and Aldering, G. and Goldhaber, G. and Knop, R. A. and Nugent, P. and Castro, P. G. and Deustua, S. and Fabbro, S. and Goobar, A. and Groom, D. E. and Hook, I. M. and Kim, A. G. and Kim, M. Y. and Lee, J. C. and Nunes, N. J. and Pain, R. and Pennypacker, C. R. and Quimby, R. and Lidman, C. and Ellis, R. S. and Irwin, M. and McMahon, R. G. and {Ruiz-Lapuente}, P. and Walton, N. and Schaefer, B. and Boyle, B. J. and Filippenko, A. V. and Matheson, T. and Fruchter, A. S. and Panagia, N. and Newberg, H. J. M. and Couch, W. J. and Project, The Supernova Cosmology},
  year = 1999,
  journal = {Astrophys. J.},
  volume = {517},
  number = {2},
  eprint = {astro-ph/9812133},
  pages = {565--586},
  publisher = {IOP Publishing},
  issn = {0004-637X},
  doi = {10.1086/307221},
  urldate = {2022-11-02},
  archiveprefix = {arXiv},
  collaboration = {Supernova Cosmology Project},
  langid = {english}
}

@article{SupernovaSearchTeam:1998fmf,
  title = {Observational {{Evidence}} from {{Supernovae}} for an {{Accelerating Universe}} and a {{Cosmological Constant}}},
  author = {Riess, Adam G. and Filippenko, Alexei V. and Challis, Peter and Clocchiatti, Alejandro and Diercks, Alan and Garnavich, Peter M. and Gilliland, Ron L. and Hogan, Craig J. and Jha, Saurabh and Kirshner, Robert P. and Leibundgut, B. and Phillips, M. M. and Reiss, David and Schmidt, Brian P. and Schommer, Robert A. and Smith, R. Chris and Spyromilio, J. and Stubbs, Christopher and Suntzeff, Nicholas B. and Tonry, John},
  year = 1998,
  journal = {Astron. J.},
  volume = {116},
  number = {3},
  eprint = {astro-ph/9805201},
  pages = {1009--1038},
  publisher = {IOP Publishing},
  issn = {1538-3881},
  doi = {10.1086/300499},
  urldate = {2022-11-02},
  archiveprefix = {arXiv},
  collaboration = {Supernova Search Team},
  langid = {english}
}

@article{Takada:2015mma,
  title = {Geometrical Constraint on Curvature with {{BAO}} Experiments},
  author = {Takada, Masahiro and Dor{\'e}, Olivier},
  year = 2015,
  month = dec,
  journal = {Phys. Rev. D},
  volume = {92},
  number = {12},
  eprint = {1508.02469},
  primaryclass = {astro-ph.CO},
  pages = {123518},
  publisher = {American Physical Society},
  doi = {10.1103/PhysRevD.92.123518},
  urldate = {2024-12-30},
  archiveprefix = {arXiv}
}

@article{Vafa:2005ui,
  title = {The {{String Landscape}} and the {{Swampland}}},
  author = {Vafa, Cumrun},
  year = 2005,
  month = oct,
  journal = {arXiv:hep-th/0509212},
  eprint = {hep-th/0509212},
  doi = {10.48550/arXiv.hep-th/0509212},
  urldate = {2022-11-09},
  archiveprefix = {arXiv}
}

@article{Vagnozzi:2020rcz,
  title = {The Galaxy Power Spectrum Take on Spatial Curvature and Cosmic Concordance},
  author = {Vagnozzi, Sunny and Di Valentino, Eleonora and Gariazzo, Stefano and Melchiorri, Alessandro and Mena, Olga and Silk, Joseph},
  year = 2021,
  month = sep,
  journal = {Phys. Dark Univ.},
  volume = {33},
  eprint = {2010.02230},
  primaryclass = {astro-ph.CO},
  pages = {100851},
  issn = {2212-6864},
  doi = {10.1016/j.dark.2021.100851},
  urldate = {2025-11-05},
  archiveprefix = {arXiv}
}

@article{vandeBruck:2019vzd,
  title = {Dark Energy, the Swampland, and the Equivalence Principle},
  author = {{van de Bruck}, Carsten and Thomas, Cameron C.},
  year = 2019,
  month = jul,
  journal = {Phys. Rev. D},
  volume = {100},
  number = {2},
  eprint = {1904.07082},
  primaryclass = {hep-th},
  pages = {023515},
  publisher = {American Physical Society},
  doi = {10.1103/PhysRevD.100.023515},
  urldate = {2023-07-05},
  archiveprefix = {arXiv}
}

@article{vandenHoogen:1999qq,
  title = {Scaling Solutions in {{Robertson-Walker}} Spacetimes},
  author = {van den Hoogen, R. J. and Coley, A. A. and Wands, D.},
  year = 1999,
  journal = {Class. Quant. Grav.},
  volume = {16},
  number = {6},
  eprint = {gr-qc/9901014},
  pages = {1843--1851},
  issn = {0264-9381},
  doi = {10.1088/0264-9381/16/6/317},
  urldate = {2024-12-30},
  archiveprefix = {arXiv},
  langid = {english}
}

@article{VanRiet:2023pnx,
  title = {Beginners Lectures on Flux Compactifications and Related {{Swampland}} Topics},
  author = {Van Riet, Thomas and Zoccarato, Gianluca},
  year = 2023,
  month = nov,
  journal = {Phys. Rept.},
  volume = {1049},
  eprint = {2305.01722},
  primaryclass = {hep-th},
  pages = {2296},
  doi = {10.1016/j.physrep.2023.11.003},
  urldate = {2023-11-30},
  archiveprefix = {arXiv}
}

@article{Verde:2019ivm,
  title = {Tensions between the Early and Late {{Universe}}},
  author = {Verde, Licia and Treu, Tommaso and Riess, Adam G.},
  year = 2019,
  month = sep,
  journal = {Nature Astron.},
  volume = {3},
  number = {10},
  eprint = {1907.10625},
  primaryclass = {astro-ph.CO},
  pages = {891},
  publisher = {Nature Publishing Group},
  issn = {2397-3366},
  doi = {10.1038/s41550-019-0902-0},
  urldate = {2022-11-02},
  archiveprefix = {arXiv},
  copyright = {2019 Springer Nature Limited},
  langid = {english}
}

@article{Virey:2008nu,
  title = {On the Determination of Curvature and Dynamical Dark Energy},
  author = {Virey, J.-M. and {Talon-Esmieu}, D. and Ealet, A. and Taxil, P. and Tilquin, A.},
  year = 2008,
  journal = {JCAP},
  volume = {12},
  number = {12},
  eprint = {0802.4407},
  primaryclass = {astro-ph},
  pages = {008},
  issn = {1475-7516},
  doi = {10.1088/1475-7516/2008/12/008},
  urldate = {2024-12-30},
  archiveprefix = {arXiv},
  langid = {english}
}

@article{Wetterich:1987fm,
  title = {Cosmology and the Fate of Dilatation Symmetry},
  author = {Wetterich, C.},
  year = 1988,
  journal = {Nucl. Phys. B},
  volume = {302},
  number = {4},
  eprint = {1711.03844},
  primaryclass = {hep-th},
  pages = {668--696},
  issn = {0550-3213},
  doi = {10.1016/0550-3213(88)90193-9},
  urldate = {2026-01-17},
  archiveprefix = {arXiv}
}

@article{Wetterich:1994bg,
  title = {The {{Cosmon}} Model for an Asymptotically Vanishing Time Dependent Cosmological 'Constant'},
  author = {Wetterich, Christof},
   doi = {https://doi.org/10.48550/arXiv.hep-th/9408025},
  year = 1995,
  journal = {Astron. Astrophys.},
  volume = {301},
  eprint = {hep-th/9408025},
  pages = {321--328},
  archiveprefix = {arXiv}
}

@article{WMAP:2012nax,
  title = {{{NINE-YEAR WILKINSON MICROWAVE ANISOTROPY PROBE}} ({{WMAP}}) {{OBSERVATIONS}}: {{COSMOLOGICAL PARAMETER RESULTS}}},
  shorttitle = {{{NINE-YEAR WILKINSON MICROWAVE ANISOTROPY PROBE}} ({{WMAP}}) {{OBSERVATIONS}}},
  author = {Hinshaw, G. and Larson, D. and Komatsu, E. and Spergel, D. N. and Bennett, C. L. and Dunkley, J. and Nolta, M. R. and Halpern, M. and Hill, R. S. and Odegard, N. and Page, L. and Smith, K. M. and Weiland, J. L. and Gold, B. and Jarosik, N. and Kogut, A. and Limon, M. and Meyer, S. S. and Tucker, G. S. and Wollack, E. and Wright, E. L.},
  year = 2013,
  month = sep,
  journal = {Astrophys. J. Suppl.},
  volume = {208},
  number = {2},
  eprint = {1212.5226},
  primaryclass = {astro-ph.CO},
  pages = {19},
  publisher = {The American Astronomical Society},
  issn = {0067-0049},
  doi = {10.1088/0067-0049/208/2/19},
  urldate = {2024-06-03},
  archiveprefix = {arXiv},
  collaboration = {WMAP},
  langid = {english}
}

@article{Yang:2022kho,
  title = {Revealing the Effects of Curvature on the Cosmological Models},
  author = {Yang, Weiqiang and Giar{\`e}, William and Pan, Supriya and Di Valentino, Eleonora and Melchiorri, Alessandro and Silk, Joseph},
  year = 2023,
  month = mar,
  journal = {Phys. Rev. D},
  volume = {107},
  number = {6},
  eprint = {2210.09865},
  primaryclass = {astro-ph.CO},
  pages = {063509},
  publisher = {American Physical Society},
  doi = {10.1103/PhysRevD.107.063509},
  urldate = {2025-11-05},
  archiveprefix = {arXiv}
}

@article{Zlatev:1998tr,
  ids = {zlatevQuintessenceCosmicCoincidence1999a},
  title = {Quintessence, Cosmic Coincidence, and the Cosmological Constant},
  author = {Zlatev, Ivaylo and Wang, Limin and Steinhardt, Paul J.},
  year = 1999,
  journal = {Phys. Rev. Lett.},
  volume = {82},
  number = {5},
  eprint = {astro-ph/9807002},
  pages = {896--899},
  publisher = {American Physical Society},
  doi = {10.1103/PhysRevLett.82.896},
  archiveprefix = {arXiv}
}

@article{Alvarez:2019ues,
    author = "{\'A}lvarez, Miguel and Orjuela-Quintana, J. Bayron and Rodriguez, Yeinzon and Valenzuela-Toledo, Cesar A.",
    title = "{Einstein Yang{\textendash}Mills Higgs dark energy revisited}",
    eprint = "1901.04624",
    archivePrefix = "arXiv",
    primaryClass = "gr-qc",
    reportNumber = "PI/UAN-2019-645FT",
    doi = "10.1088/1361-6382/ab3775",
    journal = "Class. Quant. Grav.",
    volume = "36",
    number = "19",
    pages = "195004",
    year = "2019"
}

@article{Lesgourgues:2011re,
    author = "Lesgourgues, Julien",
    title = "{The Cosmic Linear Anisotropy Solving System (CLASS) I: Overview}",
    eprint = "1104.2932",
    archivePrefix = "arXiv",
    primaryClass = "astro-ph.IM",
    month = "4",
    year = "2011"
}

@article{Blas:2011rf,
    author = "Blas, Diego and Lesgourgues, Julien and Tram, Thomas",
    title = "{The Cosmic Linear Anisotropy Solving System (CLASS) II: Approximation schemes}",
    eprint = "1104.2933",
    archivePrefix = "arXiv",
    primaryClass = "astro-ph.CO",
    reportNumber = "CERN-PH-TH-2011-082, LAPTH-010-11",
    doi = "10.1088/1475-7516/2011/07/034",
    journal = "JCAP",
    volume = "07",
    pages = "034",
    year = "2011"
}

@article{Brinckmann:2018cvx,
    author = "Brinckmann, Thejs and Lesgourgues, Julien",
    title = "{MontePython 3: boosted MCMC sampler and other features}",
    eprint = "1804.07261",
    archivePrefix = "arXiv",
    primaryClass = "astro-ph.CO",
    reportNumber = "TTK-18-15",
    doi = "10.1016/j.dark.2018.100260",
    journal = "Phys. Dark Univ.",
    volume = "24",
    pages = "100260",
    year = "2019"
}

@article{Brout:2022vxf,
    author = "Brout, Dillon and others",
    title = "{The Pantheon+ Analysis: Cosmological Constraints}",
    eprint = "2202.04077",
    archivePrefix = "arXiv",
    primaryClass = "astro-ph.CO",
    doi = "10.3847/1538-4357/ac8e04",
    journal = "Astrophys. J.",
    volume = "938",
    number = "2",
    pages = "110",
    year = "2022"
}

@ARTICLE{2011MNRAS.416.3017B,
       author = {{Beutler}, Florian and {Blake}, Chris and {Colless}, Matthew and {Jones}, D. Heath and {Staveley-Smith}, Lister and {Campbell}, Lachlan and {Parker}, Quentin and {Saunders}, Will and {Watson}, Fred},
        title = "{The 6dF Galaxy Survey: baryon acoustic oscillations and the local Hubble constant}",
      journal = "Mon. Not. Roy. Astron. Soc.",
         year = 2011,
        month = oct,
       volume = {416},
       number = {4},
        pages = {3017-3032},
          doi = {10.1111/j.1365-2966.2011.19250.x},
archivePrefix = {arXiv},
       eprint = {1106.3366},
 primaryClass = {astro-ph.CO},
       adsurl = {https://ui.adsabs.harvard.edu/abs/2011MNRAS.416.3017B}
}

@article{Ross:2014qpa,
    author = "Ross, Ashley J. and Samushia, Lado and Howlett, Cullan and Percival, Will J. and Burden, Angela and Manera, Marc",
    title = "{The clustering of the SDSS DR7 main Galaxy sample {\textendash} I. A 4 per cent distance measure at $z = 0.15$}",
    eprint = "1409.3242",
    archivePrefix = "arXiv",
    primaryClass = "astro-ph.CO",
    doi = "10.1093/mnras/stv154",
    journal = "Mon. Not. Roy. Astron. Soc.",
    volume = "449",
    number = "1",
    pages = "835--847",
    year = "2015"
}

@article{Guo:2015gpa,
    author = "Guo, Rui-Yun and Zhang, Xin",
    title = "{Constraining dark energy with Hubble parameter measurements: an analysis including future redshift-drift observations}",
    eprint = "1512.07703",
    archivePrefix = "arXiv",
    primaryClass = "astro-ph.CO",
    doi = "10.1140/epjc/s10052-016-4016-x",
    journal = "Eur. Phys. J. C",
    volume = "76",
    number = "3",
    pages = "163",
    year = "2016"
}

@article{Moresco:2016mzx,
    author = "Moresco, Michele and Pozzetti, Lucia and Cimatti, Andrea and Jimenez, Raul and Maraston, Claudia and Verde, Licia and Thomas, Daniel and Citro, Annalisa and Tojeiro, Rita and Wilkinson, David",
    title = "{A 6{\%} measurement of the Hubble parameter at $z\sim0.45$: direct evidence of the epoch of cosmic re-acceleration}",
    eprint = "1601.01701",
    archivePrefix = "arXiv",
    primaryClass = "astro-ph.CO",
    doi = "10.1088/1475-7516/2016/05/014",
    journal = "JCAP",
    volume = "05",
    pages = "014",
    year = "2016"
}

@article{Chevallier:2000qy,
    author = "Chevallier, Michel and Polarski, David",
    title = "{Accelerating universes with scaling dark matter}",
    eprint = "gr-qc/0009008",
    archivePrefix = "arXiv",
    doi = "10.1142/S0218271801000822",
    journal = "Int. J. Mod. Phys. D",
    volume = "10",
    pages = "213--224",
    year = "2001"
}

@article{Linder:2002et,
    author = "Linder, Eric V.",
    title = "{Exploring the expansion history of the universe}",
    eprint = "astro-ph/0208512",
    archivePrefix = "arXiv",
    doi = "10.1103/PhysRevLett.90.091301",
    journal = "Phys. Rev. Lett.",
    volume = "90",
    pages = "091301",
    year = "2003"
}

@article{Caldwell:2005tm,
    author = "Caldwell, R. R. and Linder, Eric V.",
    title = "{The Limits of quintessence}",
    eprint = "astro-ph/0505494",
    archivePrefix = "arXiv",
    doi = "10.1103/PhysRevLett.95.141301",
    journal = "Phys. Rev. Lett.",
    volume = "95",
    pages = "141301",
    year = "2005"
}

@article{Garcia-Serna:2023xfw,
    author = "Garc{\'\i}a-Serna, Santiago and Orjuela-Quintana, J. Bayron and Valenzuela-Toledo, C{\'e}sar A. and Ocampo-Dur{\'a}n, Hern{\'a}n",
    title = "{Reconstructing the parameter space of nonanalytical cosmological fixed points}",
    eprint = "2302.09181",
    archivePrefix = "arXiv",
    primaryClass = "gr-qc",
    doi = "10.1142/S0218271823500736",
    journal = "Int. J. Mod. Phys. D",
    volume = "32",
    number = "11",
    pages = "2350073",
    year = "2023"
}

@article{Bamba:2012cp,
    author = "Bamba, Kazuharu and Capozziello, Salvatore and Nojiri, Shin'ichi and Odintsov, Sergei D.",
    title = "{Dark energy cosmology: the equivalent description via different theoretical models and cosmography tests}",
    eprint = "1205.3421",
    archivePrefix = "arXiv",
    primaryClass = "gr-qc",
    doi = "10.1007/s10509-012-1181-8",
    journal = "Astrophys. Space Sci.",
    volume = "342",
    pages = "155--228",
    year = "2012"
}

@article{Vagnozzi:2017ovm,
    author = "Vagnozzi, Sunny and Giusarma, Elena and Mena, Olga and Freese, Katherine and Gerbino, Martina and Ho, Shirley and Lattanzi, Massimiliano",
    title = "{Unveiling $\nu$ secrets with cosmological data: neutrino masses and mass hierarchy}",
    eprint = "1701.08172",
    archivePrefix = "arXiv",
    primaryClass = "astro-ph.CO",
    doi = "10.1103/PhysRevD.96.123503",
    journal = "Phys. Rev. D",
    volume = "96",
    number = "12",
    pages = "123503",
    year = "2017"
}

@article{Vagnozzi:2020dfn,
    author = "Vagnozzi, Sunny and Loeb, Abraham and Moresco, Michele",
    title = "{Eppur {\`e} piatto? The Cosmic Chronometers Take on Spatial Curvature and Cosmic Concordance}",
    eprint = "2011.11645",
    archivePrefix = "arXiv",
    primaryClass = "astro-ph.CO",
    doi = "10.3847/1538-4357/abd4df",
    journal = "Astrophys. J.",
    volume = "908",
    number = "1",
    pages = "84",
    year = "2021"
}

@article{Dhawan:2021mel,
    author = "Dhawan, Suhail and Alsing, Justin and Vagnozzi, Sunny",
    title = "{Non-parametric spatial curvature inference using late-Universe cosmological probes}",
    eprint = "2104.02485",
    archivePrefix = "arXiv",
    primaryClass = "astro-ph.CO",
    doi = "10.1093/mnrasl/slab058",
    journal = "Mon. Not. Roy. Astron. Soc.",
    volume = "506",
    number = "1",
    pages = "L1--L5",
    year = "2021"
}

@article{Jiang:2024xnu,
    author = "Jiang, Jun-Qian and Pedrotti, Davide and da Costa, Simony Santos and Vagnozzi, Sunny",
    title = "{Nonparametric late-time expansion history reconstruction and implications for the Hubble tension in light of recent DESI and type Ia supernovae data}",
    eprint = "2408.02365",
    archivePrefix = "arXiv",
    primaryClass = "astro-ph.CO",
    doi = "10.1103/PhysRevD.110.123519",
    journal = "Phys. Rev. D",
    volume = "110",
    number = "12",
    pages = "123519",
    year = "2024"
}

@article{Elizalde:2004mq,
    author = "Elizalde, Emilio and Nojiri, Shin'ichi and Odintsov, Sergei D.",
    title = "{Late-time cosmology in (phantom) scalar-tensor theory: Dark energy and the cosmic speed-up}",
    eprint = "hep-th/0405034",
    archivePrefix = "arXiv",
    doi = "10.1103/PhysRevD.70.043539",
    journal = "Phys. Rev. D",
    volume = "70",
    pages = "043539",
    year = "2004"
}

@article{Berera:2025jbj,
    author = "Berera, Arjun and Bernardo, Heliudson and Brahma, Suddhasattwa and Calder{\'o}n-Figueroa, Jaime and O. Ramos, Rudnei and Toomey, Michael W.",
    title = "{Heterotic Warm Inflation}",
    eprint = "2509.15125",
    archivePrefix = "arXiv",
    primaryClass = "hep-th",
    reportNumber = "MIT-CTP/5905",
    month = "9",
    year = "2025",
    volume={2026},
   ISSN={1029-8479},
   url={http://dx.doi.org/10.1007/JHEP06(2026)099},
   DOI={10.1007/jhep06(2026)099},
   number={6},
   journal={Journal of High Energy Physics},
   publisher={Springer Science and Business Media LLC},
}

@article{Alexander:2022own,
    author = "Alexander, Stephon and Bernardo, Heliudson and Toomey, Michael W.",
    title = "{Addressing the Hubble and S $_{8}$ tensions with a kinetically mixed dark sector}",
    eprint = "2207.13086",
    archivePrefix = "arXiv",
    primaryClass = "astro-ph.CO",
    doi = "10.1088/1475-7516/2023/03/037",
    journal = "JCAP",
    volume = "03",
    pages = "037",
    year = "2023"
}

@article{Smith:2024ibv,
    author = "Smith, Adam and Mylova, Maria and Brax, Philippe and van de Bruck, Carsten and Burgess, C. P. and Davis, Anne-Christine",
    title = "{A Minimal Axio-dilaton Dark Sector}",
    eprint = "2410.11099",
    archivePrefix = "arXiv",
    primaryClass = "hep-th",
    doi = "10.1088/1475-7516/2025/07/023",
    month = "10",
    year = "2024"
}

@article{Burgess:2021obw,
    author = "Burgess, C. P. and Dineen, Danielle and Quevedo, F.",
    title = "{Yoga Dark Energy: natural relaxation and other dark implications of a supersymmetric gravity sector}",
    eprint = "2111.07286",
    archivePrefix = "arXiv",
    primaryClass = "hep-th",
    reportNumber = "CERN-TH-2021-192",
    doi = "10.1088/1475-7516/2022/03/064",
    journal = "JCAP",
    volume = "03",
    number = "03",
    pages = "064",
    year = "2022"
}

@article{Smith:2025uaq,
    author = "Smith, Adam and Mylova, Maria and van de Bruck, Carsten and Burgess, C. P. and Di Valentino, Eleonora",
    title = "{The Serendipitous Axiodilaton: A Self-Consistent Recombination-Era Solution to the Hubble Tension}",
    eprint = "2512.13544",
    archivePrefix = "arXiv",
    primaryClass = "astro-ph.CO",
    month = "12",
    year = "2025"
}

\end{document}